\documentclass[preprint,12pt]{aastex}
\usepackage{emulateapj5}

\slugcomment{\footnotesize {\LaTeX}ed at \number\time\ min., \today}
\shorttitle{Tracing the Mass during Low-Mass Star Formation, IV }
\shortauthors{Young et al.}

\begin{document}


\title {\bf Tracing the Mass during Low-Mass Star Formation, IV: Observations and Modeling of the Submillimeter Continuum Emission 
from Class I Protostars}
\author {Chadwick H. Young, Yancy L. Shirley, Neal J. Evans II}
\affil{Department of Astronomy, The University of Texas at Austin,
       Austin, Texas 78712--1083}
\email{cyoung, yshirley, nje@astro.as.utexas.edu}
\and
\author{Jonathan M. C. Rawlings}
\affil{Department of Physics and Astronomy, University College London,
        Gower Street, London WC1E 6BT}
\email{jcr@star.ucl.ac.uk}


\begin{abstract}
We present results from the observations and modeling of seventeen Class I cores with the Submillimetre Common Users Bolometer Array 
(SCUBA) on the James Clerk Maxwell Telescope (JCMT).  Our sample consists of cores with 64$<T_{bol}$(K)$<$270,   
$0.2<L_{obs}/$L$_\odot<12$, and $50<L_{obs}/L_{smm}<1000$.    
By modeling the transfer of radiation through 
the envelope for nine cores, we find, for a power law distribution $n(r)=n_f(r/r_f)^{-p}$, the average and standard deviation
$\langle p\rangle=1.6\pm0.4$ and a median of $p=1.8$. However, the
inclusion of a disk or other point-like component can cause the derived $p$ to be shallower by as much as 0.5.  
We discuss uncertainties due to  the interstellar radiation field (ISRF), disks, dust opacity, and outer radii in our 
modeling results.  We find no evidence for a truncated outer radius or radially variant dust properties in most sources.  Uncertainty in 
the strength of the ISRF and 
possible existence of a disk contribute the greatest uncertainty in $p$.  
In addition, we test the Shu collapse model for our sources and discuss the application of simpler analyses that derive a density
power law distribution directly from the slope of the intensity radial profile.   
The total mass of the envelope in our sample has a range of   $0.04<M_{env}/$M$_\odot<5.0$, but these masses disagree with the
virial masses derived from molecular line observations, indicating that observations of molecular lines 
do not trace the mass in some Class I cores.  We also discuss several sources individually.
In particular, IRAS 03256+3055, with its unique morphology,  is an ideal object for testing theories of fragmentation 
in the formation of low-mass protostars.  Also, we note the possibility, through some simple calculations, that IRAS 04385+2550
is a young, forming substellar object.  Finally, we discuss the nature of 
these sources in light of various evolutionary indicators and find that $T_{bol}$ and $L_{obs}/L_{smm}$ are often inconsistent
in distinguishing Class 0 from Class I cores. We note that, in this sample, the $L_{obs}/L_{smm}$
criterion redefines many of these Class I sources (by $T_{bol}$) as Class 0 sources.
 
\end{abstract}

\keywords{stars: formation, low-mass  --- ISM: dust ---
ISM: individual (CB230, L1251B, IRAS 04166+2706, IRAS 04169+2702, IRAS 04239+2436, IRAS 04248+2612, 
IRAS 04295+2251, IRAS 04361+2547, IRAS 04381+2540, IRAS 04385+2550, IRAS 03256+3055)}

\section{Introduction}
                    
The formation of low-mass stars has received considerable attention, but the 
process by which material moves from the outer envelope to the central protostar is still debated.   Theories 
differ, primarily, in their prediction of the density, $n(r)$,  and velocity, $\vec{v}(r)$, as a function of radius, $r$.  
Larson (1969) and Penston (1969) proposed a model with
initial conditions of constant density and zero velocity.  Their model encounters problems due to its prediction of 
supersonic infall velocities that are not observed in regions of low-mass star formation (Zhou 1992).  
Shu (1977) proposed a scenario where infall begins  from the inside, i.e.,  inside-out collapse.  
The initial density configuration is a singular isothermal sphere (SIS) with $n(r)\propto r^{-2}$.  
Collapse begins at the center and propagates outward
at the sound speed, $a_{eff}$, leaving behind a density distribution that approaches $n(r) \propto r^{-1.5}$.  
Some perturbations to the Shu model include
rotation (Terebey, Shu, \& Cassen 1984) and magnetic fields (Galli \& Shu 1993).  
McLaughlin \& Pudritz (1997) proposed a different model in which accretion increases with time.  Their
scenario was similar to Shu (1977)---a collapse front propagating into a static envelope---but, because of differing initial 
conditions and a polytropic equation of state, they predict a much shallower density profile in the static envelope ($n\propto r^{-1}$).  
Foster \& Chevalier (1993)  
conducted hydrodynamical calculations of the collapse of a Bonnor-Ebert sphere and found the density approaches a 
$r^{-2}$ distribution at small radii.
Other models exist with various predictions in the density distribution, $n\propto r^{-p}$.  Therefore,
observational determinations of this density distribution can help us to discern the physical processes governing the formation
of low-mass stars.

Classification of the stages in protostellar evolution has grown more detailed.  Initially, Lada et al. (1987) used the
infrared spectral index to define Classes I-III. Later, Myers \& Ladd (1993) proposed a classification of these forming stars
based on their bolometric temperature ($T_{bol}$); this classification scheme was further codified by Chen et al. (1995).  
Andr\'{e} et al. (1993)  defined a criterion to discern Class 0 from Class I cores: the ratio of the submillimeter 
to  bolometric luminosity.  In this paper, we refer to the inverse of this quantity ($L_{bol}/L_{smm}$),
a quantity that presumably increases with time. 

The luminosity of a low-mass core (protostar$+$envelope) is emitted mostly at  infrared wavelengths due to the reprocessing of 
higher energy radiation from the central protostar by dust grains in the envelope.  These heated grains also emit detectable
radiation  at submillimeter wavelengths.  Because the cloud is optically thin to this radiation,  a measure of this dust emission
renders a sampling of the entire mass in the cloud.  In the past, astronomers have used single bolometers to detect this emission, 
but resolution and sensitivity limited their progress.   
Submillimeter bolometer arrays, such as SCUBA, have allowed us to map this dust emission from the protostellar envelopes
 and, for the first time, draw conclusions about  the nature and distribution of mass within these envelopes.

In this paper, we present the observations and modeling of Class I cores.  We describe our sample selection, observations, 
data reduction, and some basic data analysis in Section 2.  In Section 3, we discuss various aspects of modeling these sources.
power law models were the primary focus, but we also present simpler top hat models of the density distribution.
Also, we report on  Shu collapse models for nine sources.  Additionally, in this third section,
we compare the application of simpler analyses to our more detailed models.  Also, we evaluate some possible scenarios where
a disk is present and how the disk affects the modeled density profile.  In Section 4, we discuss some individual sources.  Section 5 holds
a discussion of masses, classification, and the density structure of Class I protostars.  Finally, in Section 6, we summarize our 
conclusions.  The frugal reader, who wishes only for our major conclusions, can focus primarily on Sections 3, 5, and 6.

\section{SCUBA Observations}

\subsection{Sample}

In Shirley et al. (2000, hereafter, Paper I), we presented the results from observations of a sample ranging from Class $-1$ to
Class I cores.  In that sample, 
the presence of  multiple sources within the observed field was common for the Class I sources but uncommon for the Class 0 
objects.   
Therefore, Shirley et al. suggested that many Class I sources might appear as
such because of the existence of multiple sources within the IRAS beam.  This suggestion  was one motivation for observing this sample of
Class I objects:  to determine the existence and effect of multiplicity.  In addition, we hoped to present a similar treatment 
and analysis as in Papers I-III (Shirley et al., 2000; Evans et al. 2001; Shirley et al. 2002)---i.e., determining source characteristics 
and proposing possible density and temperature profiles through simulations with a one-dimensional radiative transfer code.

We have selected a sample of sixteen sources; these are listed in Table 1 with their positions in B1950.0 and J2000.0 
coordinates.   The positions were obtained, mostly, from the IRAS PSC as well as from 2 $\mu$m observations pursued by Tamura et al. 
(1991) and Moriarty-Schieven et al. (1994).

The distances to these sources range from 140 to 450 parsecs (pc), but most of the objects (13) are  in the nearby Taurus 
star-forming region at 140 pc (Elias 1978).  We adopt a distance of 450 pc  for CB230 (Launhardt \& Henning 1997), 300 pc for 
L1251B (Kun \& Prusti  1993), and 320 pc for IRAS 03256+3055 in Perseus (de Zeeuw 1999).  The distance for Perseus is based
on recent Hipparcos observations and differs from  that used in Papers I \& III (220 pc).

We classify the cores in our sample by their bolometric temperature ($T_{bol}>70$ K for Class I).  Therefore, two objects, L1251B
and IRAS 03256+3055 (with $T_{bol}<70$ K), are not included in the average values for various parameters describing Class I cores.

\subsection{Observations}

We observed this sample of Class I sources with SCUBA, the Submillimeter Continuum Users Bolometer Array (Holland et al. 1999), on the 
James Clerk Maxwell Telescope (JCMT) atop Mauna Kea, Hawaii.  Over four nights in November 1999 and February 2000, 
we mapped each of these cores at 850 and 450 $\mu$m by the 64-point jiggle map method that  fully samples a field of 
2\farcm3.  Since we expected to see very extended emission, we adopted a five-point mapping technique with 
30$\arcsec$ spacings between each of the five 64-point jiggle maps.  By this method, we removed the effects of bad bolometers and 
integrated for a longer time on the central 2$\arcmin$ of the map. 
In some sources, we increased the coverage of our maps to include extended emission.

We performed pointing and skydip measurements between each five-point map.  The average pointing RMS, 
over the four nights of observations, was $\pm$2\farcs0.    
We measured the sky opacity via the skydip method at 450 and 850 $\mu$m ($\langle \tau_{450}\rangle=1.07$ 
and $\langle \tau_{850}\rangle = 0.21$) while also recording the opacity as measured by the CSO radiometer at 225 GHz 
($\langle \tau_{CSO}\rangle=0.05$).  In our calculation of $\tau_{850}$ and $\tau_{450}$, we used the updated hot and cold load 
temperatures as given by Archibald et al. (2000).  In addition, we find a relationship between $\tau_{850}$ and $\tau_{CSO}$
that is in agreement with that presented by Archibald et al. (2000).    
We list all opacity measurements in Table~\ref{opacity}.

\begin{figure}
\figurenum{1}
\plotone{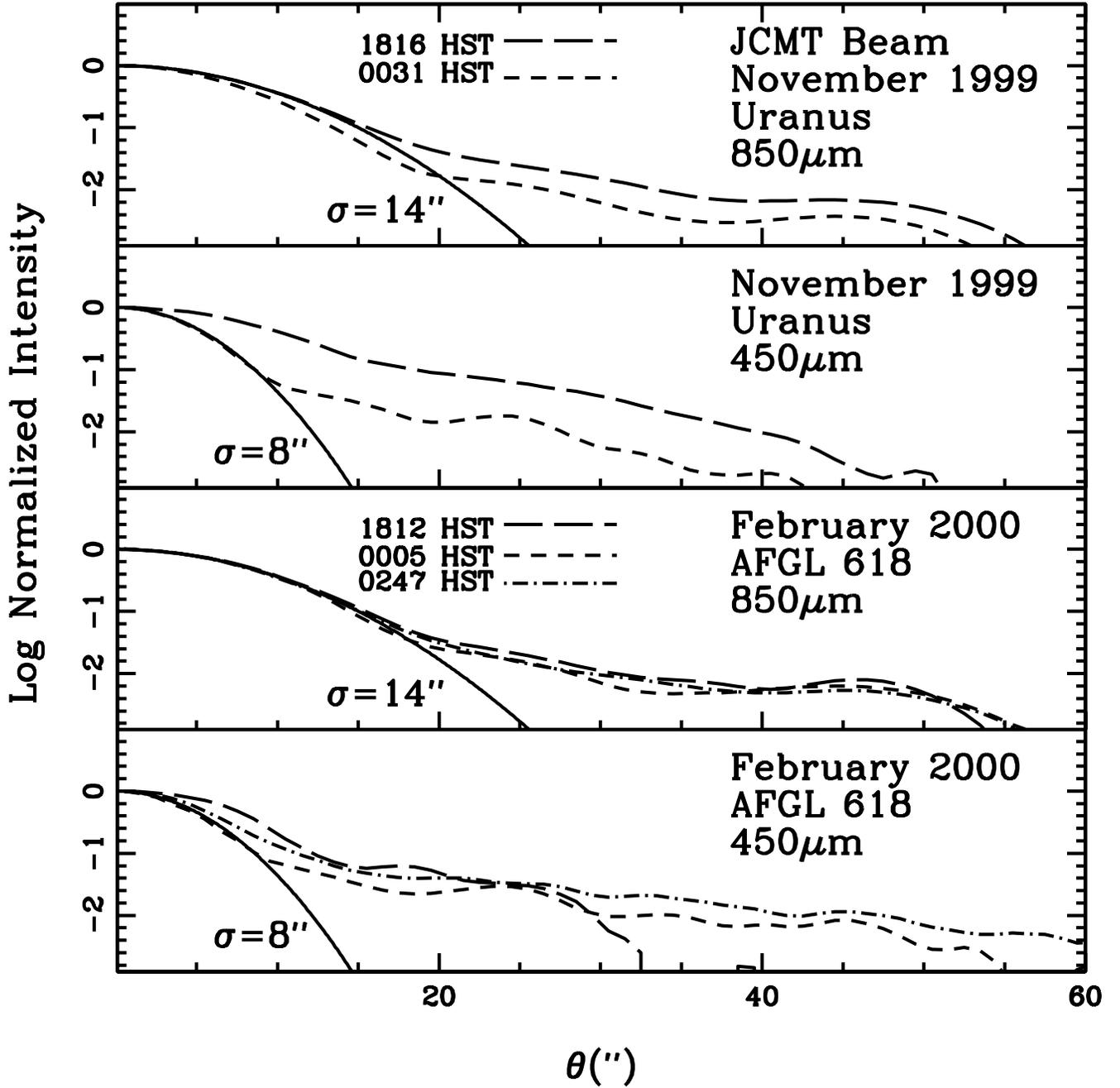}
\figcaption{\label{beams}The observed JCMT beam profiles were obtained through observations of Uranus or AFGL618 during 
	November 1999 and February 2000, respectively.  For each month, we show the beams from one representative night
	as measured at different times (given in Hawaii-Aleutian Standard Time). 
	The solid lines are Gaussian functions with a FWHM of $14\arcsec$  and $8\arcsec$  for 850 and 450 $\mu$m, 
		respectively. }
\end{figure}

Observations of Uranus (November 1999) and AFGL 618 (February 2000) were used to determine the beam profile of the JCMT.  In 
Figure~\ref{beams}, we show the radial intensity profiles of these point sources for various times during the night---our observing
shift was from about 1730-0130, Hawaii-Aleutian Standard Time (HST).  Two points should be considered in regard to these beam profiles:
the significant and variable nature of the sidelobes and the changing FWHM size of the beam ($\theta_{mb}$).
First,  Mueller et al. (2002) noted the significance of sidelobes in their treatment of dust continuum emission from 
massive star-forming regions.  They find
that disregard of these sidelobes and the assumption of a Gaussian beam can cause the derived density power law index to be 
shallower by $\sim$0.4.  Indeed, the beam profiles in Figure~\ref{beams}\ show considerable power (about $1/10$ of the maximum)
in the second and, sometimes, third sidelobes.  Additionally, we find the sidelobes become smaller as the night progresses and
the dish cools.  
Second, for both nights, the 450 $\mu$m beam profile has a large FWHM
($\sim$11$\arcsec$) at about 1800 HST, but the beamsize decreases to $\theta_{mb}\sim8\arcsec$ at about midnight (HST).
For November 1999, as shown in Figure~\ref{beams},
the 450 $\mu$m FWHM decreased by 3$\arcsec$ as the night progressed.   
However, because the S/N is much better and the beam more stable at 850 $\mu$m, we relied more heavily on it in the determination 
of certain
source properties (e.g., the deconvolved FWHM size).  The 850 $\mu$m beam profile is also variable ($\delta FWHM \sim 1-2\arcsec$), 
however, and we propagate this variability as an uncertainty in the measured properties dependent on the beamshape.  
Additionally, in our modeling, we used the observed beam profile that is
closest in time to the source observation.  In one case (CB230), where there were no beam profiles within one hour in time, 
we took the time-weighted average of two neighboring beam profiles;  in all other cases, the beam profiles were taken from
observations of Uranus or AFGL 618 that were within one-half hour of the source observation.  
While the existence of and variations in the sidelobes 
\it significantly affect \rm the analysis of radial intensity profiles, we have taken great care to remove uncertainties due
to these imperfections.

\subsection{Image Reduction \& Calibration}

\begin{figure}
\figurenum{2} 
\centering
 \vspace*{7.8cm}
   \leavevmode
   \includegraphics{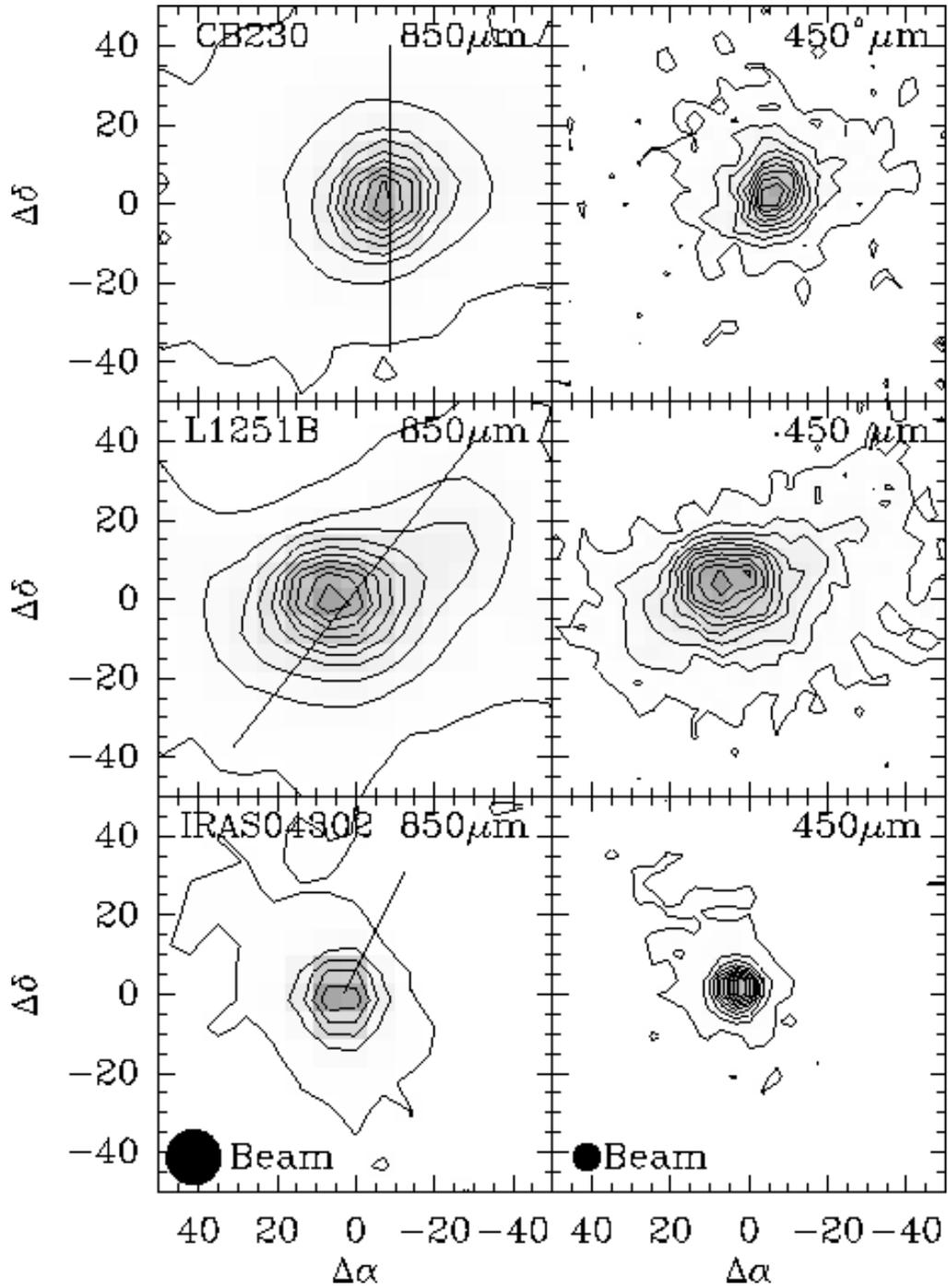}
\vskip 4.25in
\figcaption{\label{contour_a}Maps of emission at 850 and 450 $\mu$m.  For all the maps, the lowest contour is at the 2-$\sigma$ level.  
The contours increase by intervals of ten percent of the peak for CB230 (10-$\sigma$ and 2-$\sigma$ for 850 $\mu$m and 450 $\mu$m)
and L1251B (6-$\sigma$ and 2-$\sigma$ for 850 $\mu$m and 450 $\mu$m).  The contours of IRAS 04302+2247 are shown in increments of 
twenty percent (10-$\sigma$) for the 850 $\mu$m map and ten percent (4-$\sigma$) for the 450 $\mu$m map.  
 Central positions are in Table 1. Outflow directions are labeled with bold lines; references for outflows are as follows: 
	CB230, Yun \& Clemens, 1994.; L1251B, Sato et al., 1994; IRAS 04302+2247, Bontemps, et al. 1996.}
\end{figure}

\begin{figure}
\figurenum{3} 
\centering
 \vspace*{7.8cm}
   \leavevmode
   \includegraphics{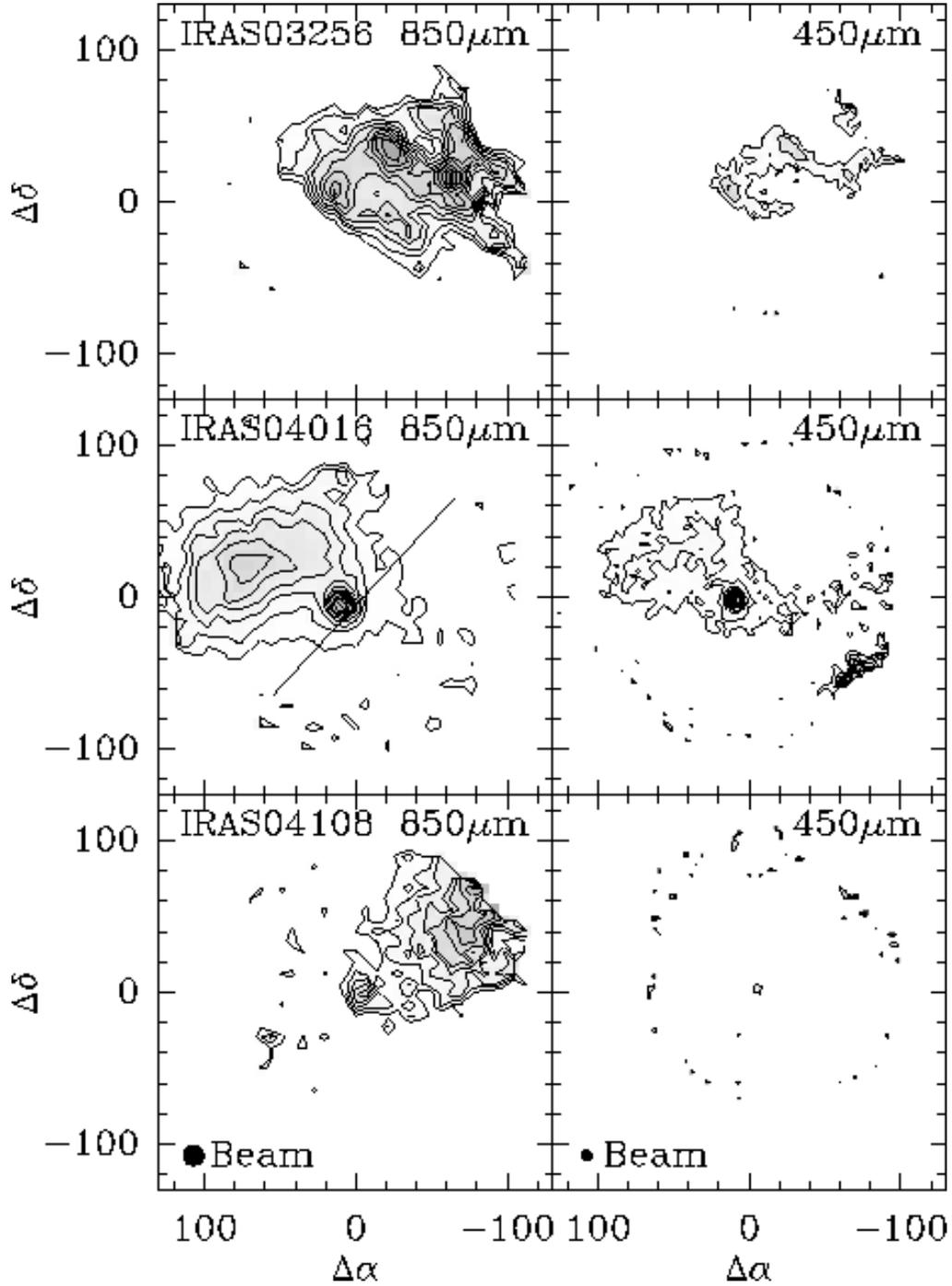}
\vskip 4.25in
\figcaption{\label{contour_b}Maps of emission at 850 and 450 $\mu$m. For all the maps, the lowest contour is at the 2-$\sigma$ level.  
		The 850 $\mu$m contours increase by intervals of ten percent of the peak for IRAS 03256+3055 (1-$\sigma$), 
		IRAS 04016+2610 
		(4-$\sigma$), 
		and IRAS 04108+2803 (2-$\sigma$).  Contours for IRAS 04016+2610 at 450 $\mu$m are drawn at levels of 10 percent 
		(2-$\sigma$).  
		The 450 $\mu$m maps for IRAS 03256+3055 and IRAS 04108+2803 have contours at 1-$\sigma$ intervals.  
            	Central positions are in Table 1.  Outflow directions are labeled with bold lines; references for outflows 
		are as follows: IRAS 04016+2610, Hogerheijde, et al. 1998.}
\end{figure}

\begin{figure}
\figurenum{4} 
\centering
 \vspace*{7.8cm}
   \leavevmode
   \includegraphics{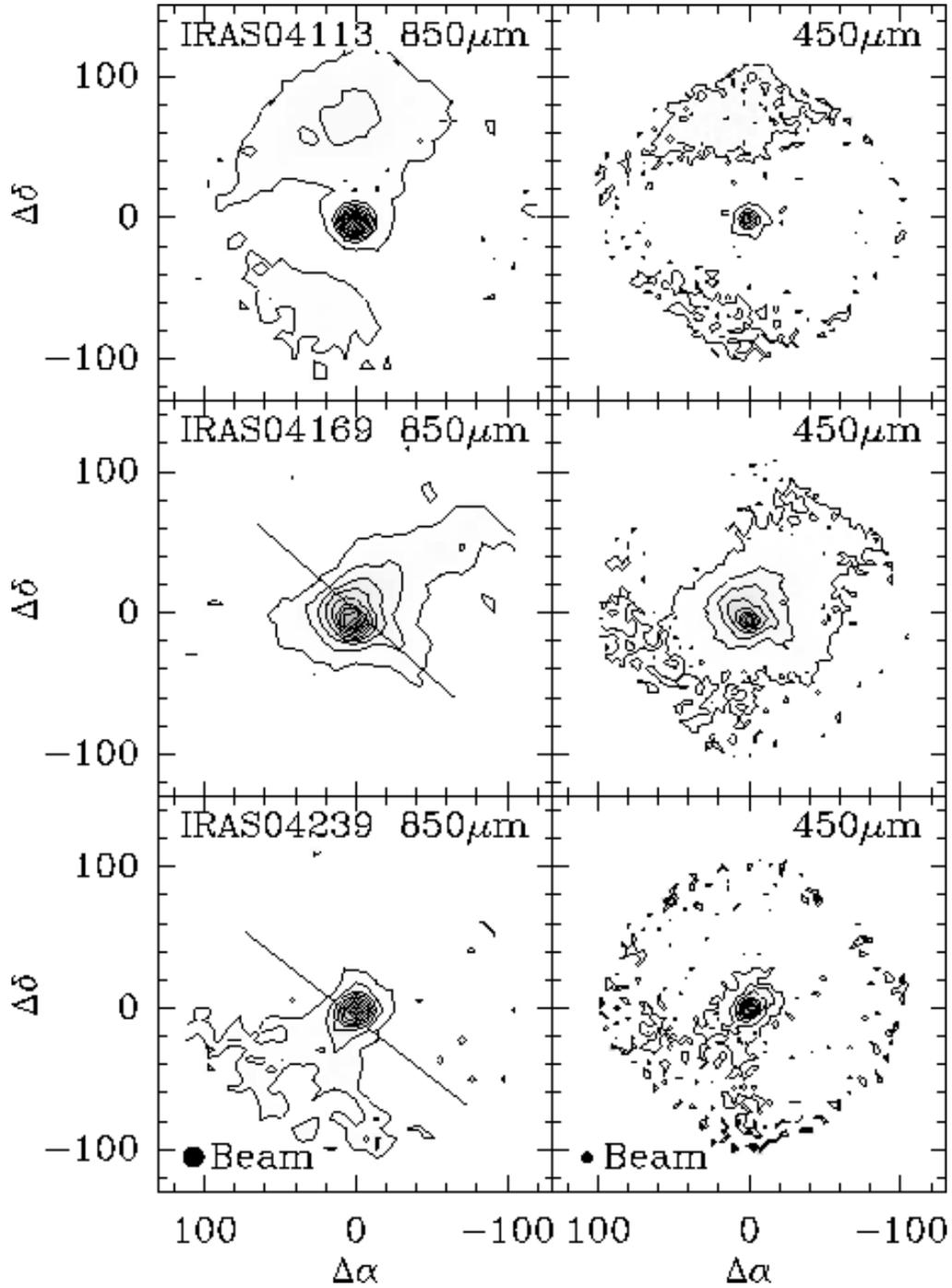}
\vskip 4.25in
\figcaption{\label{contour_c}Maps of emission at 850 and 450 $\mu$m.  For all the maps, the lowest contour is at the 2-$\sigma$ level.
The 850 $\mu$m  contours increase by ten percent of the peak (IRAS 04113+2758, 6-$\sigma$; IRAS 04169+2702, 3-$\sigma$; IRAS 04239+2436, 
2-$\sigma$).  
The 450 $\mu$m contours for IRAS 04113+2758 and IRAS 04169+2702 increase by twenty percent (4- and 6-$\sigma$, respectively) 
and by ten percent 
(2-$\sigma$) for IRAS 04239+2436.   
	    Central positions are in Table 1.   Outflow directions are labeled with bold lines; references for outflows are as 
		follows: IRAS 04169+2702, Bontemps et al., 1996.; 
	    IRAS 04239+2436, Saito et al., 2001. IRAS 04113+2758 shows no detectable outflow activity (Saito et al. 2001).}
\end{figure}

\begin{figure}
\figurenum{5} 
\centering
 \vspace*{7.8cm}
   \leavevmode
   \includegraphics{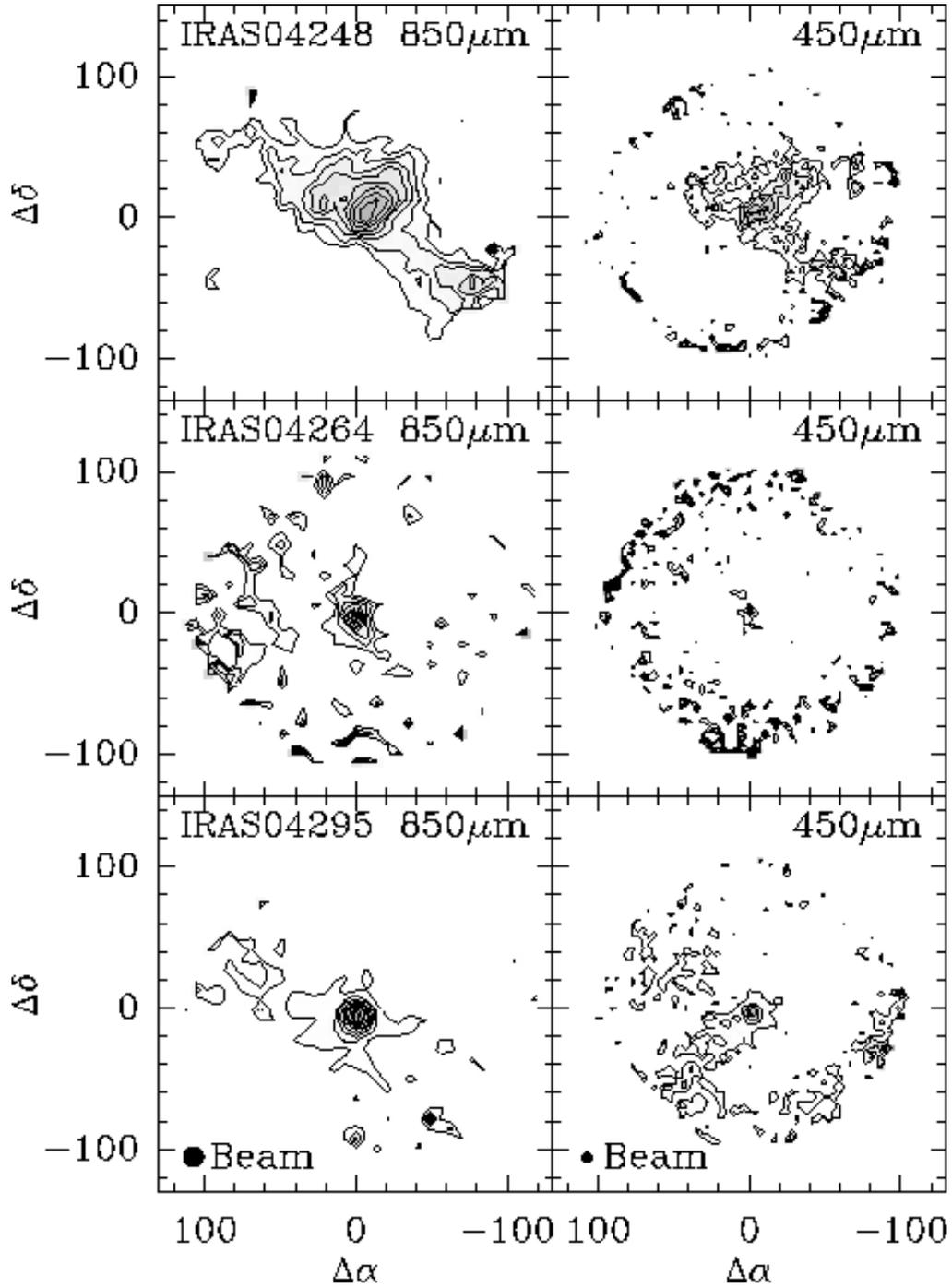}
\vskip 4.25in
\figcaption{\label{contour_d}Maps of emission at 850 and 450 $\mu$m.  For all the maps, the lowest contour is at the 2-$\sigma$ level. 
		For IRAS 04248+2612 and IRAS 04264+2433, the 850 and 450 $\mu$m contours increase by 1-$\sigma$ intervals.  
		IRAS 04295+2251
		has contours that increase by ten percent of the peak (2-$\sigma$) at 850 $\mu$m and twenty percent (3-$\sigma$) at
		450 $\mu$m.    
	    Central positions are in Table 1.}
\end{figure}

\begin{figure}
\figurenum{6} 
\centering
 \vspace*{7.8cm}
   \leavevmode
   \includegraphics{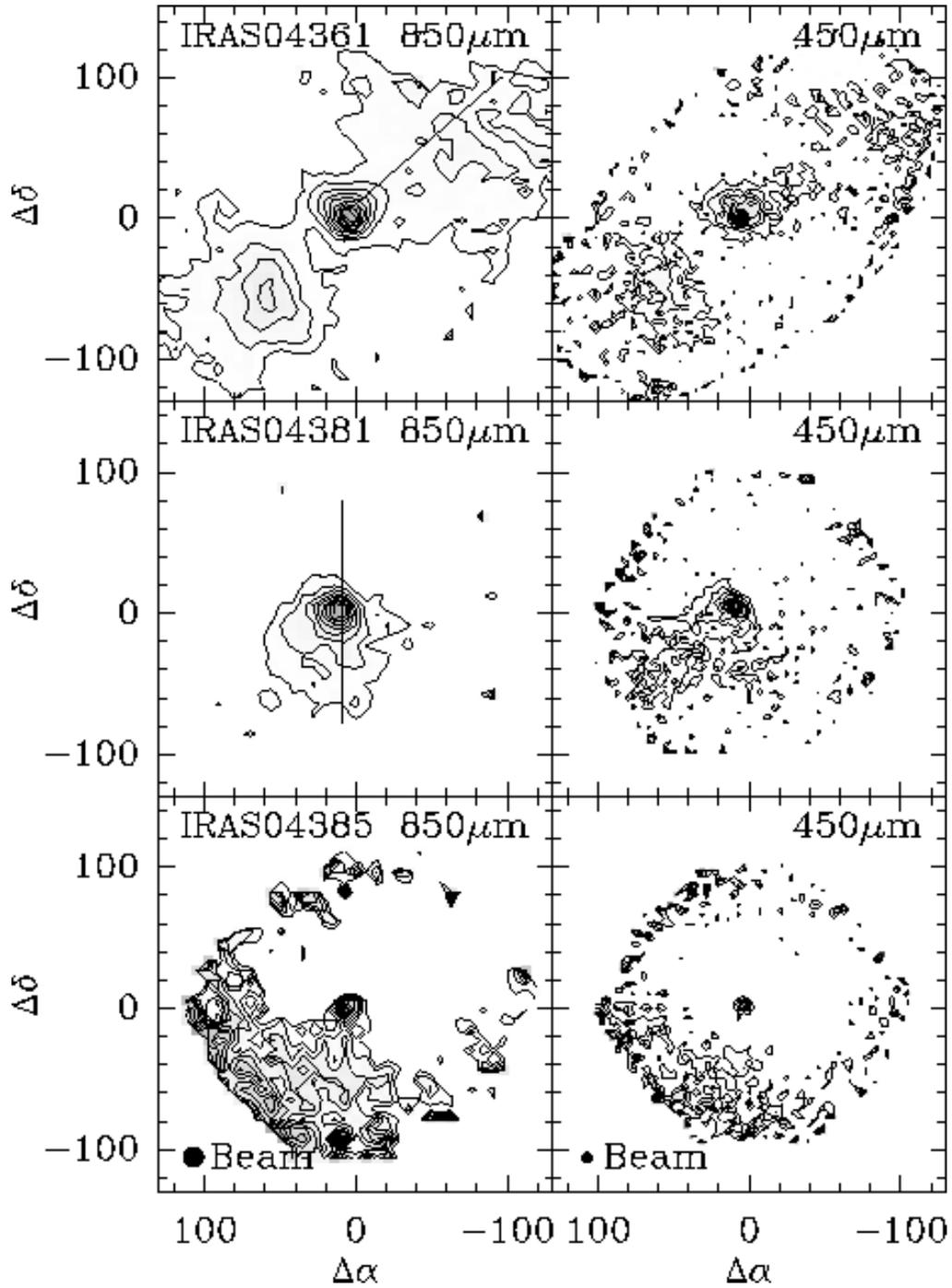}
\vskip 4.25in
\figcaption{\label{contour_e}Maps of emission at 850 and 450 $\mu$m.  For all the maps, the lowest contour is at the 2-$\sigma$ level.
		The 850 and 450 $\mu$m contours increase by ten percent of the peak for IRAS 04361+2547 
		and IRAS 04381+2540 corresponding to
		5- and 2-$\sigma$ for IRAS 04361+2547 (850 and 450 $\mu$m, respectively) and 1-$\sigma$ for IRAS 04381+2540 (both maps).
		The contours for IRAS 04385+2550 are drawn at the 1-$\sigma$ level.	         
	    	Central positions are in Table 1.  Outflow directions are labeled with bold lines; references for outflows are 
		as follows: IRAS 04361+2547 and IRAS 04381+2540, Bontemps et al. 1996.}
\end{figure}

We analyzed and reduced our data with the SCUBA User's Reduction Facility (SURF) (Jenness \& Lightfoot 1997).  The first-order 
effects of chopping had already been removed from the raw data, so our reduction entailed
flatfielding, extinction correction, removal of sky variations, and the coadding and rebinning of the maps.  
This process of data reduction is described in Paper I.  The contour maps (with
greyscale) for 850 and 450 $\mu$m are in Figures~\ref{contour_a}-\ref{contour_e}.

We calibrated our flux measurements through observations of Uranus and AFGL 618, a primary and secondary calibrator, respectively.  
We  measured total voltages within 40$\arcsec$ and 120$\arcsec$ apertures.  By dividing these values 
into the expected flux from each of these
sources\footnote{We use the \it fluxes \rm package in SURF to obtain the expected fluxes for Uranus.  In November 1999, these
were: $S_{850}=65.63$ Jy and $S_{450}=173.37$ Jy.  The fluxes for AFGL 618 were $S_{850}=4.56$ Jy and $S_{450}=11.2$ Jy as
reported on the SCUBA secondary calibrator web page.  These fluxes for AFGL 618 are comparable (within uncertainty) to the fluxes 
given, recently, by Jenness et al. (2002).}, we obtained the nightly calibration factors.  For each night, we average
the calibration factors and report this average in Table~\ref{opacity}.  In February, we observed AFGL618 seven times and found
significant variations ($\sim$40\%) in our measurement of its flux at 450 $\mu$m.  These variations are reflected in the
large standard deviation reported as uncertainties ($\delta (C_{40}^{450})$ and $\delta (C_{120}^{450})$) in Table~\ref{opacity}.  
To account for uncertainty in the flux calibration, we take the average of the calibration factors as determined from the observations
from all three nights in February and propagate the standard deviation as uncertainty in the flux measurement.  
This final average was used in the voltage-to-flux conversion for data obtained in February, 2000,
and is reported in Table~\ref{opacity}.

\subsection{Photometry}

We find the flux for each of our images by first measuring the total voltage within  40$\arcsec$ and 120$\arcsec$ apertures.  Then,
we use the calibration factors given in Table~\ref{opacity} to convert these values to flux measurements.  The fluxes at 450 and 
850 $\mu$m  are given in Table~\ref{flux}.  Statistical and calibration errors dominate our uncertainty (calculated
as in Paper I,  Equation 2).  In the modeling of these sources, we use the \it normalized \rm intensity ($I_\nu^{norm}(\theta)$), 
so uncertainty in calibration has no effect on analysis of the radial profile.  

In our February 2000 observations, we had maps with a high signal-to-noise ratio, but the calibration errors 
approach $100\%$ at 450 $\mu$m (e.g., L1251B).  In Table~\ref{flux}, we include the S$/$N ratio to give an 
indication of the quality of our maps.  Additional photometry information is given in Table~\ref{sed}.
The peak intensity centroids are given in Table~\ref{flux}.  In addition,
for IRAS 04016+2610 and 04361+2547, we give the flux and centroid for  neighboring condensations.  
In Table~\ref{src_prop}, we give the calculated values for $T_{bol}$, $L_{obs}$, and $L_{obs}/L_{smm}$.  

\subsection{Morphology}

We define an aspect ratio for each of the sources by dividing the major and minor axis lengths as measured at the 2-$\sigma$ contour.  
These values are reported in Table~\ref{src_prop}.  For some cores (e.g., IRAS 03256+3055, 04016+2610 and 04108+2803), the major and 
minor axes were not discernable.  
In some cases, the cores are not ellipsoidal but, instead, have extensions in different directions.  Also,
the aspect ratio defined by the outer contours does not always reflect the morphology of the inner contours.
The average aspect ratio was 1.3$\pm$0.4 with the most aspherical source being IRAS 04248+2612 with an aspect ratio of 2.3. 

For each modeled source, we determine the deconvolved source size ($\theta_{dec}$) from the FWHM size 
of the beam  ($\theta_{mb}$) and source ($\theta_{src}$) intensity profiles:  $\theta_{dec}=\sqrt{\theta_{src}^2-\theta_{mb}^2}$.
The values for $\theta_{dec} / \theta_{mb}$ are in Table~\ref{src_prop}; the reported uncertainties are based on the 
variability of $\theta_{mb}$ ($\delta \theta_{mb}=1$-$3\arcsec$).  In most cases ($\sim$70\%) , we resolve the emission, 
but the observed source size is sometimes comparable to the FWHM of the beam.  For example, IRAS 04295+2251, with  
$\theta_{dec} / \theta_{mb}=0.7$, has an observed intensity profile that, at the FWHM, is only 1.3 times the FWHM of the beam (i.e.,
$\theta_{src}=20\arcsec$ and $\theta_{mb}=16\arcsec$).  In Figure~\ref{beams}, we showed that the FWHM beamwidth
does change slightly as the telescope cools.  However, this mostly affects the 450 $\mu$m beam, and we derive our deconvolved
source sizes from the 850 $\mu$m profiles.  Any source with $\theta_{dec} / \theta_{mb}>1$ is definitely resolved.  
Of course, those sources with $\theta_{dec} / \theta_{mb}=0$ are clearly unresolved.  For those sources with 
$0<\theta_{dec} / \theta_{mb}<1$, we have more closely analyzed their profiles.  IRAS 04113+2758 was observed in November 1999 when
the FWHM of the beam was quite variable ($\pm3\arcsec$), so we do not consider this a resolved source. 
IRAS 04016+2610, 04239+2436, and 04295+2251 were all observed in February 2000, when the 850 $\mu$m beam was fairly stable.
We consider each of these sources to be resolved. Further, we determine $\theta_{dec} / \theta_{mb}$ from the FWHM of the
intensity profiles, but our data is sensitive out to 3-5 beamwidths for most sources---i.e. we do resolve the outer envelope emission
from these cores.  As we will show in Section 3.1, the
real discerning power of our observations is not in the inner regions but in the observations of emission from the outer
regions of the envelope.

Also, several sources showed additional emission within the maps.  For example, IRAS 04016+2610, 04108+2803, and 04361+2547
show sources of continuum emission that have comparable (or even greater) fluxes than the observed Class I protostar.  
Statistically, however, distinct, neighboring condensations of dust are not common in our sample---only about 20\%.  In Paper I, 
we suggested Class I cores might be classified as such because of multiple sources within a beamwidth;  in that sample, multiple sources
were found in 9$/$16 of the observed Class 0 and Class I cores.  However, the occurrence of 
multiplicity is much less (3$/$16) in this sample of Class I cores.

\section{Models}

\subsection{Power Law Models}

As in Paper III, we have modeled the dust emission from a subset of these Class I cores.  We have  selected the following sources
for modeling based on their symmetric morphology and quality of data:  CB230, L1251B, IRAS 04166+2706, IRAS 04169+2702, IRAS 04239+2436, 
IRAS 04248+2612, IRAS 04295+2251,
IRAS 04361+2547, and IRAS 04381+2540.  We use a modification of the radiative transfer package by Egan, Leung, \& Spagna (1988), 
\it CSdust3 \rm ---a ray-tracing, 1-dimensional radiative transfer code. This program has several input 
parameters: luminosity of the central source,   opacity of the dust grains as a function of wavelength,  strength of
the interstellar radiation field (ISRF), and the density distribution of material surrounding the source. 

Primarily, we test power law density distributions, i.e.,
\begin{equation}
n(r)=n_f\left( \frac{r}{r_f}\right)^{-p}; r \in [r_i,r_o]
\end{equation}
where $n_f$ is the density at a fiducial radius, $r_f=1000$ AU.  The density has a lower limit of $1000$ cm$^{-3}$, roughly simulating 
the ambient molecular cloud. 
We choose these models because they are simple and, a priori, make
no assumptions about the theory incorporated by the collapse of material.  In a later section, we do discuss the Shu collapse model
(Shu 1977) because it has only one free parameter and is easily tested.  Other theories of 
low-mass star formation (e.g., Larson (1969),  McLaughlin \& Pudritz (1997); Foster \& Chevalier (1993);  Henriksen, Andr\'{e}, \& 
Bontemps (1997); Whitworth \& Ward-Thompson (2001); etc.) often differ primarily in the value of $p$.

For each of the model parameters, we have tested a reasonable range on two test cases: CB230
and IRAS 04295+2251.  We selected these sources as test cases because of their high S$/$N maps and azimuthally symmetric morphology.
The results from these tests are in Tables~\ref{cb230}\ and \ref{iras04295}.  From these two test cases,
we have also determined the uncertainty in $p$ caused by the uncertainty in other parameters (see Table~\ref{error}).

The luminosity of the central source ($L_{int}$) is determined by matching the modeled ($L_{mod}$) to the observed 
bolometric luminosity ($L_{obs}$).   At short wavelengths ($\lambda \leq 60\mu m$), the cloud becomes optically thick,
and the SED becomes more sensitive to opacity and geometry.  Therefore, for $L_{mod}$ and $L_{obs}$ we use only those
flux points with $\lambda>60$ $\mu$m.  For a given mass distribution (constrained by the intensity profile), 
we are able to determine the internal luminosity,  $L_{int}$, required to produce $L_{obs}$.  This luminosity, $L_{int}$, may 
include both the luminosity of the protostar as well as accretion luminosity---basically, 
any source of luminosity within the inner radius.
In most cases, $L_{int}$ is less than $L_{obs}$ because the interstellar radiation field contributes a portion of the observed 
bolometric luminosity, $\sim$0.2 L$_\odot$.  This is in accord with our results in Papers II and III.  

We use dust opacities given by Ossenkopf \& Henning (1994) for protostellar cores.  They simulated the coagulation of dust grains
(with an initial standard MRN size distribution) over $10^5$ yr and calculated the opacities for 
different initial gas densities.  For the modeling of our sample, we compare two
``types'' of dust from Table 1 of Ossenkopf \& Henning.  In Papers II \& III, we  have shown
that the dust with thin ice mantles provides the best-fit models for  preprotostellar and Class 0 sources.  
The opacities for this dust are given in the fifth column of Table 1 in Ossenkopf \& Henning (1994) and, hence, 
it is named ``OH5'' dust.  In addition, we model 
sources with the opacities from column 2 of Table 1 (OH2 dust), representing dust grains 
without ice mantles. In the two test cases, we find that OH5 dust fits the SED best, but 
the choice of dust type introduces a negligible source of uncertainty in $p$ (see Table~\ref{error}).

We use the interstellar radiation field,  given in Paper II, based on  Black (1994) and Draine (1978).  
For ultraviolet to far-infrared wavelengths, we introduce a factor, $s_{ISRF}$,  to  scale the ISRF. 
For the two test cases, $s_{ISRF}=0.3$ fits the data best,
in accord with the ISRF found to fit Class $-1$ and $0$ cores in Papers II \& III.  
For this sample, uncertainty in the ISRF contributes a large source of uncertainty in 
$p$: $\delta p \sim 0.4 $ (see Table~\ref{error}).

\it CSdust3 \rm requires the input of an inner and outer radius as boundary conditions to be used in solving the
radiative transfer.  We have developed some practical criteria in determining these radii.
We set the inner radius to 100 AU for all sources---this corresponds to $0\farcs7$ for the Taurus sources, $0\farcs2$ for CB230,
and $0\farcs3$ for L1251B. 
The outer radius should be, at least, the sum of the chop throw (120$\arcsec$)  and the radius of the model.
For example, with CB230 (at 450 pc), we used an outer radius of 100,000 AU which accounts for the 120$\arcsec$ chop
(54,000 AU) and for modeling out to 100$\arcsec$ (45,000 AU).   The outer radius for L1251B is 72,000 AU; for Taurus sources, 33,000 AU.  
We do choose a lower limit for the density ($n=10^3$ cm$^{-3}$), but, for most cores, the outer radius lies within this limit.
In Table~\ref{error}, we show the effect of varying the inner and outer radii on the derived value of $p$.  Changes in the inner
radius cause a negligible change in $p$; uncertainty in the  outer radius causes an uncertainty $\delta p=0.3$ in the modeling of CB230
and $\delta p=0.2$ for IRAS 04295+2251.   

By assigning values to these various parameters and simulating the transfer of radiation in the envelope, we calculate the 
temperature  for the assumed density distribution.  Then, as described in Paper III,  we simulate the observations 
including the effects of chopping and convolution of the data with the observed beam profile.  The selection of this 
beam profile was discussed in Section 2.2; we took great care in using the appropriate beam since it does vary with time. 
While a Gaussian beam will cause the derived value of $p$ to be steeper by $\sim$0.4, using an obseved beam profile from a different
night causes negligible uncertainty (see Table 8).  We attempted modeling of IRAS 04302+2247, a source that was obviously unresolved.
In this case, we were only able to determine that $p>3.5$, which, of course, we expect in the modeling of a point source.

We constrain the total amount of mass in the envelope by normalizing the modeled SED to the observed 850 $\mu$m flux, and the
luminosity of the internal source is constrained by the observed luminosity, $L_{obs}$.  Then, we find the best-fit value for $p$,
which we select by the lowest $\Sigma\chi^2$---the sum of the $\chi^2$ values for the 450 and 850 $\mu$m profiles 
(see Equation 4 in Paper III for the definition of $\chi^2_\lambda$).  In most cases, the best-fit model based on $\Sigma\chi^2$
is identical to that selected by $\chi^2_{850}$ (cf. L1251B, IRAS 04361, and IRAS 04381 in Table~\ref{power}); because the
uncertainty is much smaller for the 850 $\mu$m data,  $\chi^2_{850}$ is much larger than  $\chi^2_{450}$ and, hence, dominates 
the sum.  Even though the intensity profiles show Nyquist sampled data, we use, as in Papers II and III, only points spaced by a 
full beam to calculate $\chi^2_\lambda$. The derived value of $p$  is virtually unaffected by the envelope mass or the 
luminosity of the internal source.  
We list the best-fit models in Table~\ref{power} along with those models where $p=\pm0.1$ and  $p=\pm0.5$.  
In Figures~\ref{model_a}-\ref{model_i}, 
we show the modeled and observed SEDs, radial profiles, and temperature distributions for each of our modeled sources.

\subsection{Top Hat Models}

\begin{figure}
\figurenum{16}
\plotone{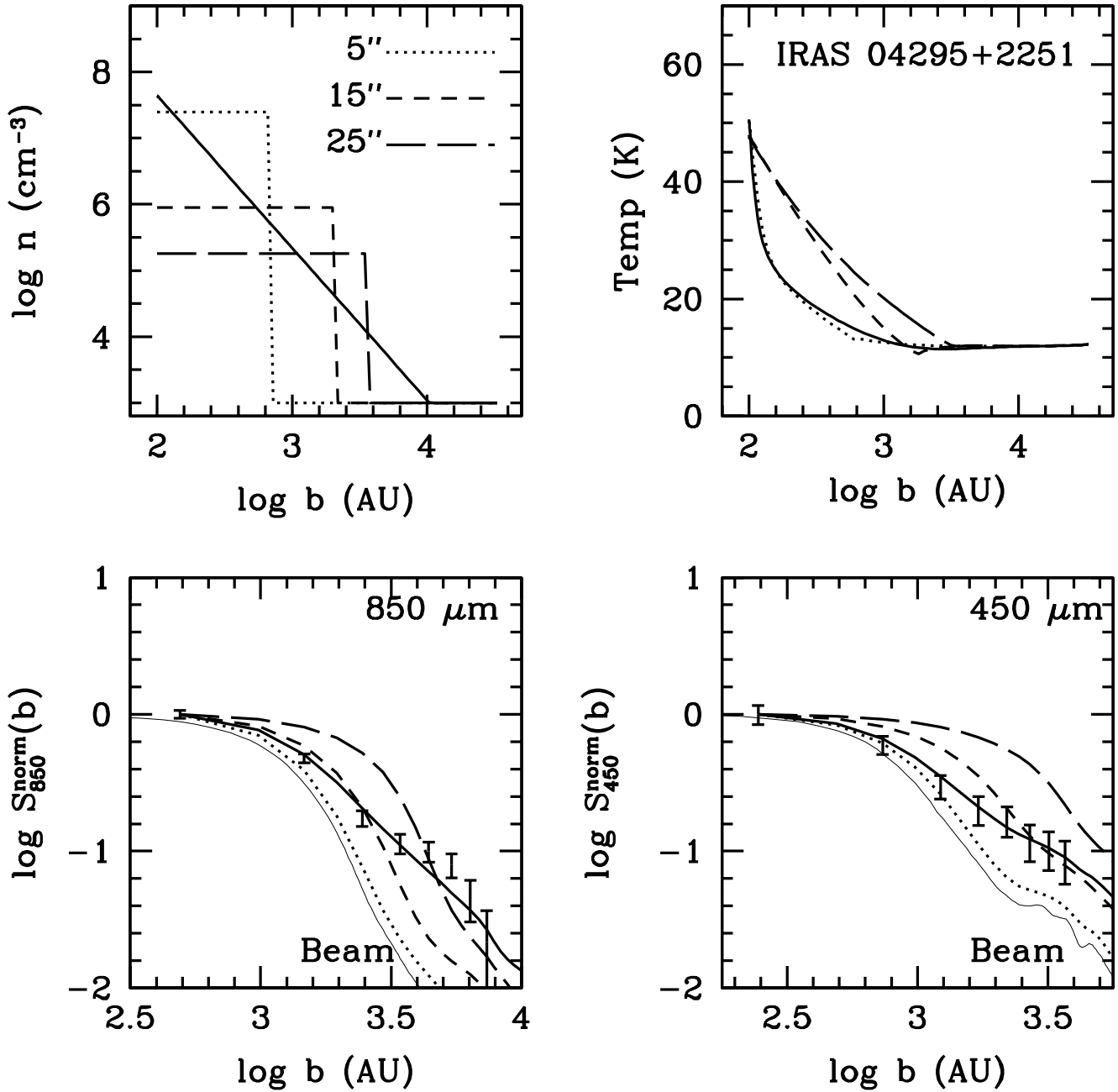}
\figcaption{\label{tophat} We show the modeled intensity profiles for the 
best-fit power law of IRAS 04295+2251 ($p=2.3$, solid line) along with a 
series of  top hat functions.  The top hat functions (dashed and dotted lines) have steps at 5\arcsec, 15\arcsec, and 25\arcsec\ 
where the density drops to 10$^3$ cm$^{-3}$.  The bottom panels also have the observed beam profile shown by the light, solid line.}
\end{figure}

We consider simple models of the density distribution in order to investigate the discerning capabilities of our methods---i.e. 
are we able to discern a power law distribution in the density from a top hat distribution?   We define a top hat density distribution
as an envelope model where the density is constant from the inner radius to some ``break'' radius; then, the density drops to 
10$^3$ cm$^{-3}$ in the region from this break radius to the outer radius.
It should be most difficult to discern a power law from top hat distribution with the source for which we derive the steepest power 
law, $p=2.3$ for IRAS 04295+2251. In Figure~\ref{tophat}, we show the modeled intensity profiles for three top hat distributions with 
breaks in the density at 5\arcsec, 15\arcsec,
and 25\arcsec\ (corresponding to 700, 2100, and 3500 AU).  The maximum density in the top hat functions is set so that the total mass is
equal to that in the derived power law distribution.  Also, in this figure, we show the best-fit power law model along with 
the observed intensity profile for IRAS 04295+2251.  This intensity profile shows a ``bump'' at about 5000 AU (or 35\arcsec) 
that is coincident with additional emission within the field-of-view. We find the 5\arcsec\ and 15\arcsec\ top hat functions do 
not have enough emission at large radii to reproduce the 850 $\mu$m profile.  The 25\arcsec\ top hat has emission in the outer
regions that is not seen in either the 850 or 450 $\mu$m observed profile.   Likewise, the intensity profiles of other sources
in our sample are not easily reproduced by  simple top hat functions. Indeed, power law models, in contrast to the
top hat functions, match the shape of the observed intensity profiles quite well.

\subsection{Shu Collapse Models}

In addition to modeling power law density distributions for our sample of nine cores, we have modeled these sources with the
Shu inside-out collapse model (Shu 1977).  This model begins with a singular isothermal sphere with a density distribution
that is proportional to $r^{-2}$.  Through some perturbation, collapse begins inside the cloud and proceeds outward at the sound
speed ($a_{eff}$).  As collapse ensues, the cloud's density distribution is described by an outer, static region with 
$n\propto r^{-2}$ and a region within the front of collapse (or infall radius, $r_{inf}$) whose density approaches 
$n\propto r^{-1.5}$ (indicative of freefall).  The beauty of this theory is that there is only one free parameter: time.  
As time progresses, the front of collapse (or infall radius, $r_{inf}$) propagates outward at $a_{eff}$.

We determine $a_{eff}$ from the observed linewidths in Table~\ref{mass}.  Assuming a gas temperature of 10K, we are able to separate
the turbulent and thermal components ($a_{turb}$ and $a_{th}$) of these linewidths and determine the effective sound speed 
($a_{eff}=\sqrt{a_{th}^2+a_{turb}^2}$).  For each source, $a_{eff}$ is in Table~\ref{mass}.

We list the parameters for our Shu collapse models in Table~\ref{shu}.
For all modeled sources, the spectral energy distribution is not fitted well by the Shu model. The Shu model overestimates all fluxes with
$\lambda>100$ $\mu$m, indicating this model predicts too much mass within the cores.  We may have overestimated  the sound speed 
in these cores and, hence, also the mass (see Section 6.1).  Also, variations in dust opacity and geometry
can greatly affect the SED (e.g., Moore \& Doty 2002).  Consequently, in choosing the best-fit Shu models,  we give little regard 
to the values for $\chi^2_{SED}$.  Instead, we focus on the nature of the intensity profiles wherein lies the strength of our analysis.

\begin{figure}
\figurenum{17}
\plotone{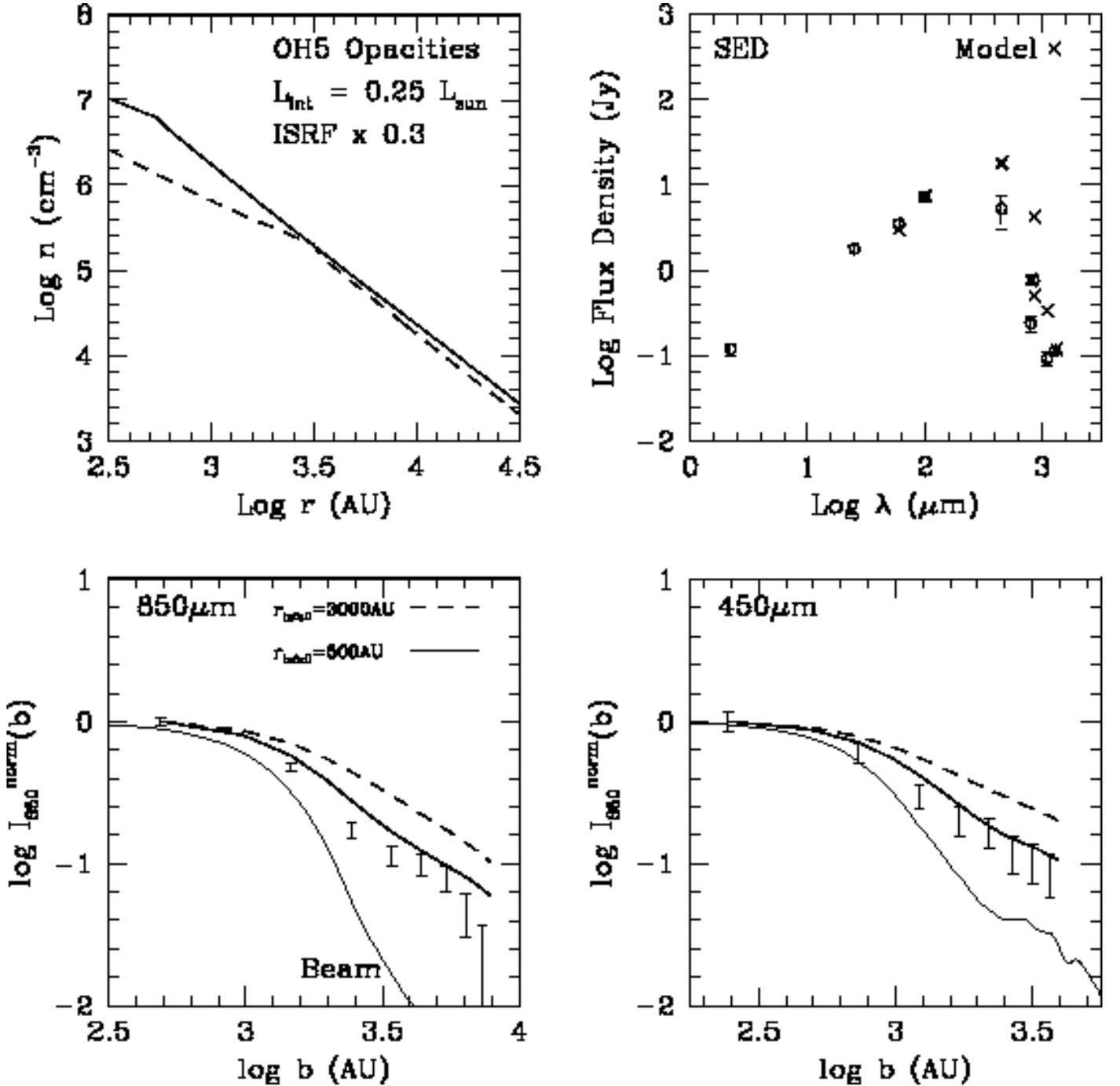}
\figcaption{\label{04295shu} Shu collapse models for IRAS 04295+2251 with two different infall radii, 500 (solid) and 3000 AU (dashed). 
In the upper left panel, we show the modeled density profile.   The modeled spectral energy distribution is marked by
crosses in the upper right panel, the observed fluxes are circles with errorbars.  The two bottom plots have the 
modeled and observed radial profiles---the best-fit model is a solid dark line.  
Also shown in these plots are the beam profiles used in the beam convolution (solid light line).}
\end{figure}

The beam at 850 $\mu$m ($\theta_{mb}=15\arcsec$) corresponds to 2100, 4500, and 6800 AU for Taurus, L1251B, and CB230, respectively.  
For all but two sources (IRAS 04169+2702 and 04381+2540), the infall radius of the best-fit Shu model falls within these radii.   
Basically, then, we are fitting straight power laws with an index of $p=2$.  In Figure~\ref{04295shu}, we show the results  of two models
for  IRAS 04295+2251.  The best-fit model is with $r_{infall}=500$ AU, but we include a model with $r_{infall}=3000$ AU. The 
larger infall radius causes the modeled profile to flatten. 

\begin{figure}
\figurenum{18}
\plotone{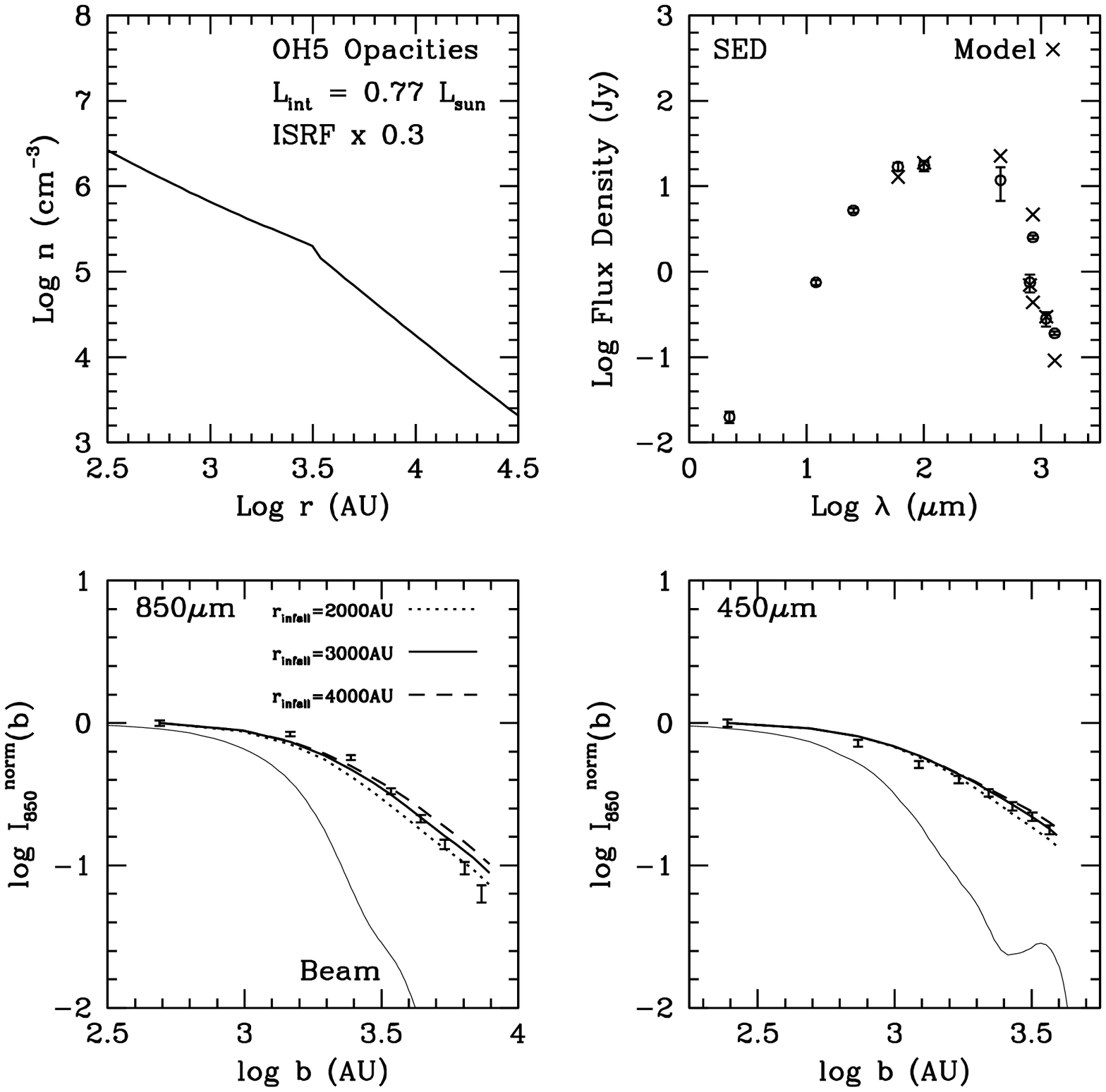}
\figcaption{\label{04169shu} Shu collapse model for IRAS 04169+2702. In the upper left panel, we show the modeled density profile.  
The modeled spectral energy distribution is marked by
crosses in the upper right panel, the observed fluxes are circles.  The two bottom plots have the modeled and observed 
radial profiles---the best-fit model is a solid dark line.  Also shown in these plots are the beam profiles used in the beam 
convolution (solid light line).}
\end{figure}

For IRAS 04169+2702, the best-fit infall radius is slightly larger  than the JCMT beamwidth 
($r_{infall}=3000$ AU),  and the best-fit Shu model has improved $\chi^2$ values over the power law model ($p=1.5$).  
In Figure~\ref{04169shu}, we show the results of this model.  IRAS 04169+2702 was fitted with a power law of $p=1.5$, so one might
conclude that the Shu model should fit with an infall radius outside our modeling range.  However, the Shu model predicts
an inner region that \it approaches \rm  $n\propto r^{-1.5}$.  The density distribution just within the infall radius is
actually significantly shallower than $r^{-1.5}$, so the steep outer region is required to fit the data. 

Our basic conclusion for these Shu collapse models is that most sources in this sample exhibit no evidence for a break in the 
power law density distribution on the scales we can probe with SCUBA ($\sim1000-8000$ AU).  

\subsection{Simple Analysis}

As in Paper I, we offer the results of a simple analysis involving the slope, $m$, of the outer part of the intensity 
profile (Adams 1991). Assuming the temperature distribution in the envelope to be described by a power law, 
$T_d(r)=T_d(r_f)(r/r_f)^{-q}$ where $r_f$ is a fiducial radius, one can determine the density power law index, which we call
$p_m$, by the following relation:  $p_m=m-q+1$ (see Equations 5-8 in Paper I; assume $q=0.4$).    
In Paper I, we argued that this method is of dubious validity for two reasons:  failure of the Rayleigh-Jeans 
approximation and deviation of the temperature profile from a power law due to heating from the ISRF and the central source.  
In order to compare with our more detailed modeling, we have performed such an analysis here to determine $p_m$. 

\begin{figure}
\figurenum{7} 
\plotone{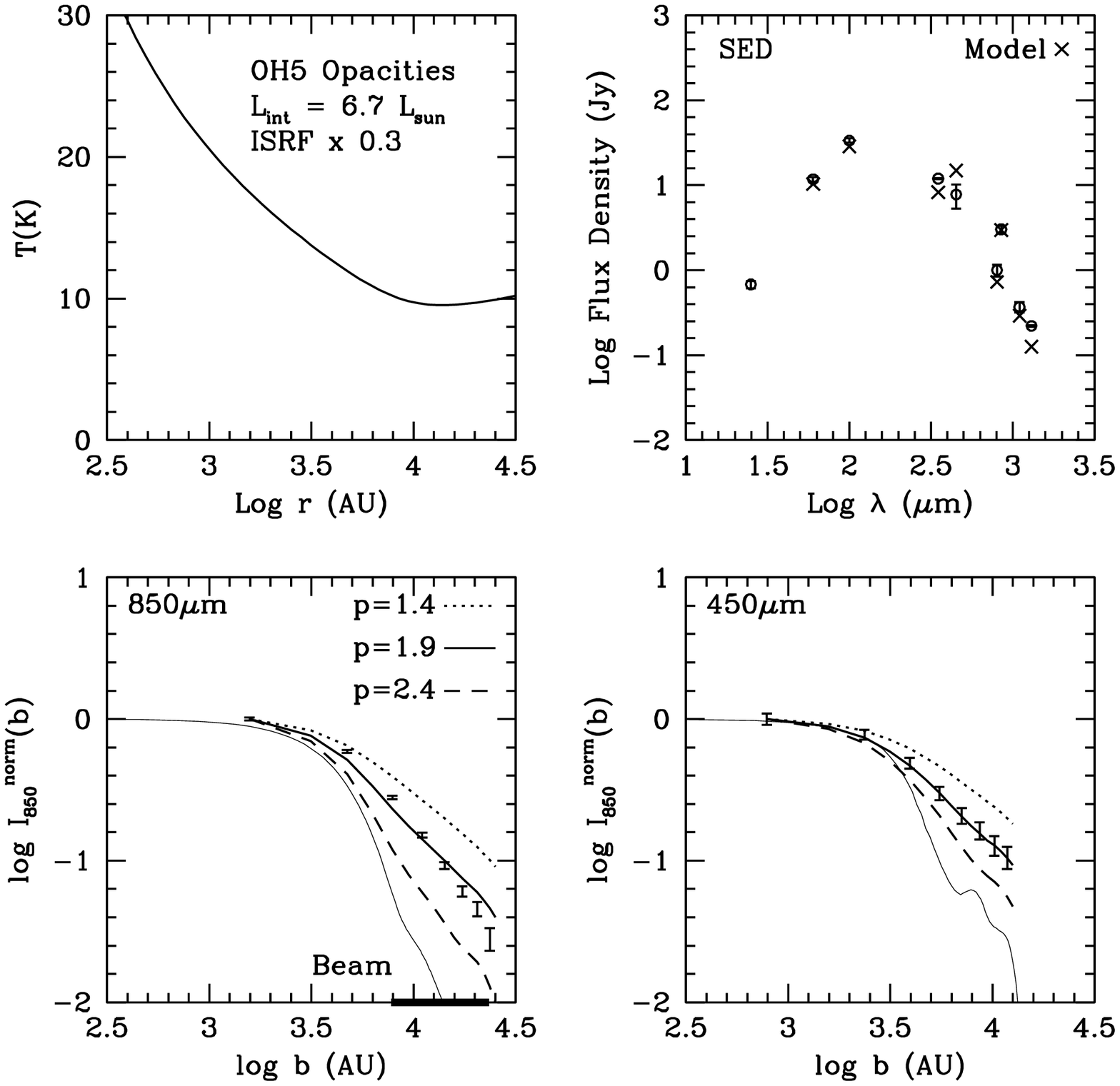} 
\figcaption{\label{model_a}Model of CB230.  In the upper left panel, we show the modeled temperature profile.  
The modeled spectral energy distribution is marked by
crosses in the upper right panel, the observed fluxes are circles.  The two bottom plots have the modeled and observed radial 
profiles---the best-fit model is a solid dark line.  Also shown in these plots are the beam profiles used in the beam convolution 
(solid light line).
The range of fit for $m$ (in the calculation of $p_m$) is shown by the bold line on the x-axis.}
\end{figure}

\begin{figure}
\figurenum{8} 
\plotone{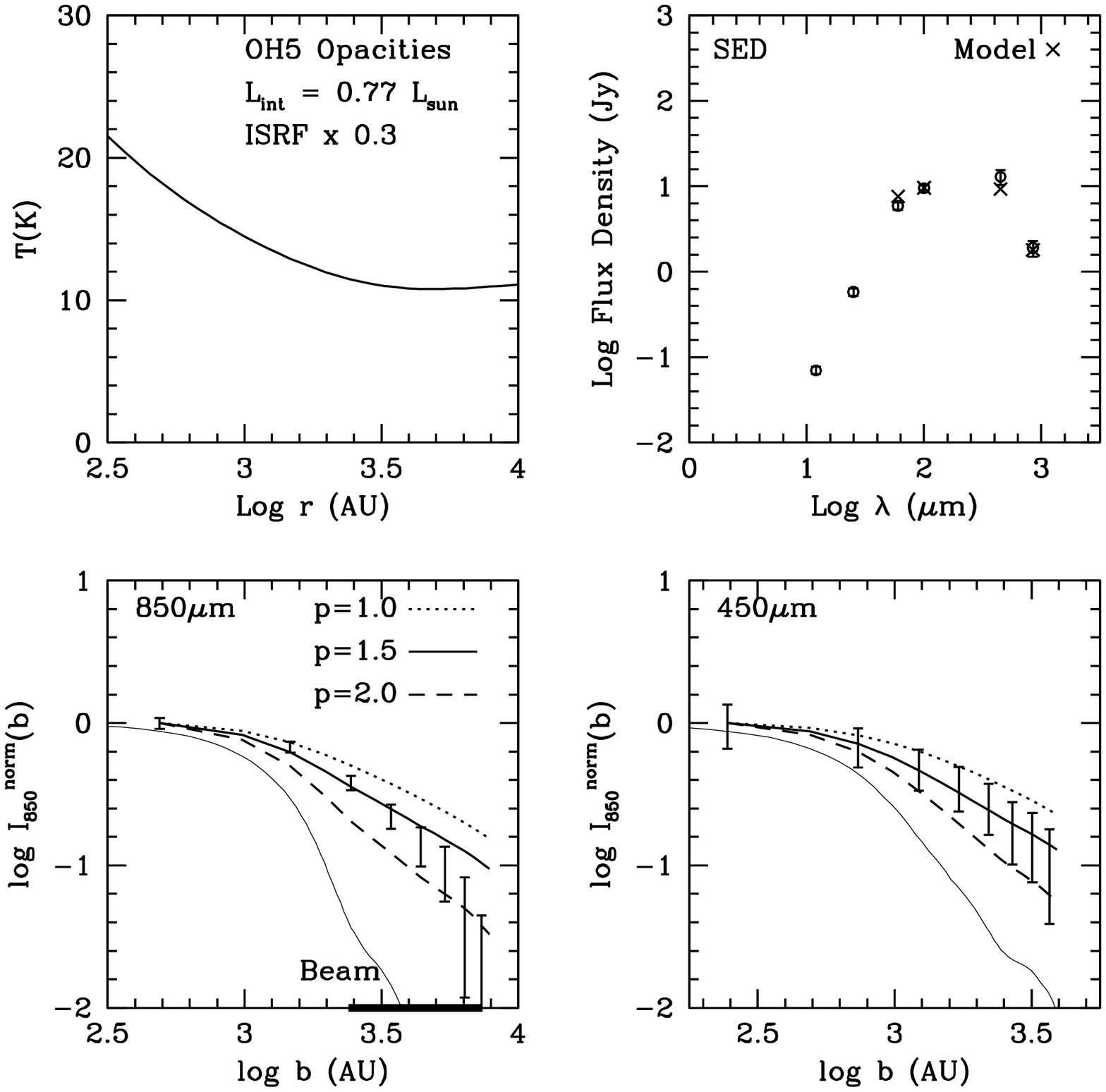} 
\figcaption{\label{model_b}Model of IRAS 04166+2706.  In the upper left panel, we show the modeled temperature profile.  The 
modeled spectral energy distribution is marked by
crosses in the upper right panel, the observed fluxes are circles.  The two bottom plots have the modeled and observed radial 
profiles---the best-fit model is a solid dark line.  Also shown in these plots are the beam profiles used in the beam convolution 
(solid light line).
The range of fit for $m$ (in the calculation of $p_m$) is shown by the bold line on the x-axis.}
\end{figure}

\begin{figure}
\figurenum{9} 
\plotone{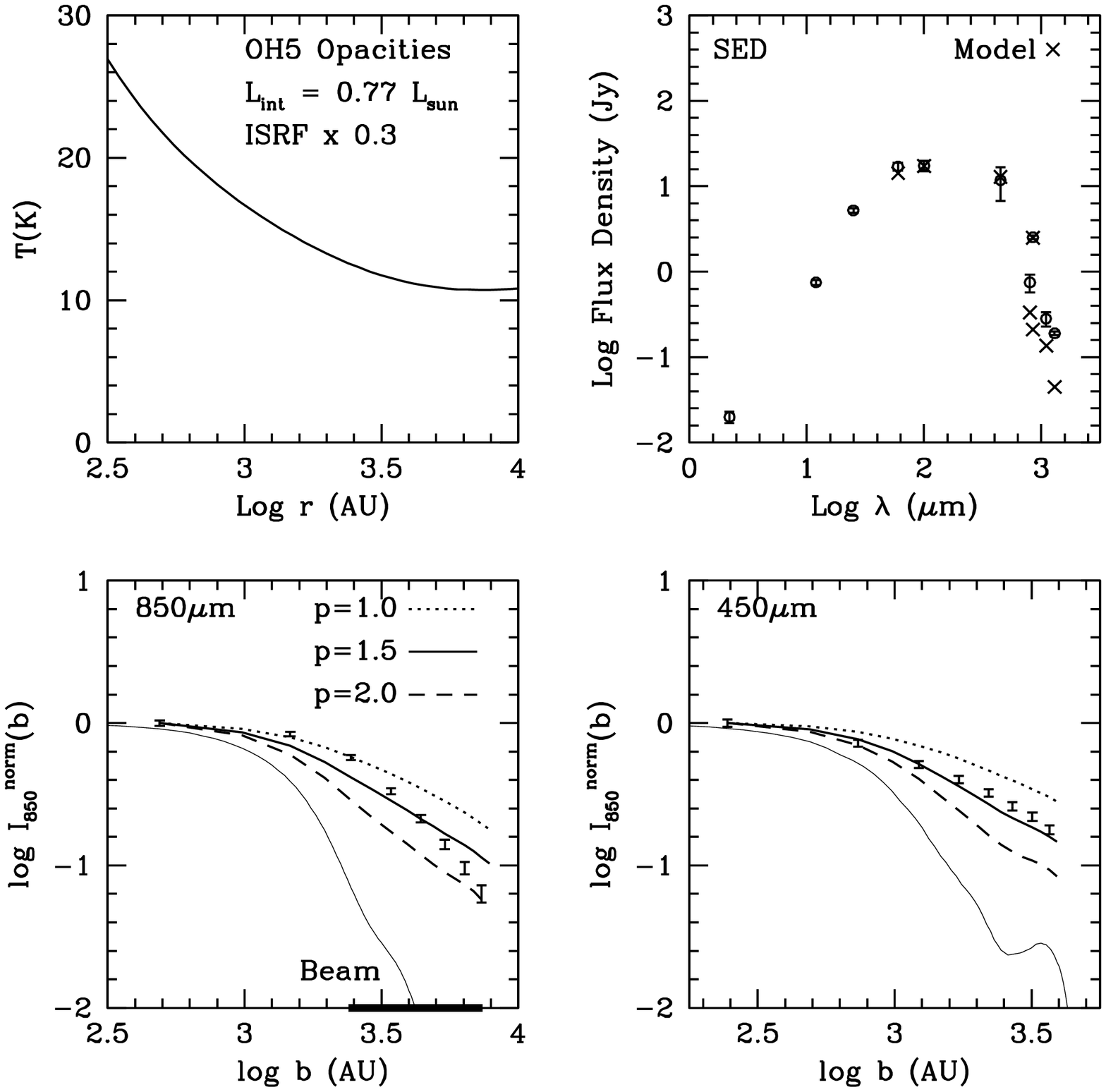} 
\figcaption{\label{model_c}Model of IRAS 04169+2702. In the upper left panel, we show the modeled 
temperature profile.  The modeled spectral energy distribution is marked by
crosses in the upper right panel, the observed fluxes are circles.  The two bottom plots have the modeled and observed radial 
profiles---the best-fit model is a solid dark line.  Also shown in these plots are the beam profiles used in the beam convolution.
The range of fit for $m$ (in the calculation of $p_m$) is shown by the bold line on the x-axis.}
\end{figure}

\begin{figure}
\figurenum{10}
\plotone{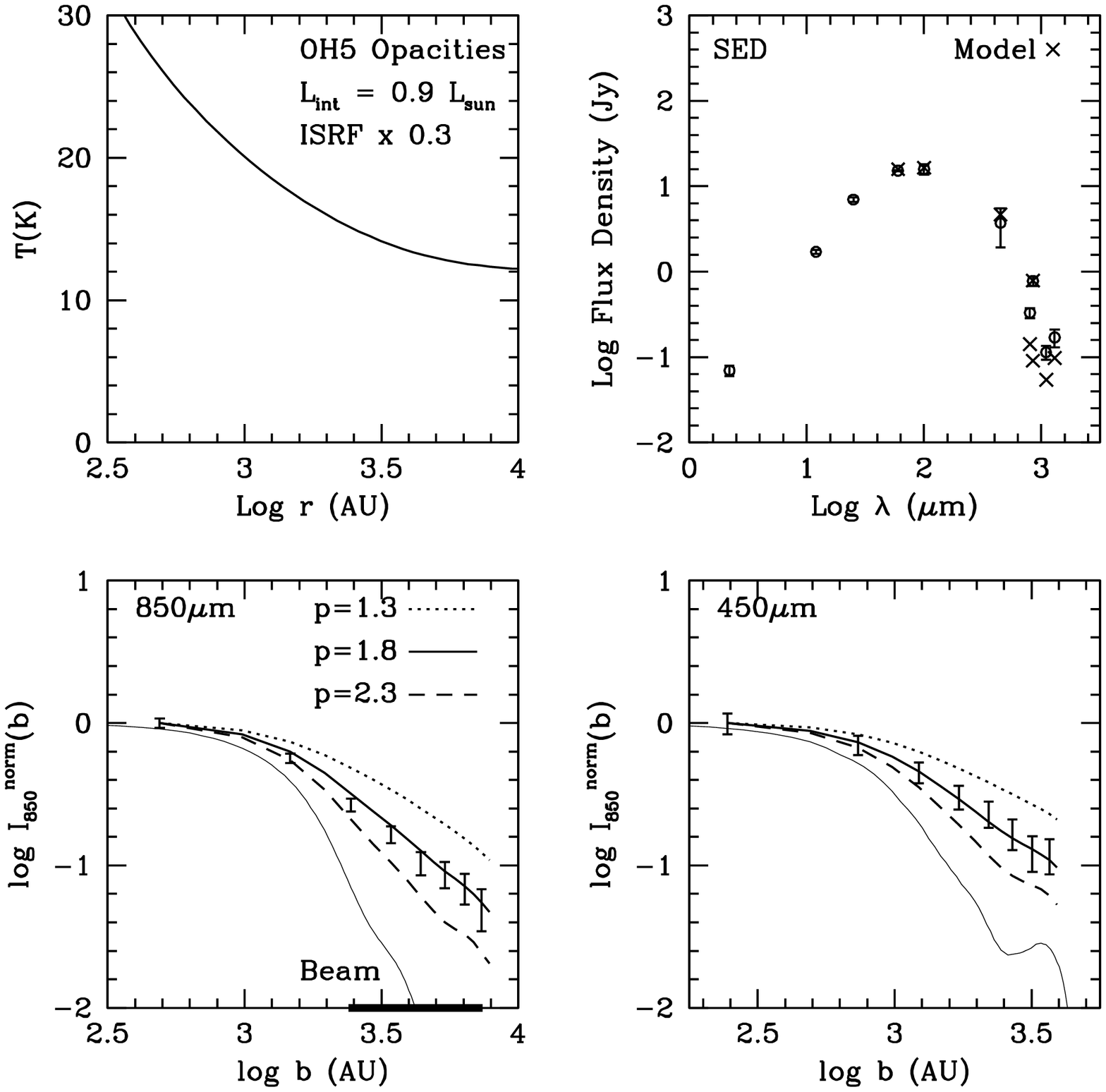}
\figcaption{\label{model_d}Model of IRAS 04239+2436. In the upper left panel, we show the modeled temperature profile.  
The modeled spectral energy distribution is marked by
crosses in the upper right panel, the observed fluxes are circles.  The two bottom plots have the modeled and observed radial 
profiles--- the best-fit model is a solid dark line.  Also shown in these plots are the beam profiles used in the beam convolution 
(solid light line).
The range of fit for $m$ (in the calculation of $p_m$) is shown by the bold line on the x-axis.}
\end{figure}

\begin{figure}
\figurenum{11}
\plotone{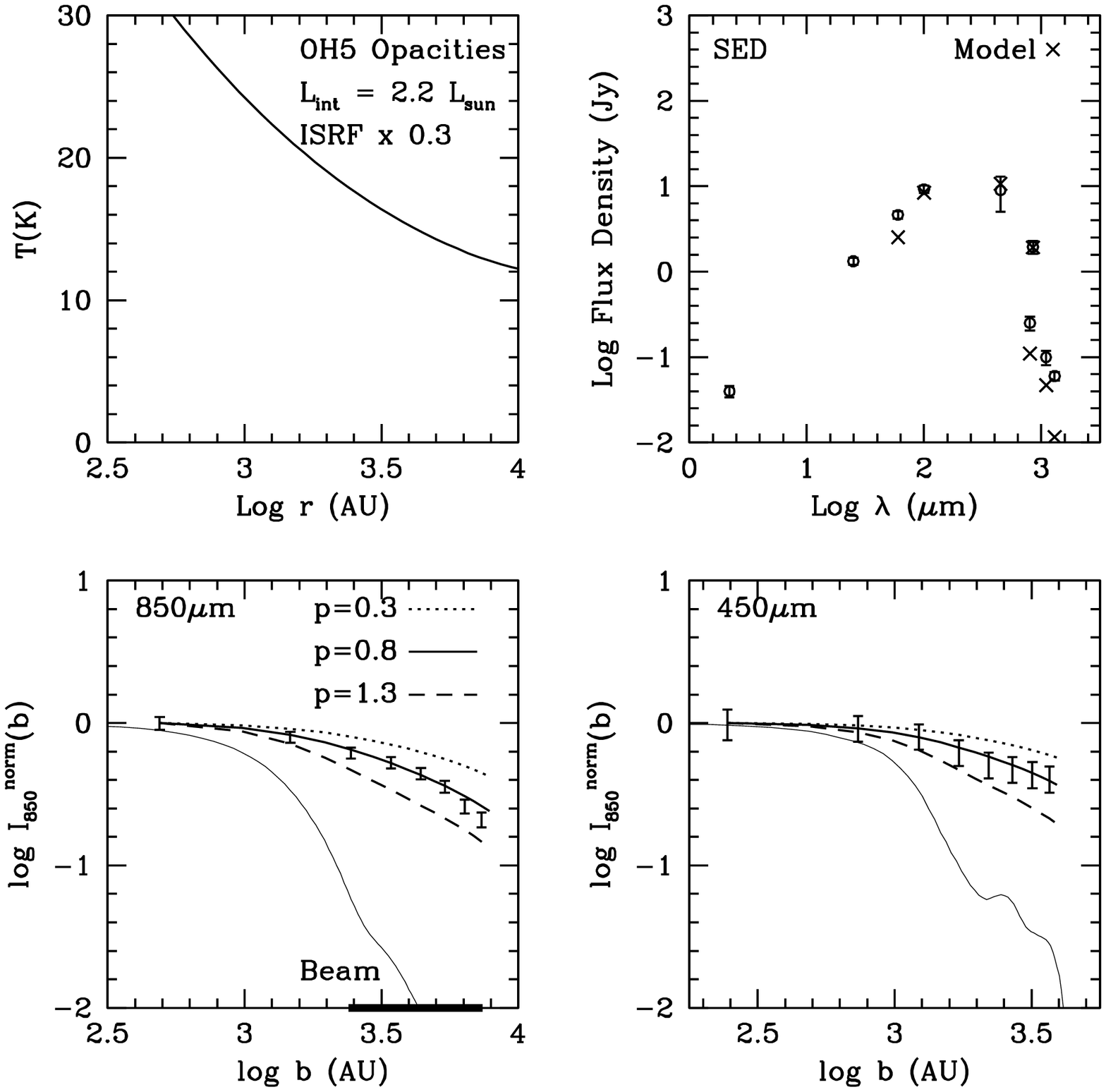}
\figcaption{\label{model_e}Model of IRAS 04248+2612.  In the upper left panel, we show the modeled 
temperature profile.  The modeled spectral energy distribution is marked by
crosses in the upper right panel, the observed fluxes are circles.  The two bottom plots have the modeled and observed radial 
profiles---the best-fit model is a solid dark line.  Also shown in these plots are the beam profiles used in the beam convolution 
(solid light line).
The range of fit for $m$ (in the calculation of $p_m$) is shown by the bold line on the x-axis.}
\end{figure}

\begin{figure}
\figurenum{12}
\plotone{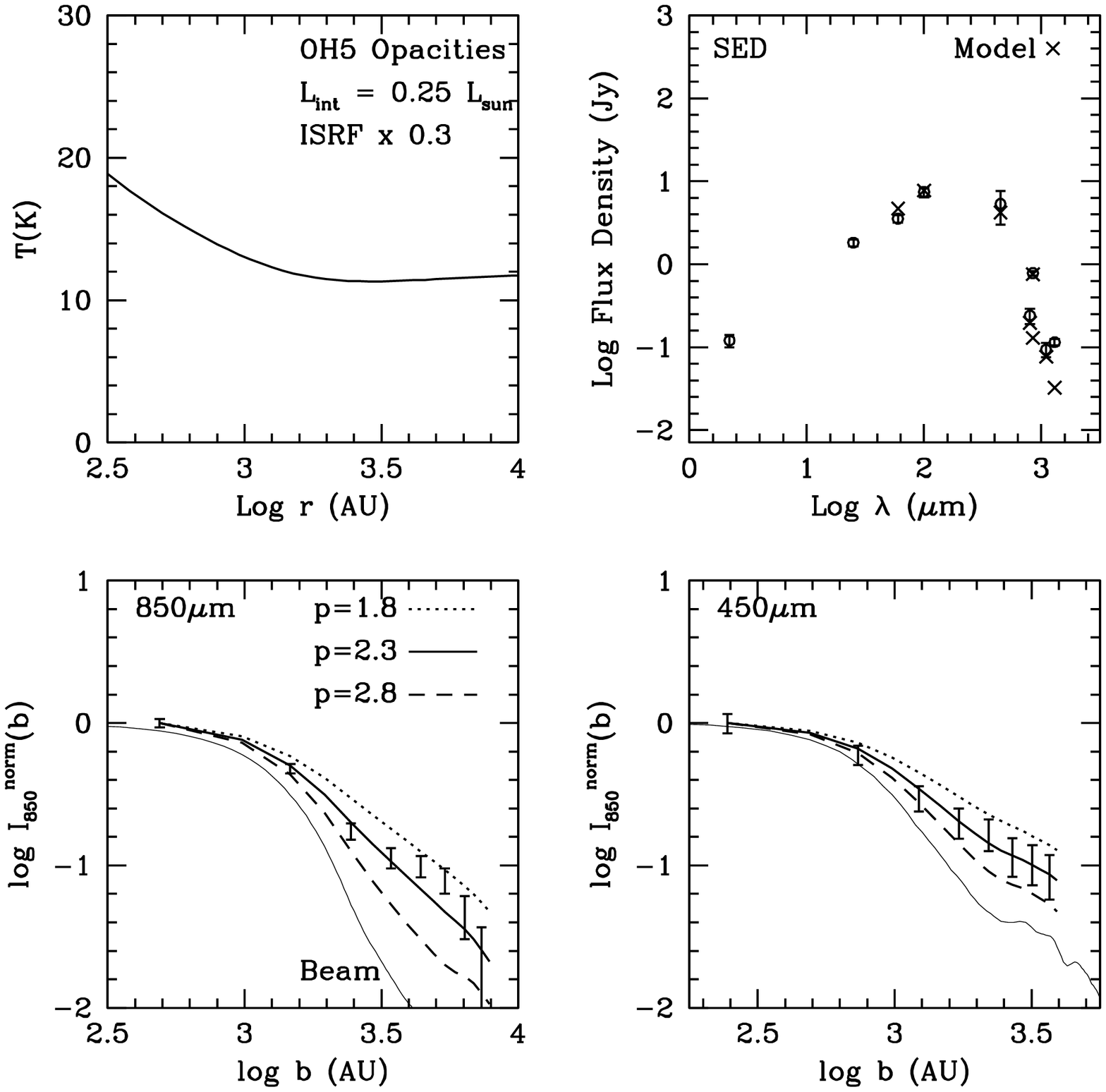}
\figcaption{\label{model_f}Model of IRAS 04295+2251. In the upper left panel, we show the modeled 
temperature profile.  The modeled spectral energy distribution is marked by
crosses in the upper right panel, the observed fluxes are circles.  The two bottom plots have the modeled and observed radial 
profiles---the best-fit model is a solid dark line.  Also shown in these plots are the beam profiles used in the beam convolution.}
\end{figure}

\begin{figure}
\figurenum{13}
\plotone{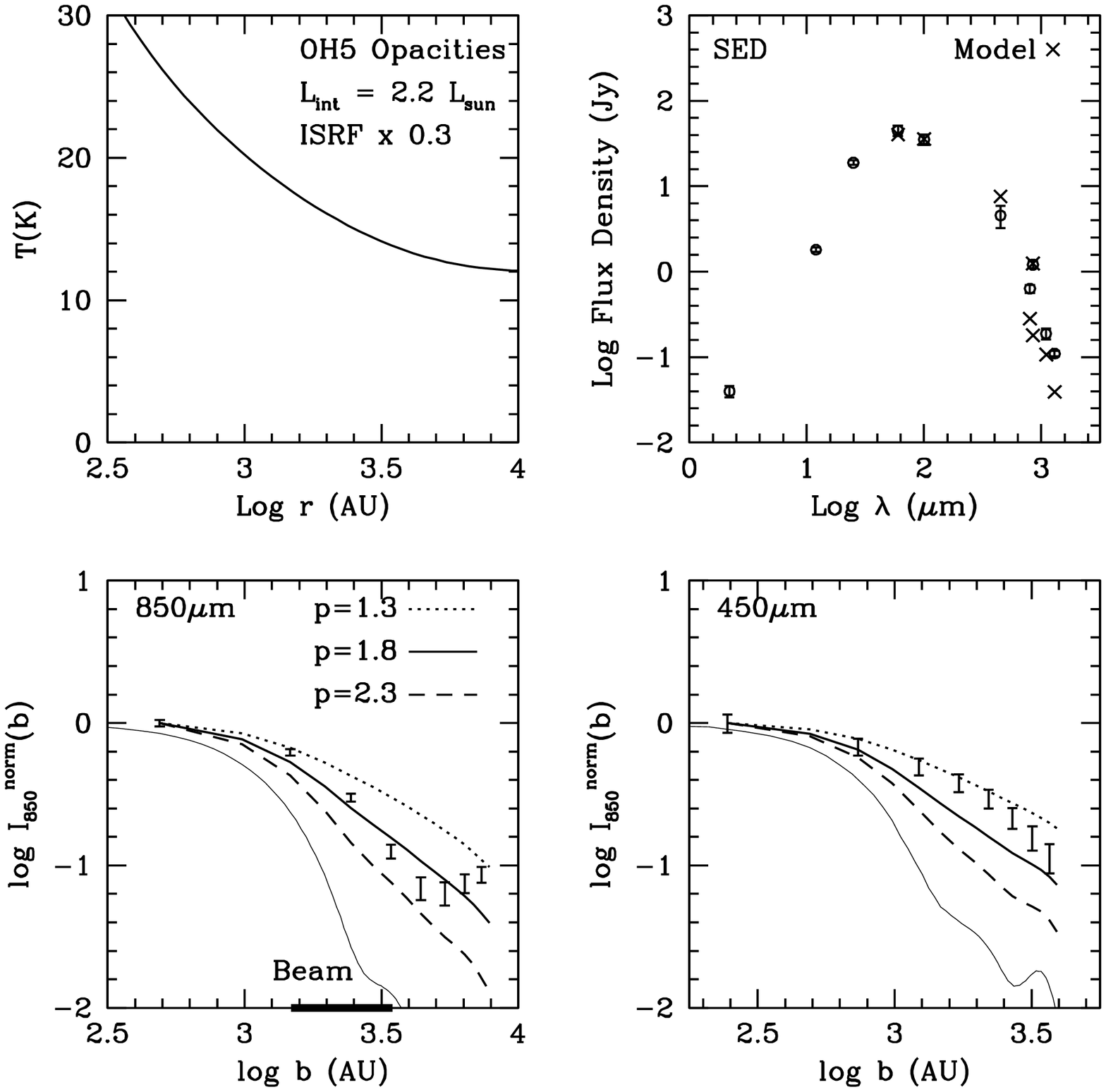}
\figcaption{\label{model_g}Model of IRAS 04361+2547.  In the upper left panel, we show the modeled 
temperature profile.  The modeled spectral energy distribution is marked by
crosses in the upper right panel, the observed fluxes are circles.  The two bottom plots have the modeled and observed radial 
profiles---the best-fit model is a solid dark line.  Also shown in these plots are the beam profiles used in the beam convolution 
(solid light line).
The range of fit for $m$ (in the calculation of $p_m$) is shown by the bold line on the x-axis.}
\end{figure}

\begin{figure}
\figurenum{14}
\plotone{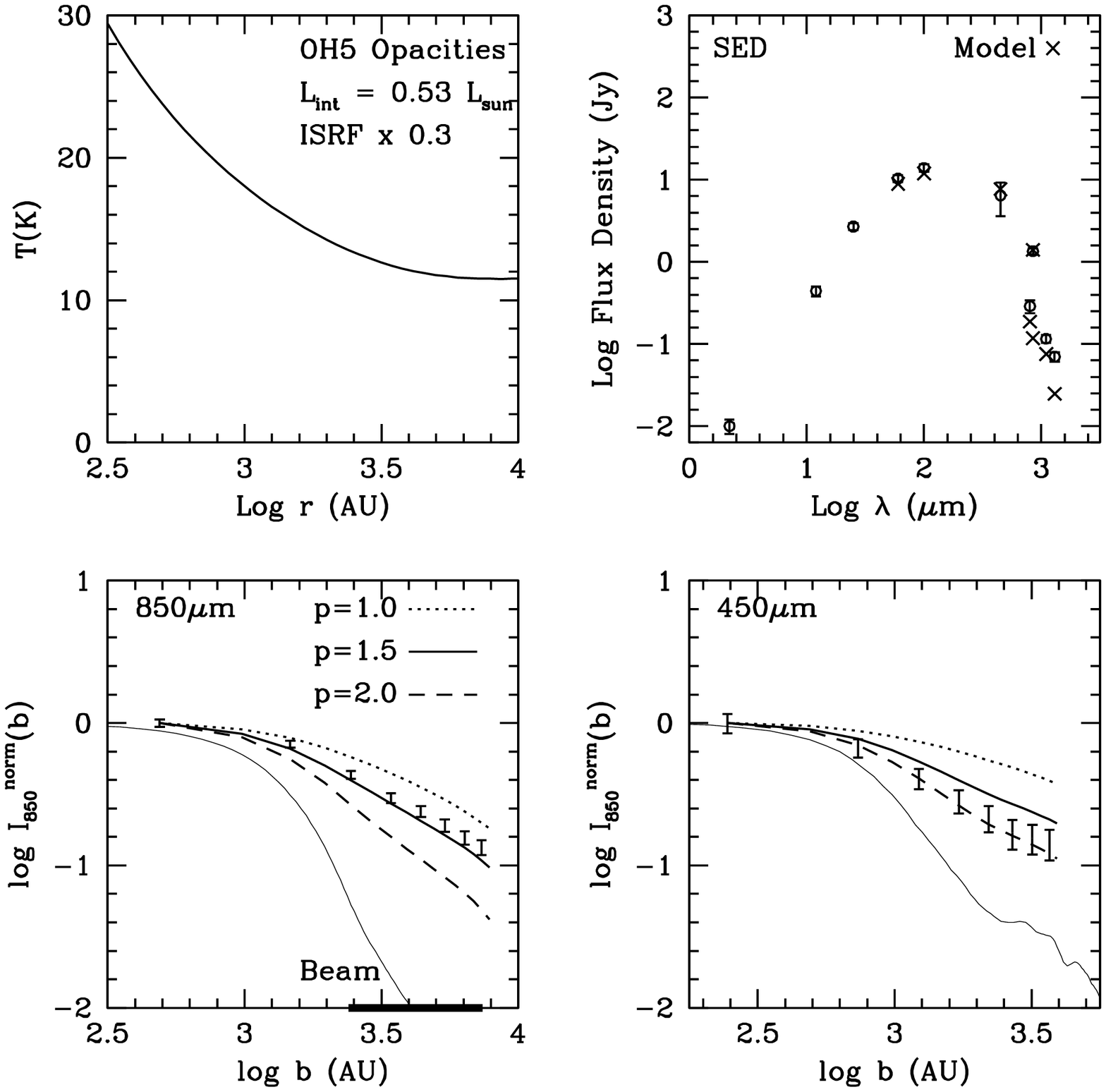}
\figcaption{\label{model_h}Model of IRAS 04381+2540.  In the upper left panel, we show the modeled 
temperature profile.  The modeled spectral energy distribution is marked by
crosses in the upper right panel, the observed fluxes are circles.  The two bottom plots have the modeled and observed radial 
profiles---the best-fit model is a solid dark line.  Also shown in these plots are the beam profiles used in the beam convolution 
(solid light line).
The range of fit for $m$ (in the calculation of $p_m$) is shown by the bold line on the x-axis.}
\end{figure}

\begin{figure}
\figurenum{15}
\plotone{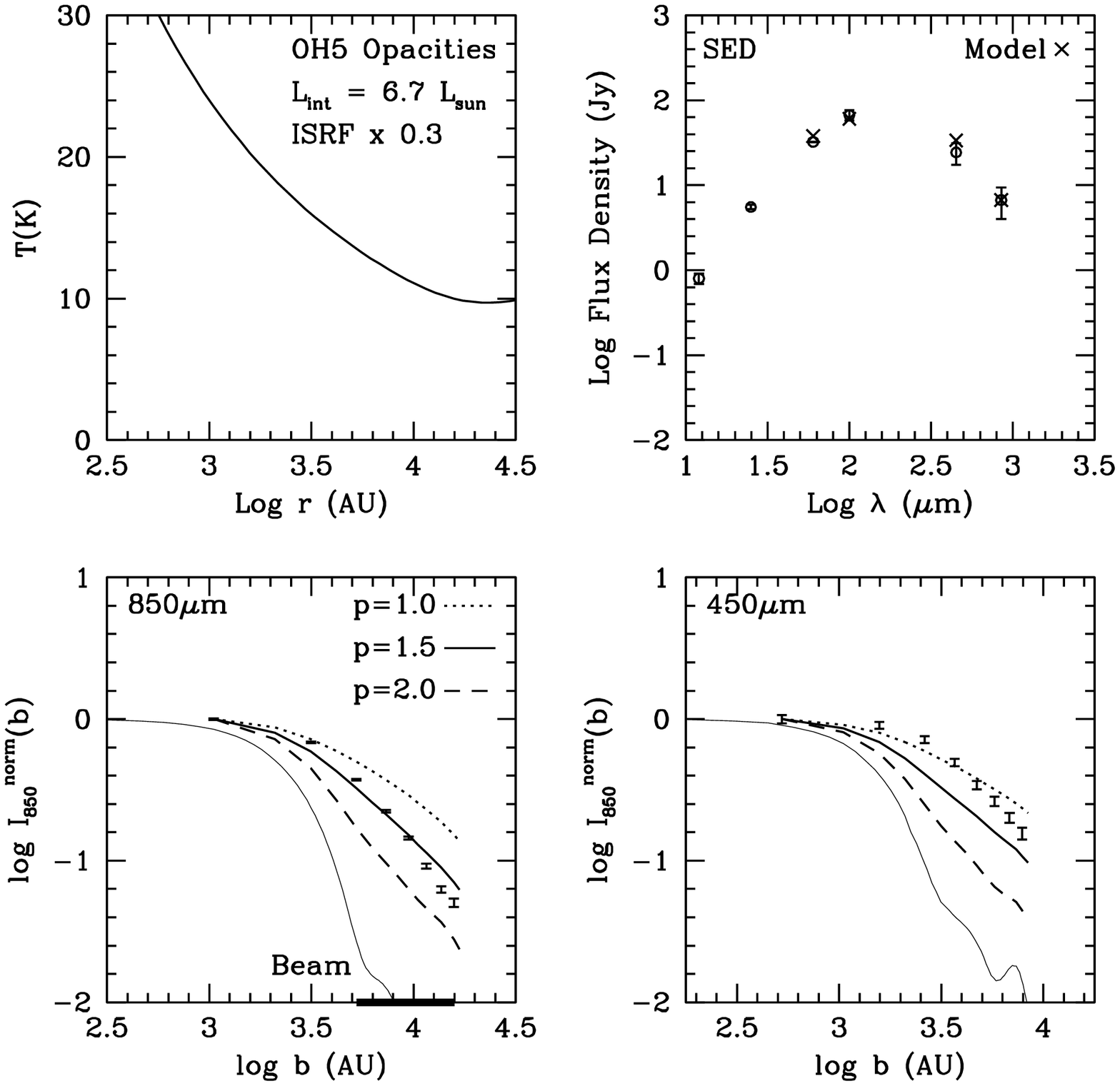}
\figcaption{\label{model_i}Model of L1251B. In the upper left panel, we show the modeled 
temperature profile.  The modeled spectral energy distribution is marked by
crosses in the upper right panel, the observed fluxes are circles.  The two bottom plots have the modeled and observed radial 
profiles---the best-fit model is a solid dark line.  Also shown in these plots are the beam profiles used in the beam convolution 
(solid light line).
The range of fit for $m$ (in the calculation of $p_m$) is shown by the bold line on the x-axis.}
\end{figure}

In Figures~\ref{model_a}-\ref{model_i}, we show the range over which we fit for $m$ by the bold line on the x-axis of the 
850 $\mu$m intensity profile.
The derived values of $p_m$ from the linear fits to the 850 $\mu$m profiles are in Table~\ref{bestpower}; we did not calculate $m$ for the
450 $\mu$m data.  The average difference between $p_m$ and $p$ is $\langle p_m-p \rangle=0.6\pm0.3$; i.e., the simple analysis
tends to predict a much steeper density profile.  This discrepancy is discussed in greater detail in  Paper I (Section 4.2),
but, in short, the failure of Rayleigh-Jeans approximation causes $p_m$ to be steeper than the actual distribution;
exclusion of the ISRF causes $p_m$ to be shallower than the actual distribution.

If we apply this correction to the sources analyzed in Paper I by this
method, we find $\langle p_m^{cor} \rangle=1.5\pm0.5$ (compared to $\langle p \rangle = 1.6\pm 0.4$ reported in this paper).  

Additionally, we apply the correction to the results of the large 1.3 mm survey by Motte \& Andr\'{e} (2001).  
At this wavelength, $h\nu /k=$11 K, so the Rayleigh-Jeans approximation (where $h \nu / kT \ll 1$) is better, but still
 not completely valid.  
Also, Motte \& Andr\'{e} chose values for $q$ that ranged from $-0.2$ to $0.4$ for their
sample of Class 0 and I cores.  From Figures~\ref{model_a}-\ref{model_i}, it is clear that a negative 
power law (indicative of external heating)
does not describe the temperature profiles for these evolved protostars, and $q=0$ can be ruled out for all but one source,
IRAS 04295+2251.  Therefore, we assume $q=0.4$ and calculate $p_m$ from their values for $m$: $\langle p_m \rangle = 1.8 \pm 0.4$,
and, with our rough correction, $\langle p_m^{cor} \rangle = 1.2 \pm 0.5$.  However, because the Rayleigh-Jeans approximation
is less invalid at these longer wavelengths, this correction factor is probably too large and, hence, $p_m^{cor}$ should be slightly
larger.  
  
\subsection{Disk Contribution}

\begin{figure}
\figurenum{19}
\plotone{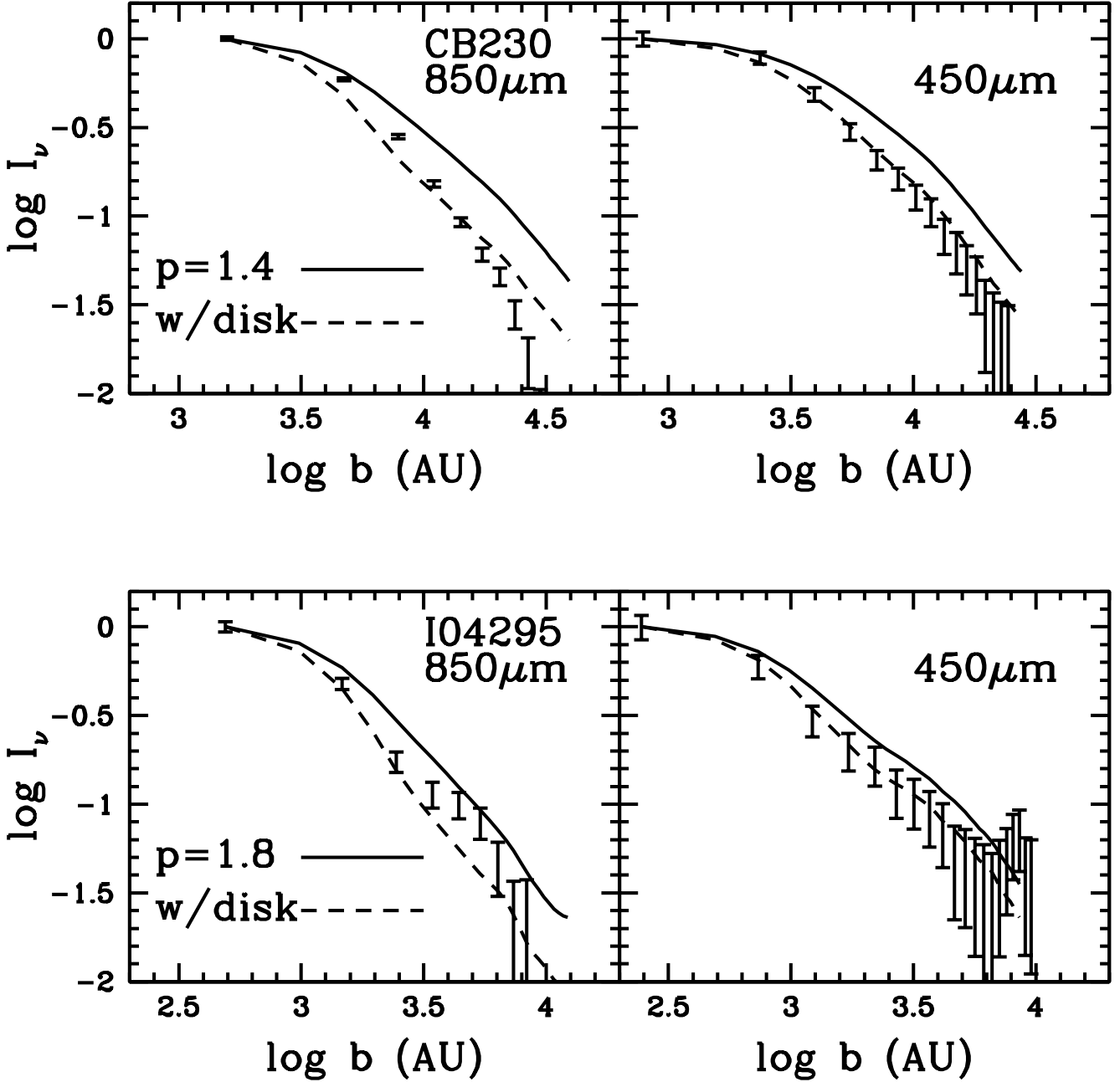}
\figcaption{\label{disk} The observed radial intensity profiles are shown as error bars for CB230 and IRAS 04285+2251.  The best-fit
models, assuming all emission originates in only the envelope, are $p=1.9$ and $p=2.3$ for CB230 and IRAS 04295+2251, respectively.  
The solid line represents a model of the envelope that has a shallower density profile by 0.5 ($p=1.4$ and $p=1.8$).   
The dashed line results from
adding a point source to this modeled envelope emission.  This point source can cause the modeled intensity profile from the  
envelope$+$disk to be considerably
steeper, and, hence,  the derived density power law for the envelope is shallower.  
We assume moderate fluxes for the disk and find the effect to 
be a decrease in $p$ by 0.5, but the effect could possibly be much more significant.}
\end{figure}

We provide a simple analysis to determine the effect of the dust emission from a disk on our models.  We assume the emission
from a disk to be included only in the central beam of our observations.  Mundy et al. (1996) found the disk for HL Tau to 
be $<$180 AU, and, since our beam subtends 2000-7000 AU, this assumption is reasonable.  For the two test cases, CB230 and IRAS 
04295+2210, we determine the effects of including a point source within the central beam that contributes one-half
of the measured flux.  
In Figure~\ref{disk}, we show the results of adding this point source to the modeled intensity profile.
The addition of this point source causes the modeled intensity profile to steepen---hence, the best-fit intensity profile,
when a point source is considered, can be matched by a \it shallower \rm density profile.  We find the best-fit density power law to be
shallower by $p=0.5$  when an unresolved component contributes one-half the flux measured within the central JCMT beamwidth.
Therefore, until the submillimeter emission from disks is better constrained, we attribute an additional source of 
uncertainty in our calculation of $p$: $\delta p=-0.5$.  However, this effect could be much more important
for Class I cores, where the envelope is less substantial, than for Class 0 cores.  

Observing submillimeter emission from disks requires submillimeter interferometers (e.g., SMA, ALMA),
but we make some simple estimates of the disk emission for one source, IRAS 04361+2547.  This source was observed in detail by
Terebey et al. (1990, 1993) at near-infrared and millimeter wavelengths from which  they  suggest the existence of a
$\sim$1000 AU disk or, at least, some dense component near the central position of this protostar.  At $\lambda=2.6$ mm, they
report $S_\nu = 21\pm7$ mJy (1993)---observations at the Owens Valley Radio Observatory (OVRO) with a synthesized
beam of $7\farcs7\times6\farcs6$.  Further, they claim that these flux measurements are dominated by the emission from
dust and not free-free emission.  
With our power law model for this source, we predict $S_\nu \sim 2$ mJy (at $\lambda=2.7$ mm); i.e., the observed
flux does require the existence of an additional source of flux other than the envelope.  Indeed, we consistently underestimate
the long-wavelength flux ($\lambda>1$ mm) with our model of the envelope for all sources, further suggesting the 
existence of some unresolved component in these cores. 

\begin{figure}
\figurenum{20}
\plotone{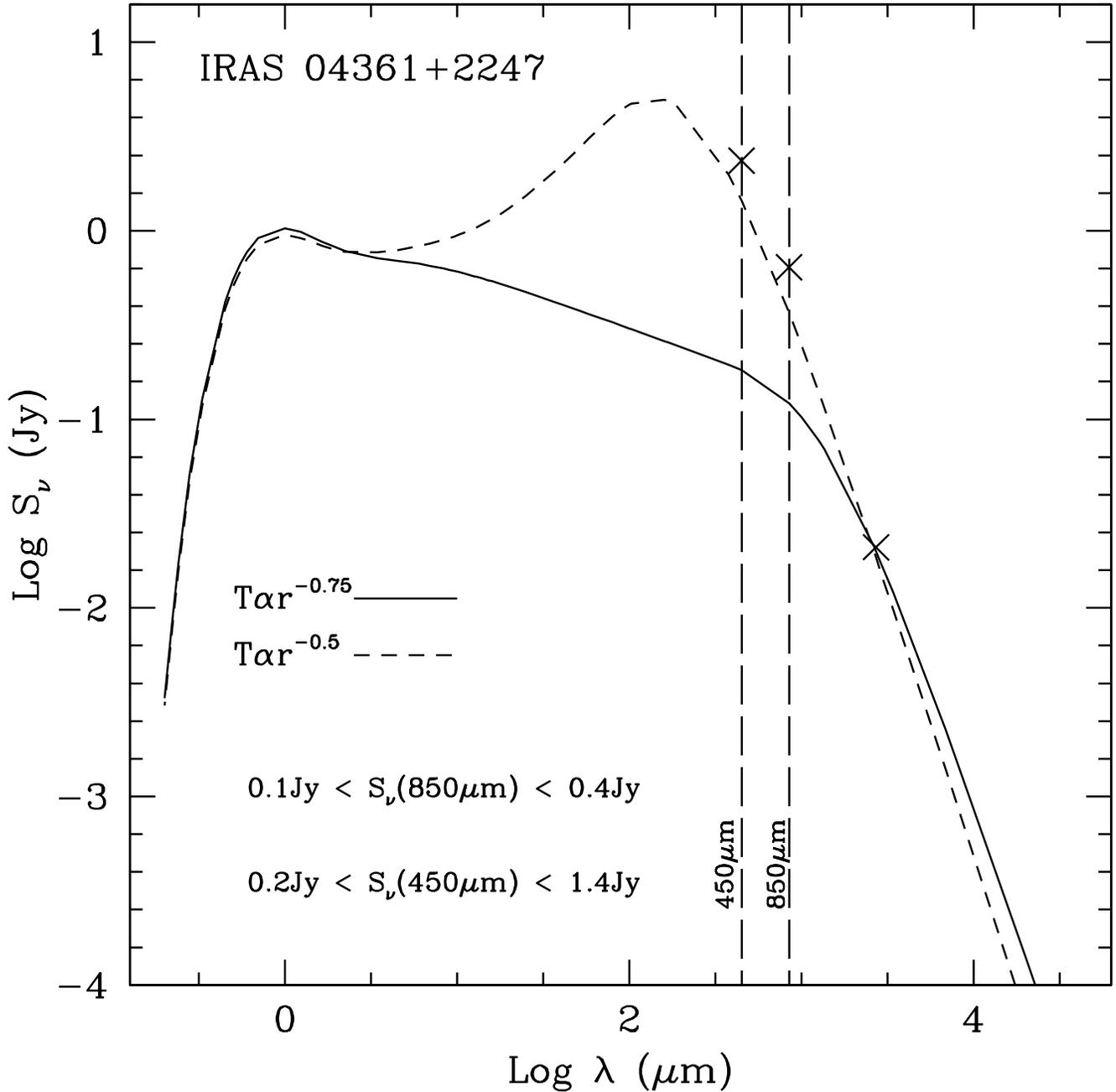}
\figcaption{\label{sed_disk} The modeled SED of a disk surrounding a protostar.  The surface density profile of the 
dust within this disk is a  power law
($\Sigma \propto r^{-1.5}$).  The solid line shows a model where the temperature power law is steep ($T \propto r^{-0.75}$) while the
dashed line model has a shallower profile ($T \propto r^{-0.5}$).  The latter is intended to simulate the effects of flaring.  
The crosses are observed 2.6 mm flux density from Terebey et al. (1993) and the submillimeter fluxes in this paper (for 40$\arcsec$ 
aperture).  The ranges for $S_\nu$ at 850 and 450 $\mu$m as derived
from the two models are given---these ranges are reasonable upper and lower limits on the dust emission from the disk at these 
wavelengths.}
\end{figure}

We adopt a simple disk model (Butner et al. 1994) in order to estimate the 850 and 450 $\mu$m flux for IRAS 04361+2547.  The
model consists of a central protostar surrounded by a disk whose surface density is described by a power law 
($\Sigma\propto r^{-1.5}$).  We assume
a dust opacity of $\kappa_\nu\propto \nu^\beta$ where $\beta=1$ and $\kappa$(850 $\mu$m)$=0.018$ cm$^2$gm$^{-1}$.  
In Figure~\ref{sed_disk}, we
show the results of two calculations.  The solid line represents a temperature profile in the disk that follows a power law,
$T \propto r^{-0.75}$.  The dashed line shows a disk with  $T \propto r^{-0.5}$;  the outer edges of the disk are hotter and, thus,
crudely simulate a flared disk.  From the model of the ``flared'' disk, we set reasonable limits for the flux at 850 and 450 $\mu$m: 
$S_\nu$(850 $\mu$m)$ < 0.4$ Jy  and $S_\nu$(450 $\mu$m)$ < 1.4$ Jy.  For each wavelength, these upper limits are 60\%\ of the observed
flux in a 40$\arcsec$ aperture. 

We have completed a similar analysis for IRAS 04108+2803.  Terebey et al. (1993) found $S_\nu$(2.6 mm)$=15$ mJy 
($\theta_{mb}=8\farcs5\times7\farcs4$) for this Class I source.  Through the simple disk model, we set the following upper limits:
$S_\nu$(850 $\mu$m)$ < 0.24$ Jy  and $S_\nu$(450 $\mu$m)$< 0.84 $ Jy. These upper limits correspond to 30$\%$ more than the observed 
850 $\mu$m flux  (within errors) and 74$\%$ of the observed 450 $\mu$m flux.  IRAS 04108+2803 is not resolved by our observations,
so it is conceivable that all of our observed flux is from a cental point source.  Likewise, the observed $2.7$ mm
flux for IRAS 04016+2610 (6.8 mJy)(Hogerheijde et al. 1997) sets an upper limit on the submillimeter fluxes 
originating from the presumed disk: $S_\nu$(850 $\mu$m)$ < 0.15$ Jy  and $S_\nu$(450 $\mu$m)$ < 0.73$ Jy, 25\%\ and 94\%\ 
of the observed
850 and 450 $\mu$m fluxes (40$\arcsec$ aperture), respectively.  The central component of IRAS 04016+2610 (L1489) was not resolved 
by our observations, so, again, it is possible that all of the submillimeter flux is from an unresolved central component.  Indeed, 
since the nearby condensation to IRAS 04016+2610 lies partially within the 40$\arcsec$ aperture, our reported flux values probably 
overestimate the flux of this Class I protostar.

Chandler \& Richer (2000) investigated the problem of disk contribution to the submillimeter continuum flux and concluded that the
existence of a disk in  Class 0 sources, where the envelope is more substantial, has little effect on the derived value for
$p$.  Further, they claim the disk does become a significant contributor in Class I objects.  The envelope is less prominent in 
these cores, and a higher fraction of flux coming from the disk (i.e., by their formalism,  decreased $f_{env}$) causes a 
lower value for $p$.  We conclude similarly, claiming that the inclusion of a disk causes
the value for $p$ to be decreased by 0.5; however, this effect could be significantly greater.    
Firmer conclusions await constraints from submillimeter interferometric observations.

\section{Individual Sources}

\subsection{CB230}
CB230 (also IRAS 21169+6804) is the most luminous source we observed.  It is also the most distant at 450 pc 
(Launhardt \& Henning 1997).  Yun \& Clemens (1994) observed a molecular outflow
in the $^{12}$CO 2-1 line for which the direction is shown on the 850 $\mu$m contour map in Figure~\ref{contour_a}.  

The submillimeter dust emission from CB230 is quite round with an aspect ratio of about 1. 
There does, however, appear to be a slight extension in the direction perpendicular to the outflow axis (oriented N-S).  Due to the
observed symmetry, we chose this source to model as a test case (see Table~\ref{cb230}).  The best-fit power law for CB230 
is $p=1.9$.  We find good fits for both the
850 and 450 $\mu$m radial profiles, and the spectral energy distribution is matched well except for the 450 $\mu$m flux.  
This discrepancy is
probably due to poor calibration at 450 $\mu$m, but, since the focus of our modeling (i.e., the radial distribution) is independent of
the absolute flux, we have not pursued this in great detail.

\subsection{IRAS 04295+2251}

IRAS 04295+2251 is the least luminous source we have modeled, with $L_{obs}=0.6$ L$_\odot$ and $L_{int}=0.3$ L$_\odot$.  It is clearly
a Class I core by the bolometric temperature criterion with $T_{bol}=270$ K, but it is Class 0 by its submillimeter luminosity---
$L_{obs}/L_{smm}=50$.  The core is resolved at 850 $\mu$m, but its deconvolved size is the least of 
any of our modeled sources 
($\theta_{dec} / \theta_{mb}=0.8$).  This source also has the highest derived value for $p$ (2.3).  We discuss the correlation between
$p$ and $\theta_{dec} / \theta_{mb}$ in a later section.   
From the modeled density profile, we found the envelope mass to be
0.10 M$_\odot$. We chose this source as one of our test cases (see Table~\ref{iras04295}) because of its symmetric morphology (aspect
ratio = 1) and high quality data (S$/$N=40 at 850 $\mu$m).

IRAS 04295+2251 is the only object in our sample that was detected by 
the NRAO VLA Sky Survey (NVSS) at 21 cm (Condon et al. 1998).  While it is not uncommon to find continuum radio emission 
($\lambda>3$ cm) from Class  I cores (Lucas et al. 2001), reports of strong emission at such a long wavelength are 
rare (cf. Skinner 1993).

\subsection{IRAS 04166+2706}

IRAS 04166+2706 is a protostar associated with the B213 dark
cloud in Taurus with a well-detected outflow in the northeast to southwest
direction (Bontemps et al. 1996).  
The observations of IRAS04166+2706 were originally presented
in Paper I; but, since the core was classified as a Class I source
(using the $T_{bol}$ criterion), we  model it here.  The submillimeter
emission is characterized by circularly symmetric contours except
for an extension perpendicular (southeast) to the outflow direction 
in the lowest contour interval (10\%\ of the peak or $3\sigma$; Paper I).
We report the fluxes in a 120$\arcsec$ aperture here:
$S_{850} = 1.9 \pm 0.6$ Jy and $S_{450} = 12.9 \pm 3.9$ Jy.  
The bolometric luminosity is 0.5 L$_\odot$
and the bolometric temperature is 75 K.

\subsection{IRAS 04248+2612}

IRAS 04248+2612 is about one arcminute north of the starless core  Barnard 217 (B217) and is coincident with HH31 IRS2.
Padgett et al. (1999) observed IRAS 04248+2612 with HST$/$NICMOS in the near-infrared bands.  Their images showed a clear bipolar 
structure, oriented NW-SE, that they claim is a reflection nebulosity, near-infrared light escaping by way of an outflow cavity.  
In addition,  
they observe a dark lane aligned perpendicular to this nebulosity.  Our submillimeter maps (which exclude B217), show two sets of
extensions:  1) on a small scale of $\sim$50$\arcsec$, in the direction of the bipolar nebulosity (NW-SE) and 2) on a larger 
scale ($\sim$200$\arcsec$), aligned with the dark lanes observed by Padgett et al. (1999) in a NE-SW direction.  This large-scale 
structure is probably indicative of a filamentary-like distribution of mass in the cloud.
The small-scale structure (within the central $30\arcsec$) is probably due to heating of the dust caused, in some way, by an outflow.     

Myers et al. (1988) report no  outflow activity in their
$^{12}$CO 2-1 observations, for which the noise in one channel was $\sim$0.2 K, but Moriarty-Schieven et al. (1992) find that the 
$^{12}$CO 3-2 line observed at the central position shows low velocity line wings ($\Delta v\sim4.5$ km s$^{-1}$).  Unfortunately, 
no molecular maps clarifying this outflow question have been published.  

Despite the asymmetric morphology of this source, we have attempted to model the dust emission.  The best-fit power law is the lowest
of any source, $p=0.8$, and renders quite a good fit (see Figure~\ref{model_e}).  Of course,  the extended emission is included in the 
azimuthal radial average, so one might argue that this is the source of the modeled, flattened density distribution.  However,
we excluded the emission from the large-scale, NE-SW extensions and found that the best-fit power law distribution 
is still very low: $p=1.0$.  IRAS 04248+2612 is clearly different from all of our other sources, both in morphology and modeled density
distribution.  Its aspect ratio (see Table~\ref{src_prop}) is 2.3, the highest of any source.  In Paper III, we showed that
there does seem to be some correlation between aspect ratio and density distribution---i.e., those sources with a high aspect 
ratio tend to have flatter density distributions.  It is possible that, in IRAS 04248+2612, we are observing  the formation of a star 
under different physical processes and initial conditions than with the other cores.

\subsection{IRAS 03256+3055}

IRAS 03256+3055 is at 320 pc in the Perseus star-forming region.  Unfortunately, this object has received little attention
observationally, but we offer the first map of its dust emission (Figure~\ref{contour_b}).  Ladd, Myers, \& Goodman (1994) reported 
undetectable NH$_3$ emission towards this object, but other reports of molecular line emission are found in the literature 
(Mardones et al. 1997, Gregersen et al. 2000).
We give its SED information in  Table~\ref{sed}; this consists of a 60 $\mu$m flux and our submillimeter fluxes.  Dent, Matthews,
\& Ward-Thompson (1998) report a very low 800 $\mu$m flux---0.125 Jy as compared to the 2.45 Jy we measure at 850 $\mu$m.  However, their
observations were with a single bolometer with a beamsize of $\sim$16$\arcsec$ while the emission from IRAS 03256+3055 spans 
$\sim$100$\arcsec$.  
 
The luminosity of this object is 0.7 L$_\odot$, and, from its poorly sampled SED, we also derive a bolometric temperature of 16 K.  
This bolometric temperature is much lower than that reported by Jennings et al. (1987)(74 K) and, subsequently, Mardones et al. (1997), 
but  they used IRAS upper limits on the flux at 12, 25, and 100 $\mu$m in their calculations. 
The envelope mass of IRAS 03256+3055 is 1.7 M$_\odot$ and is shared between, at least, three distinct 
dust condensations and significant extended emission.  

Nakamura \& Li (2002) have conducted axisymmetric calculations concerning the formation of stellar groups via fragmentation.  At least,
in morphology, their results are strikingly similar to what we see in IRAS 03256+3055---a ring-like structure with multiple condensations
of dust.  IRAS 03256+3055 provides an ideal laboratory in which to test the feasibility of such calculations as that presented by
Nakamura \& Li (2002) and, similarly, Boss (1993).

\subsection{IRAS 04016+2610, 04108+2803, 04361+2547, and 04385+2250}

These four cores are located in Taurus and associated with L1489, L1495, TMR1, and TMC1. 
(respectively, for IRAS 04016+2610, 04108+2803, 04361+2547, and 04385+2250).  
They are all Class I cores by their bolometric temperatures (i.e., $T_{bol}>70$ K), but the submillimeter luminosity
of IRAS 04108+2803 ($L_{obs}/L_{smm}=111$) identifies it as a Class 0 core.  

Each of these cores has a neighboring condensation as shown in Figures~\ref{contour_b}\ \&~\ref{contour_e}.  The 
flux condensation associated with IRAS 04016+2610 has been noted as a Class $-1$ core by some (e.g., Hogerheijde \& Sandell 2000), and  
we agree with this classification.  Hogerheijde et al. (1998) observed an outflow associated with this Class I core that is not coincident
with the flux condensation.

IRAS 04361+2547 does exhibit a peak of submillimeter flux that is coincident with the outflow direction (outflow observed by Bontemps 
et al. 1996).  Motte \& Andr\'{e} observed
this source  at 1.3 mm, but the condensation is not obvious at this longer wavelength indicative of the dust in this area being somewhat
warmer than, for example, the Class $-1$ core associated with IRAS 04016+2610, which was detected by Motte \& Andr\'{e} (2000).  
Perhaps, the dust is heated directly by the outflow, or it is swept up and then heated by radiation from the central protostar
that escapes by way of the outflow cavity.  Regardless, IRAS 04361+2547 is probably not a Class $-1$ core.  

IRAS 04108+2803 and 04385+2250 are both surrounded by significant extended emission, but, in contrast to 04361+2547 and 04016+2610,
this emission is not distinct from the central core.  It could be that these cores have evolved by some asymmetrical collapse or that they
are part of some larger star-forming complex.  In the next section, we discuss IRAS 04385+2250 in the context of 
forming substellar objects.  

\section{Discussion}
\subsection{Envelope Mass}

\begin{figure}
\figurenum{21}
\plotone{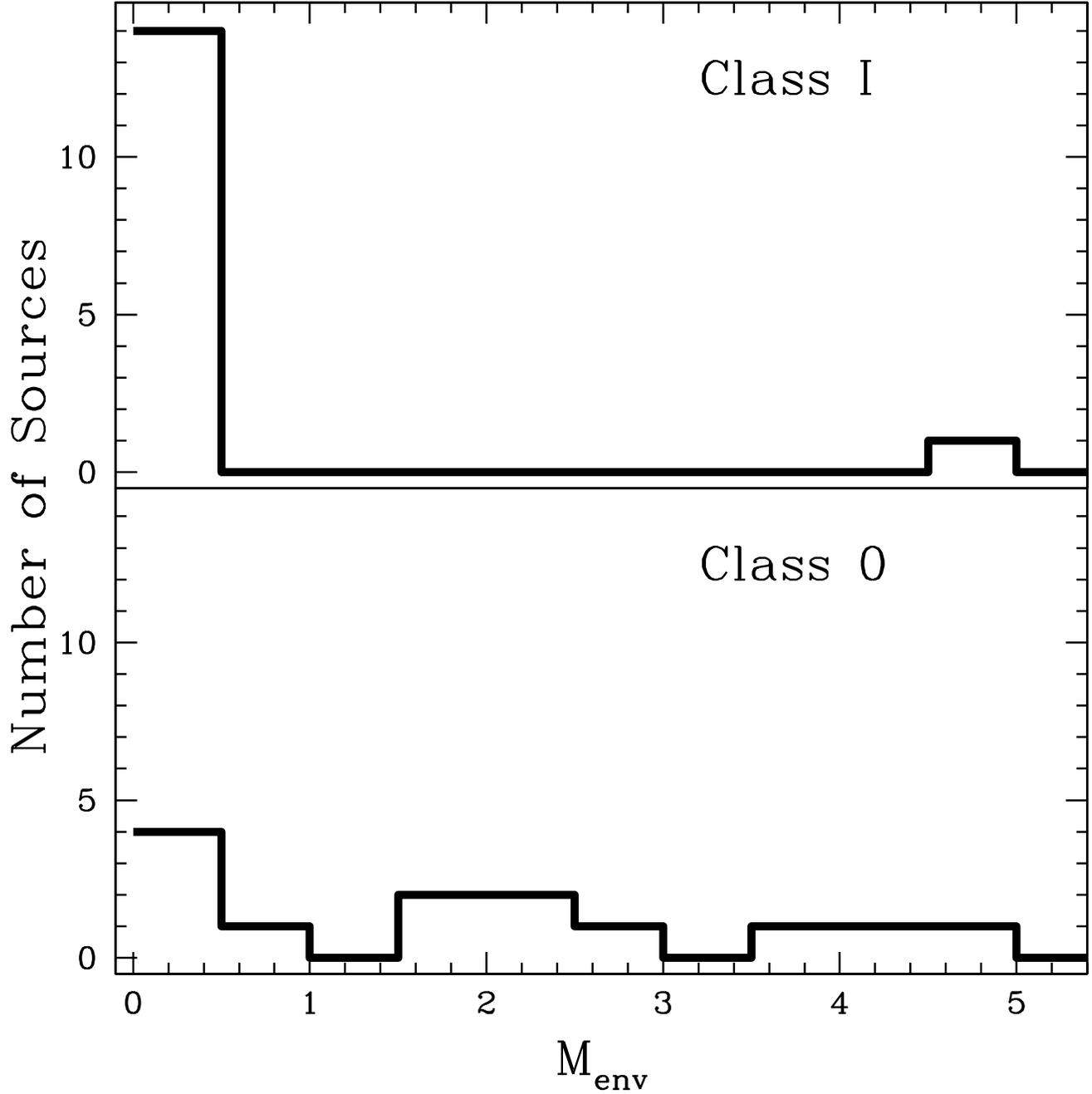}
\figcaption{\label{mdust} The histogram of envelope masses for Class I and Class 0 cores from this paper and Paper III.  We show
		the mass derived from the modeled density profile for the modeled sources and the isothermal mass (assuming 100:1 
		gas-to-dust ratio) as discussed in the text.  All Taurus cores in this sample have $M_{env}<0.5$ M$_\odot$.  Those
		sources with  $M_{env}>0.5$ M$_\odot$ have $T_{bol} < 75$ K.  }
\end{figure}

For the modeled sources, we have calculated the envelope mass by integrating the  best-fit power law density distribution 
to an outer radius of 60$\arcsec$ ($M_{env}$ per Equation 6 in Paper III). 
Because the cloud is optically thin to submillimeter wavelength radiation, the following expression is
is often used to estimate the envelope mass (Hildebrand 1983):
\begin{equation}
M_{env}^{iso} = \frac{S_\nu D^2}{B_\nu(T_{iso}) \kappa_\nu};
\end{equation}
where $S_\nu$ is the flux density at 850 $\mu$m in a 120$\arcsec$ beam (Table~\ref{flux}), $B_\nu$ is the Planck function, 
$\kappa_\nu$ is the opacity per gram of gas and dust at 850 $\mu$m ($\kappa_\nu= 1.8\times10^{-2}$ cm$^{-2}$gm$^{-1}$ for OH5 dust), 
$T_{iso}$ is the isothermal dust temperature, and $D$ is the distance to the source.  Of course, the dust is not isothermal, 
but we use our models to estimate the isothermal temperature ($T_{iso}$) that makes Equation 2 agree with the masses derived
from the more detailed models (see Equation 7, Paper III).  We use the resulting
average $T_{iso}$ (16$\pm$4 K) for the modeled sources to calculate the isothermal envelope mass ($M_{env}^{iso}$) for the
unmodeled sources.
The values for $M_{env}$ and  $M_{env}^{iso}$ are in Table~\ref{mass}.
In Figure~\ref{mdust}, we show a histogram of these calculated masses  for the 
Class I sources in this paper and the Class 0 cores in Papers I and IV.  The envelopes of Class I cores are 
clearly less massive than Class 0 sources.  Only CB230, of the Class I cores, has $M>0.5$ M$_\odot$, and it is very close to 
the Class 0 boundary with $T_{bol}=74$ K.
The remaining cores have  $M< 0.5$ M$_\odot$  and are in Taurus.  

\begin{figure}
\figurenum{22}
\plotone{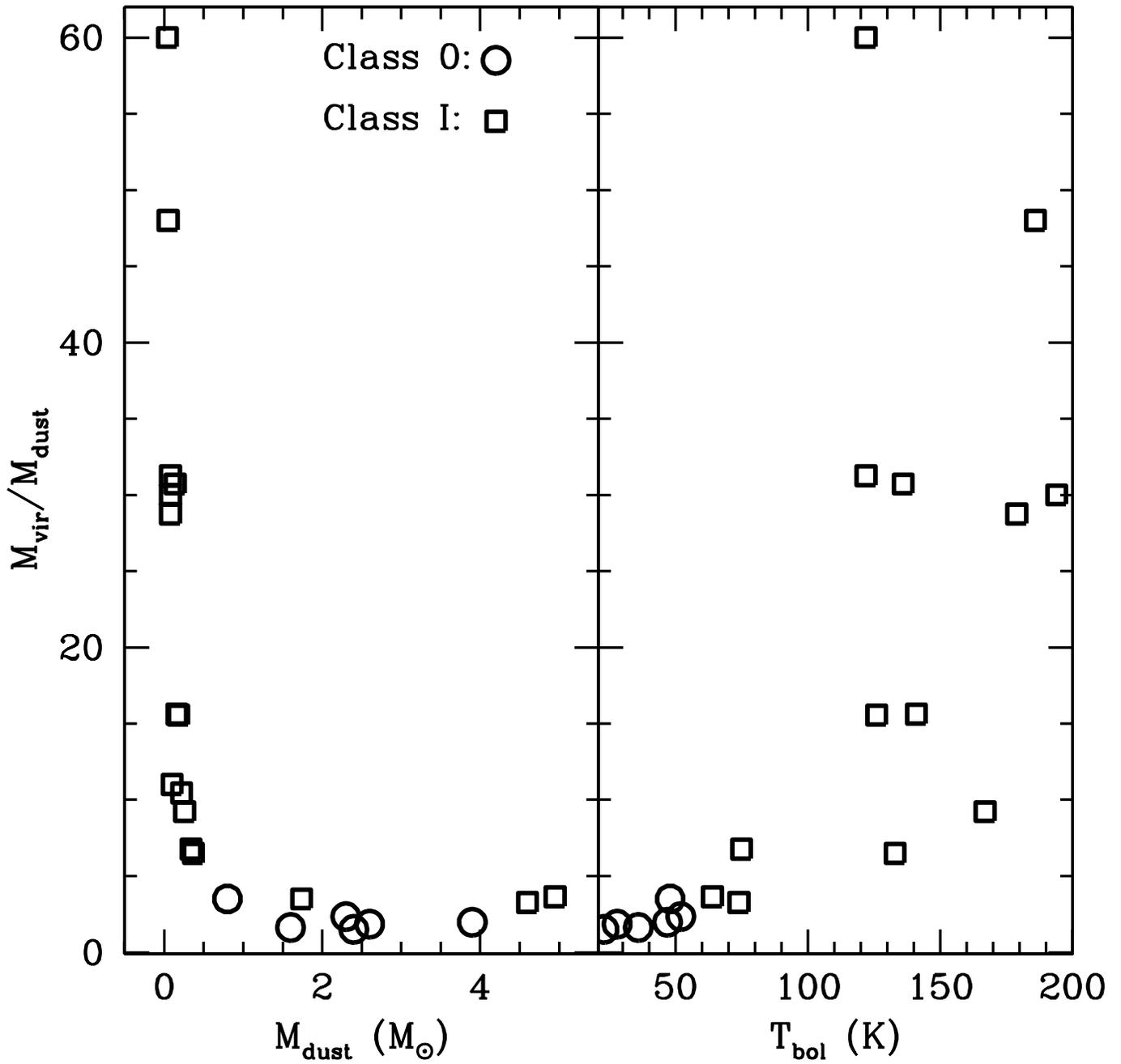}
\figcaption{\label{mvir}We plot the ratio of virial to dust mass ($M_{vir}/M_{dust}$) against $M_{dust}$ and $T_{bol}$. Class I cores
		are represented by squares; Class 0 cores (from Paper III) are circles. }
\end{figure}

In Table~\ref{mass}, we also give the FWHM linewidth, associated reference,  and the calculated virial mass.  The virial mass,
$M_{vir}^{\theta_{ap}}$, is corrected for the best-fit power law density distribution as in Paper III (Equations 8 \& 9).  
For those sources not modeled, we adopt the average $p$ (1.6) to calculate the virial mass.  To be consistent with our dust
mass calculations, we adopt an aperture of $\theta_{ap}=120\arcsec$, which corresponds, in Taurus, to 16,800 AU.    
We plot the ratio of virial to dust mass ($M_{vir}/M_{dust}$) against $M_{dust}$ and $T_{bol}$ in 
Figure~\ref{mvir}.  For $T_{bol} > 70$ K,
these two calculations of the envelope mass begin to diverge.  Reasonable variations in the dust opacity can introduce a factor
of ten uncertainty (Ossenkopf \& Henning 1994), but these variations cannot account for values of $M_{vir}/M_{dust}\sim60$.  
One scenario is that the molecular tracers used, mostly N$_2$H$^+$, do not  trace the material surrounding Class I
cores.  Caselli et al. (2002) mapped two Class I cores coincident with our sample: IRAS 04016+2610 and IRAS 04248+2612 (listed
as L1498 and B217, respectively, by Caselli et al.).  In both cases, the nearby preprotostellar core is observed to have strong
emission in these tracers, but the tracer shows no peak on the Class I core.  
Simply, observations of N$_2$H$^+$ do not trace the  mass in these Class I cores.

\subsection{Luminosity and Stellar Mass}
     
As discussed in the previous section, the masses for the envelopes of these cores are very low.  
Likewise, the observed luminosities are very low.
For those cores in Taurus, we find a range of bolometric luminosities from 0.2 to 4.6 L$_\odot$ with an average of 
1.4$\pm$1.3 L$_\odot$.  These luminosities are upper limits on the luminosity of the central protostellar
system because of the ISRF contribution.  
For the most luminous sources, L1251B and CB230, the ISRF contributes negligibly to the bolometric luminosity, but, for 
those cores in Taurus, the ISRF accounts for a considerable fraction of $L_{obs}$.  
We estimate the  total contribution of the ISRF to the observed luminosity to be about 0.2 L$_\odot$.

Based on the standard picture of a Class I protostar in which the observed luminosity results from accretion, 
we can calculate the mass of this protostar by a simple relationship (e.g., Stahler 1994):
\begin{equation}
M_\ast=\frac{L_{acc}R_\ast}{G \dot{M}}.
\end{equation}
Here, $M_\ast$ is the stellar mass, $L_{acc}$ is the luminosity due to accretion, $G$ is the gravitational constant, 
$R_\ast$ is the stellar radius, and $\dot{M}$ is the accretion
rate given as $\dot{M}=n_oa_T^3/G$, where $a_T$ is the isothermal sound speed and $n_o$ is a numerical constant of order unity 
(Stahler 1994).
Assuming a minimum isothermal sound speed of 0.2 km s$^{-1}$, we calculate $\dot{M}=2\times 10^{-6}$ M$_\odot$yr$^{-1}$.
Then assuming $R_\ast=3$ R$_\odot$, an overestimate of the stellar radius based on the models of Stahler (1988), 
we can calculate an upper limit on the stellar mass from this relationship.
For the least luminous of the cores in this sample (IRAS 04385+2250; $L_{obs}=0.2$ L$_\odot$), the mass, based on Equation 3, is
$M_\ast=0.01$ M$_\odot$ or, since this in the substellar realm, $M_\ast=10$ M$_{Jup}$.  
The envelope mass for this core is 0.04 M$_\odot$,
and, assuming a star formation efficiency of 25\%\ (Wilking et al. 1989), the final mass of this object would be $M_\ast=0.02$ M$_\odot$ 
(or 20 M$_{Jup}$) in about $10^4$ yr (assuming constant  $\dot{M}=2\times 10^{-6}$ M$_\odot$yr$^{-1}$).  

This analysis holds many shortcomings and uncertainties.  
The mass accretion rate ($\dot{M}$) is highly uncertain, possibly episodic, and perhaps  $1-2$ orders of magnitude smaller 
than that used in this calculation (e.g., Muzerolle et al. 1998).  
Such an accretion rate would increase $M_\ast$ by a factor of $10-100$.   On the other hand, the 
observed luminosity is probably some combination of the disk accretion luminosity and luminosity of the central protostar.    
Indeed, Muzerolle et al. (1998) find, in some cases, the accretion luminosities contribute $\sim10$\%\ to the total observed
luminosity.  If only 10\%\ of the total luminosity results from accretion (as estimated by Muzerolle et al. 1998), the calculated stellar 
mass is a factor of 10 smaller.

We can also estimate the stellar mass by other means.  If we assume an age ($\tau$), we can place IRAS 04385+2250 on 
calculated isochrones.  The models of Siess et al. (2000), with $\tau=10^6$ yr, predict $M_\ast\sim0.1$ M$_\odot$. 
Burrows et al. (2001) give $M_\ast\sim0.2$ M$_\odot$ for $\tau=10^6$ yr.  These models probably
overestimate $M_\ast$, however, because they assume no accretion and have not considered the initial conditions governed by
infall and collapse of the protostellar envelope.  Also, the age of embedded Class I cores has been estimated to be 
$\sim10^5$ yr (e.g., Myers et al. 1987, Kenyon et al. 1990);
most stellar evolution models begin at $\tau=10^6$ yr and are highly uncertain for ages less than $10^6$ yr.  
Siess et al. (1999) included accretion  ($\dot{M}=10^{-7}$ M$_\odot$yr$^{-1}$) in their calculations of pre-main sequence 
evolutionary tracks, and, for $\tau=4.5\times10^6$ yr, their models would indicate 
$M_\ast<0.1$ M$_\odot$; unfortunately, these models do not include objects with masses lower than 0.1 M$_\odot$ as 
required to accomodate IRAS 04385+2250.  Additionally, the application of these models to Class I cores is quite precarious;
these models are intended for T Tauri stars.  

Observations of young, forming (Class 0/I) brown dwarfs have yet to be firmly established, but the existence of older substellar 
objects with disks (e.g. Natta \& Testi 2001) suggest that brown dwarfs do form by similar mechanisms as T Tauri stars.
If so, we, indeed, probably have observed brown dwarfs within their natal envelopes.  Then, it should not be surprising that, in this 
sample of low-luminosity Class I cores, a fraction are substellar.  Indeed, Kenyon \& Hartmann (1995) concluded similarly, i.e. the
``luminosity problem'' for Class I objects is solved if such objects are substellar.  Of course, these authors also present another
viable solution, that the accretion radius is larger in these young objects than we expect.

\subsection{Classification: L$_{obs}$, T$_{bol}$, and L$_{smm}$}
 
Currently, the 
class system (Class $-1$---III) implies a sequence of stages that are occupied at some point by all evolving protostars (e.g., 
Andr\'{e} \& Montmerle 1994). Hence, statistical arguments have been constructed by comparing the numbers of sources in each stage  to
establish the relative time spent in each epoch (e.g., Andr\'{e} \& Montmerle 1994).  
As discussed previously,
three criteria for classification have been put forth:  $\alpha_{NIR}$ (Lada et al.  1987), $T_{bol}$ (Myers \& Ladd 1993), and
$L_{obs}/L_{smm}$ (Andr\'{e} et al. 1993).  These values are given in Table~\ref{src_prop} for our sample of sources.    

For  the selection of the SED points from which these values were calculated (see Table~\ref{sed}), we required that the aperture 
be large enough to include all emission:  $\theta_{mb}>20\arcsec$.  This restriction causes the exclusion of the near-infrared (NIR)
fluxes, but we also calculate $T_{bol}$  with the NIR data because the bolometric temperature is most affected by the NIR data.  
The NIR flux, no matter 
how weak, causes this mean frequency to be skewed to higher values.  Indeed, if we do not include the NIR flux in our calculation
of $T_{bol}$, 35\%\ of our Class I sources become Class 0 by the $T_{bol}$ criterion.  Of course, Class I sources are 
identified by their NIR emission 
and are even called, by some,  NIR protostars.  However, detection of the NIR light is often dependent on orientation and, 
hence, could feasibly 
be obscured in an otherwise Class I object.

An additional complication in distinguishing between Class I and 0 cores is found in the luminosity scheme developed by Andr\'e
et al. (1993).  They define a Class I source as one  for which $L_{obs}/L_{smm}>200$, assuming that,
as a protostar evolves, the envelope should be cleared away and, hence, this ratio should increase.  
Originally, this indicator was based on measurements within a small aperture (20$\arcsec$) because of nearby sources.  However,
for our isolated cores, it is clear that such a small aperture does not sample the total submillimeter flux emitted by the object. 
Therefore, we use only large aperture measurements ($>20\arcsec$) in the calculation of these values (Table~\ref{src_prop});  for
our submillimeter fluxes, we used the 40$\arcsec$ aperture fluxes in the calculation of $L_{obs}/L_{smm}$ so that nearby condensations
would not contribute to the submillimeter flux.

\begin{figure}
\figurenum{23}
\plotone{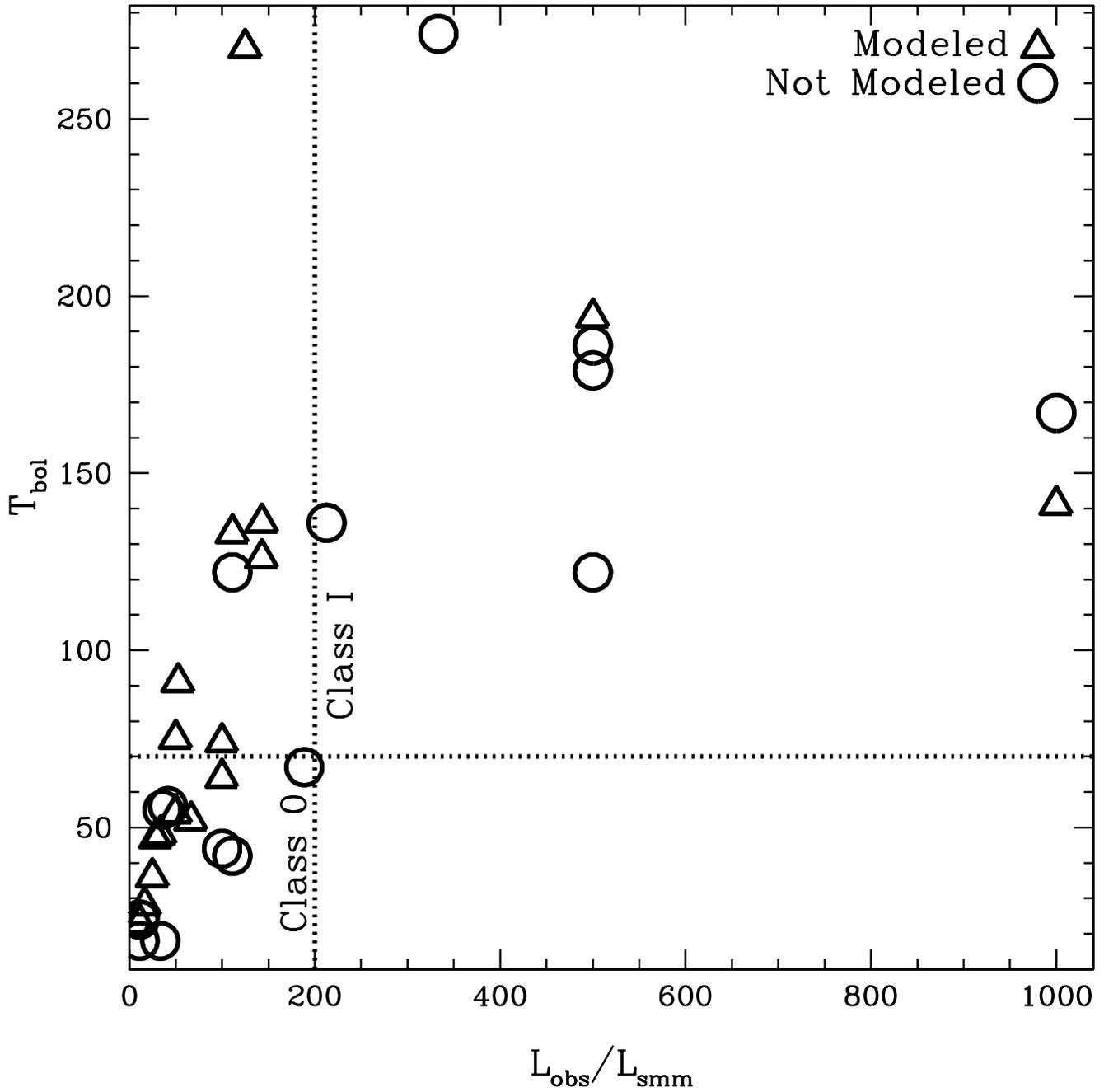}
\figcaption{\label{class} The bolometric temperature ($T_{bol}$) is plotted against $L_{obs}/L_{smm}$.
The dashed lines represent the boundary between Class 0 and I.  Class I 
objects have $T_{bol}>70$ K (Myers \& Ladd 1983) and $L_{obs}/L_{smm}>200$ (Andr\'e et al. 1993). }
\end{figure}

In Figure~\ref{class}, we show a plot of $T_{bol}$ versus $L_{obs}/L_{smm}$ that includes all sources from Paper I and this
paper.   Class I cores are characterized by $T_{bol}>70K$  (Myers \& Ladd 1993, Chen et al. 1995) or $L_{obs}/L_{smm}>200$
(Andr\'{e} et al. 1993).  
We have subdivided the plot into four regions based on these boundaries.  Class I cores should be found only in the upper right-hand
portion of the plot, and Class 0 cores should be in the lower left-hand portion.  
This is clearly not the case.  While there is a noticeable correlation between the two 
evolutionary indicators, the divisions between the classes are not so obvious.  Indeed, all of the cores in this paper's sample are
Class I by the $T_{bol}$ criteria (except for L1251B and IRAS 03256+3055);  half of these Class I cores, however,  
are Class 0 by the $L_{obs}/L_{smm}$ criterion.  
The problem arises in that the $T_{bol}$ observable is greatly affected by geometry whereas $L_{obs}/L_{smm}$ 
is not so affected.  Visser, Richer, \& Chandler (2002) also found such a discrepancy between these classification criteria.  
They find that the submillimeter luminosity criterion ($L_{obs}/L_{smm}>200$) renders far fewer Class I cores than does the bolometric 
temperature.  Further, they infer lifetimes for the evolutionary stages based on the numbers of Class 0 and I cores.
In order to create such statistical arguments concerning the lifetimes of Class $-$1-III objects, one
must recognize that different numbers of Class 0 or I objects will be found depending on the system of classification employed. 

\subsection{Density Structure}

\begin{figure}
\figurenum{24}
\plotone{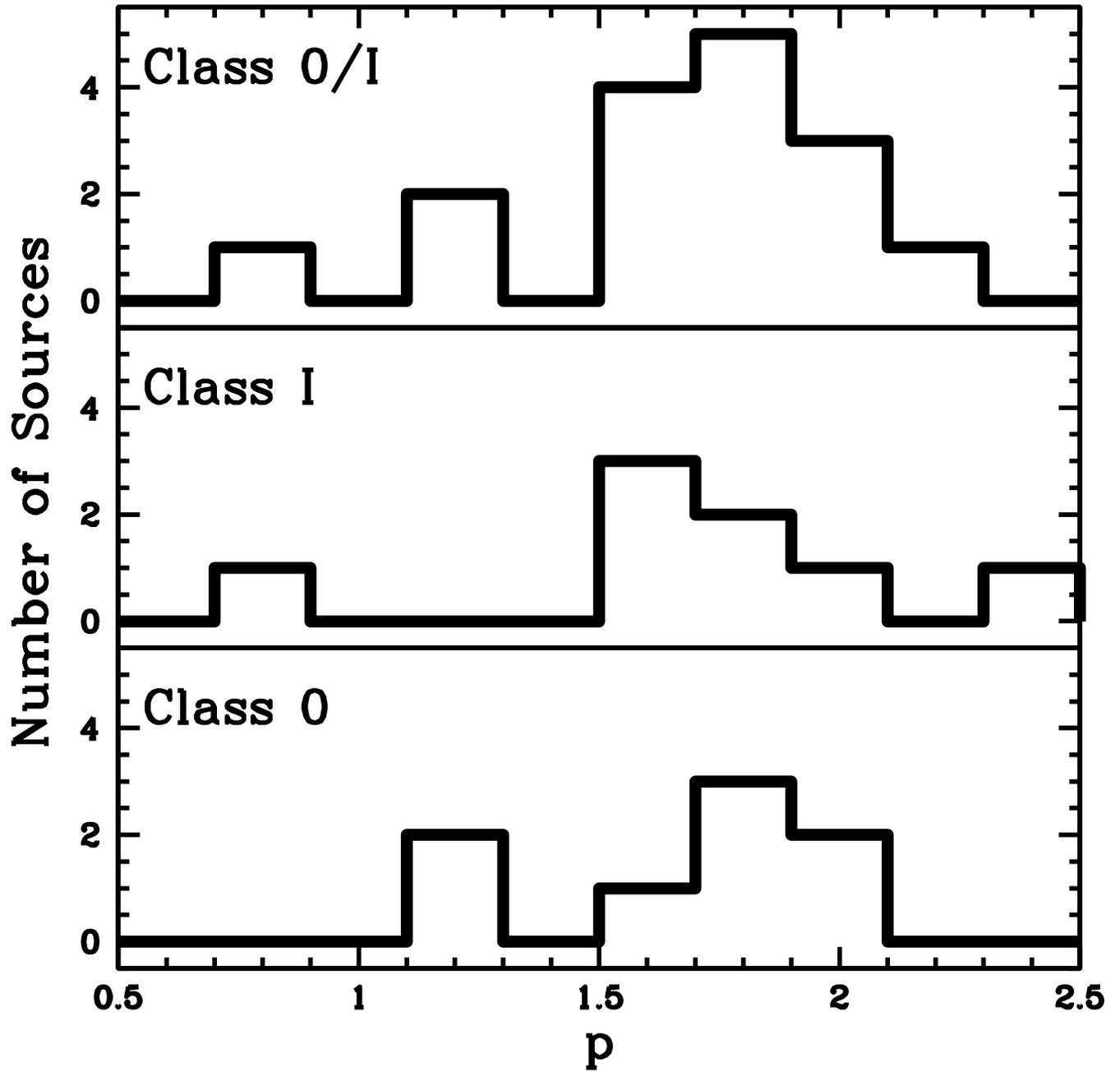}
\figcaption{\label{hist_p} All modeled sources from Paper III and this paper are represented in  
		the top histogram.  Class 0 and I (from Paper III
		and this paper) sources are shown in the bottom and middle histograms, respectively.}
\end{figure}

The primary purpose of our modeling is to derive the actual distribution of matter in the envelopes of these forming protostars.
For the Class I cores in this paper, we find the density power law exponent 
$\langle p\rangle=1.6\pm0.4$ (mean $\pm$ the standard deviation of the sample); 
however, the presence of a disk or point-like component could cause $p$ to decrease by as much as $0.5$.  The median for this sample
of Class I cores is $p=1.8$. In Paper III, we reported $\langle p\rangle=1.6\pm{0.3}$ for the Class 0 sources. 
For the Class 0 and I objects in Papers III and IV, we find $\langle p\rangle=1.6\pm0.4$, and the median 
is $p=1.8$.  If we leave out those Class 0 or I sources with aspect ratios $>1.5$, a total of four sources from Papers III and IV, 
we find $\langle p\rangle=1.8\pm0.2$, and the median is also $p=1.8$.
We include these modeled sources in a histogram (Figure~\ref{hist_p}) both separately and together.  The value for $p$ is clearly peaked
between 1.5-2.0, seemingly inconsistent with models that predict very shallow distributions---e.g., McLaughlin \& Pudritz (1997).
However, these models should not be excluded altogether, for we do see evidence, in some sources, of a shallower profile, and
the inclusion of a disk allows for significantly shallower density profiles.  

\begin{figure}
\figurenum{25}
\plotone{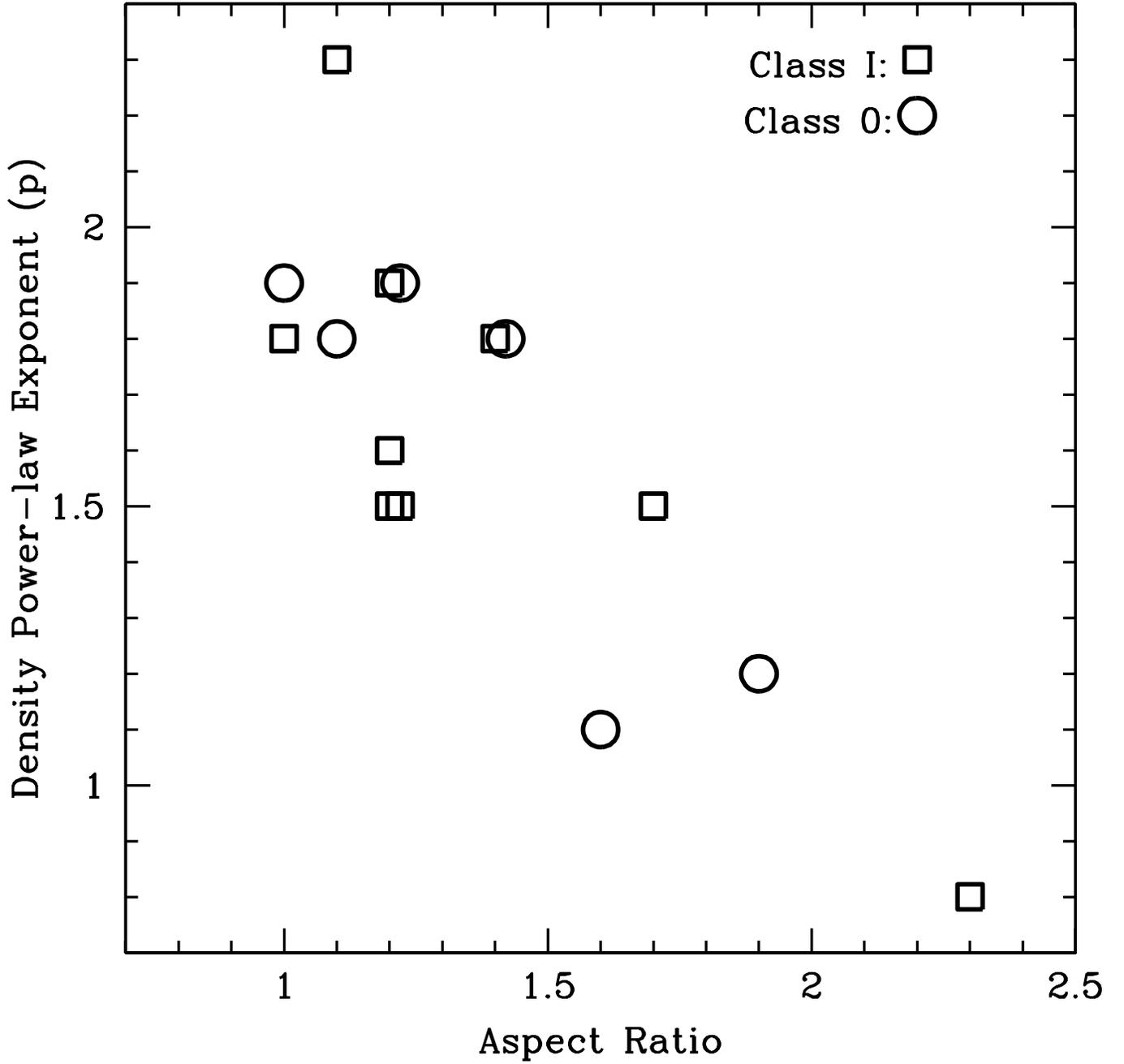}
\figcaption{\label{aspect}The aspect ratio (given in Table~\ref{src_prop}) is on the horizontal axis and the modeled density power law 
exponent ($p$) is on the vertical axis.  More aspherical objects are best modeled with shallow density profiles---an effect
that  is possibly real and not just an artifact of our one-dimensional model.}
\end{figure}

In Figure~\ref{aspect}, for all modeled sources (from Paper III and IV), we show the aspect ratio as a function of $p$.  A clear
correlation exists between these two values such that the more aspherical sources exhibit shallower profiles.  Of course,
the azimuthal averaging of an aspherical object could result in a shallow intensity profile (and, hence, shallower density profile),
but we claim this is a small effect.  In Section 4.4, we discussed the modeling of IRAS 04248+2612 and how the derived density
power law changes negligibly with exclusion of the extended emission in the NE-SW extensions.  We also found this to be true 
in aspherical, Class 0 sources---see Paper III.
 
\begin{figure}
\figurenum{26}
\plotone{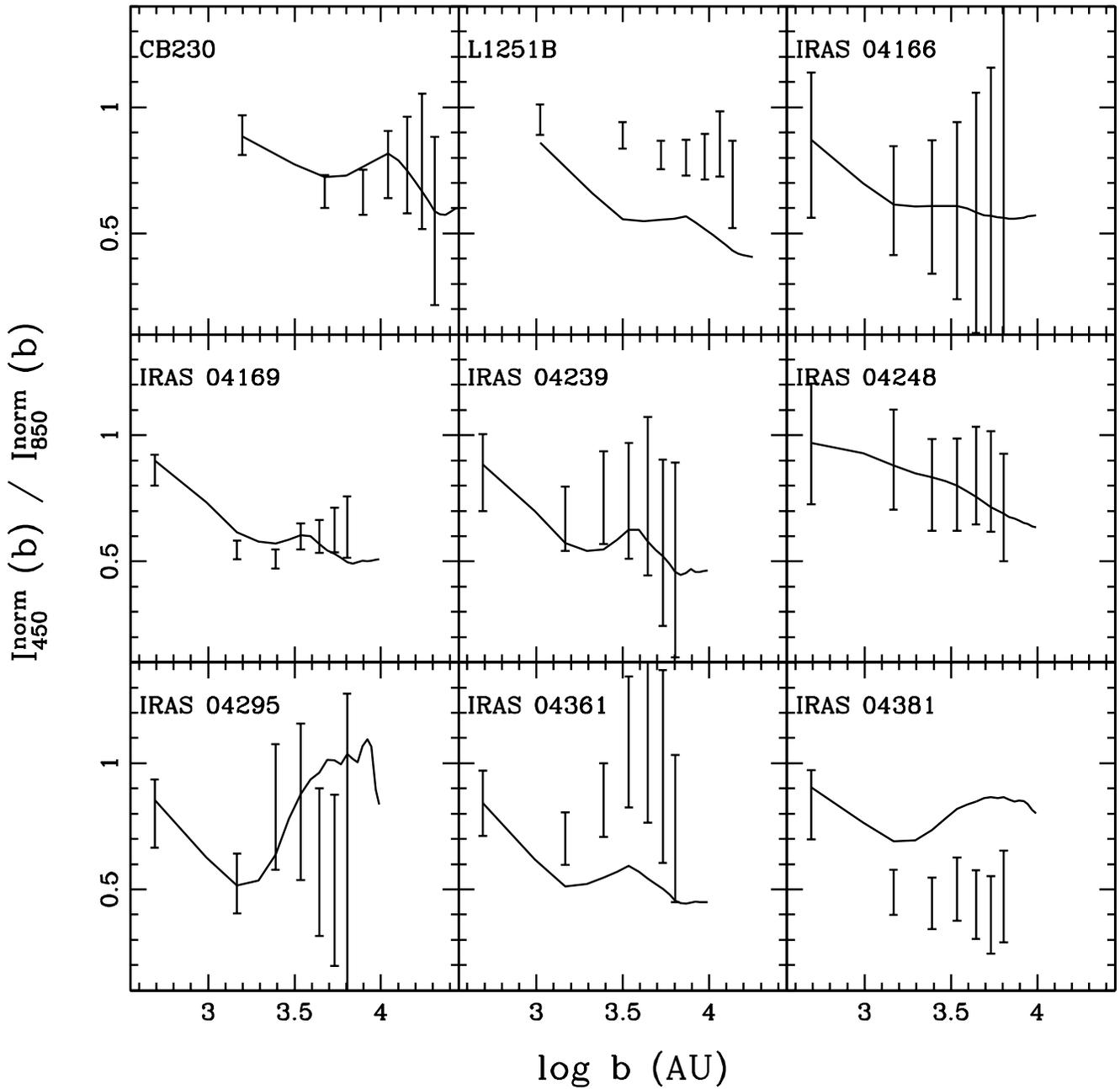}
\figcaption{\label{spect}The error bars represent the observed ratio of 450 to 850 $\mu$m flux.  The solid line represents 
		the modeled value.}
\end{figure}

As in Paper III, we find little evidence for variation of dust properties within the envelope.  We compare the normalized
intensity ratios ($I^{norm}_{450\mu m}/I^{norm}_{850\mu m}$)  in Figure~\ref{spect}\  and conclude that, for most sources, 
the observed variations in this ratio are well-fitted by including the effects of external heating by the ISRF and 
convolution of the actual beam.  For L1251B, IRAS 04361+2547, and IRAS 04381+2540, we are unable to simultaneously 
match the observed 850 and 450 $\mu$m intensity profiles.  The dust opacity may vary with radius in these sources,
but it could also be that we have not appropriately modeled the temperature in these cores, perhaps, due to 
asymmetric heating, externally, by the ISRF or, internally, by a disk$+$protostar or binary central source.

\begin{figure}
\figurenum{27}
\plotone{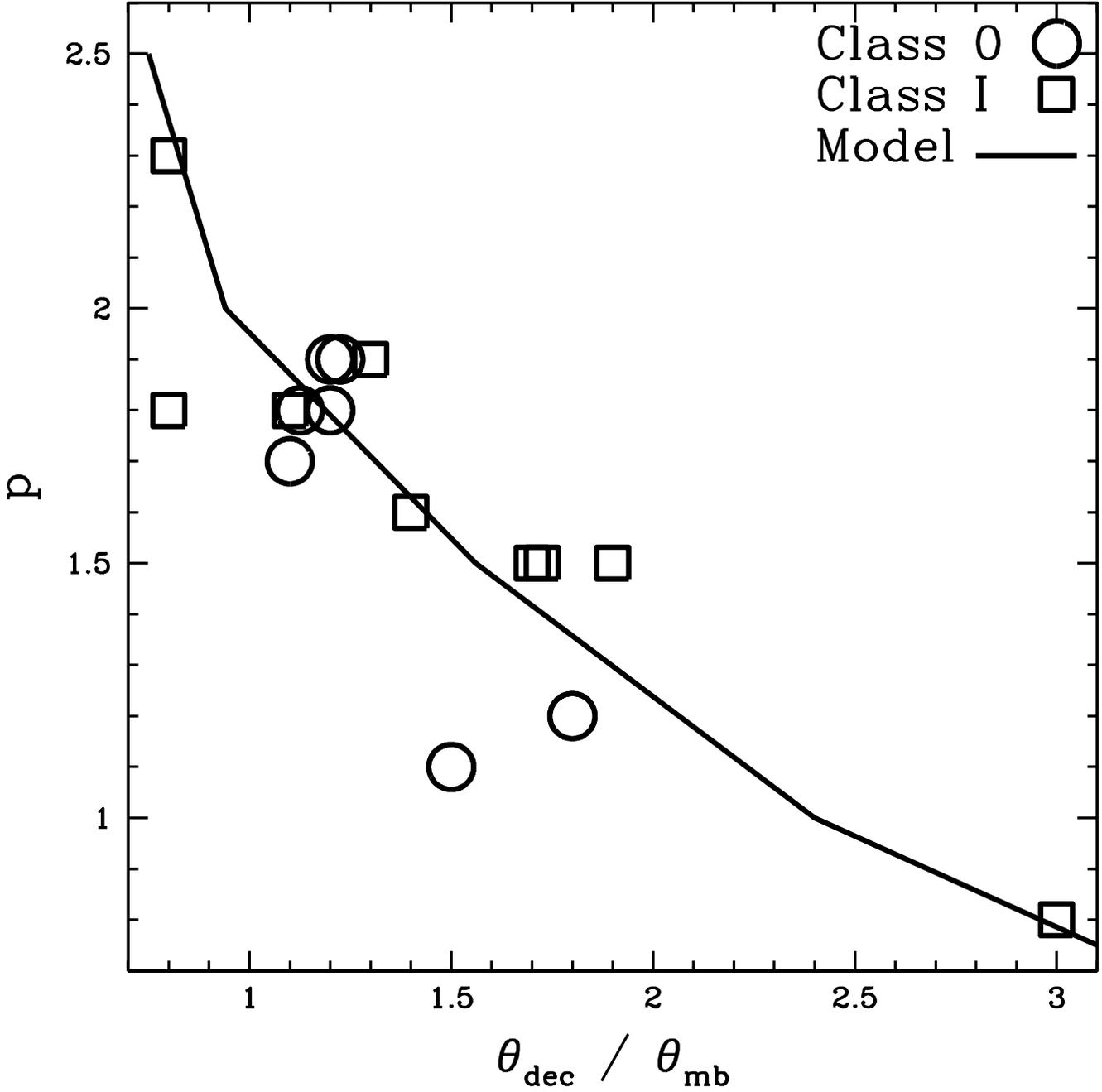}
\figcaption{\label{beams_p} The derived value for $p$ is plotted against the ratio of deconvolved source size and beam size (at FWHM). 
		The solid line represents models of the dust emission with $0.5 < p < 2.5$.  
		We note an obvious trend of increasing $p$ for the less well-resolved sources.  This trend is possibly an
		effect due to the cores compact nature, but we also note the possibility that we are observing the outer parts of the
		envelope in these ``smaller'' cores. }
\end{figure}

While power law density distributions do not have a physical size, we define an observational size from the deconvolved FWHM 
of our modeled sources (Table~\ref{src_prop}).  In Figure~\ref{beams_p}, we show the derived
value for $p$ plotted against the effective size ($\theta_{dec}/\theta_{mb}$) for the modeled Class 0 and I cores 
(from Papers III \& IV).  
Additionally, the solid line shows a series of models for $0.5 < p < 2.5$ for which the deconvolved model size is measured with
the 15$\arcsec$ beam.  The models show that steeper power laws are clearly less well-resolved in agreement with the observed trend. 

The trend in Figure~\ref{beams_p}, while not surprising,  is quite striking; smaller sources exhibit a larger value for $p$.  We 
stress this point for three reasons. First, Terebey, Chandler, \& Andr\'{e} (1993) addressed this resolution issue---that steeper
power law density distributions are less well-resolved.   However, one must be very careful when analyzing the intensity profiles
of marginally resolved sources because uncertainty in the beamshape can be $\sim1-3\arcsec$ (see Section 3.2).  
Indeed, there are examples of marginally resolved sources for which the authors have derived very steep density power laws ($p>2.5$),
density distributions that have little theoretical basis.  
 
Also, we propose an intriguing possibility implying this trend is not entirely 
systematic and might be indicative of some physical quality.  Perhaps, we are observing only the outer envelope 
of these less well-resolved cores.  That is, these ``smaller''
sources are not small because they are more evolved and have collapsed over time, but they are small because they began with an initial 
preprotostellar core that was lower in mass and, hence, has a smaller outer radius than the other cores in our sample.  
In the Shu collapse model, we would expect a steeper
density profile in the outer, static envelope, and it might be that, indeed, we are seeing just that.  

A third scenario is also possible.  Perhaps, the ``smaller'' cores, which we interpret to have steeper density profiles, have a
greater disk contribution.  This point source of emission, as shown in a previous section,  results in the derivation of a steeper
power law distribution in the density.  In this case, the envelope mass distribution could be considerably shallower that we 
predict here.  Indeed, if the disk contribution is more significant as the core size decreases, the trend for the envelope density
distribution shown in Figure~\ref{beams_p} could be shifted such all cores harbor an envelope described by $1.0<p<1.5$, a possibility
that would suggest the validity of the theory presented by McLaughlin \& Pudritz (1997).

We compare these results to those recently presented by J{\o}rgensen, Sch\"{o}ier, and van Dishoeck (2002) (hereafter, JSD) 
and find that we reach similar conclusions.  
These authors model two sources coincident with our sample: IRAS 04361+2547 and IRAS 04381+2540, which they call TMR1 and TMC1, 
respectively.  Our derived values of the density power law exponent for these two sources are 
consistent with those of JSD.   However, their modeling differs from ours in the assigning of two parameters: 
the ISRF and outer radius; additionally, they do not simulate the effects of chopping.  
Exclusion of the ISRF should cause JSD to derive a shallower power law exponent to fit the 
observed data.  Also, JSD adopt much smaller outer radii than we do in this paper---in some cases, a factor of four smaller.  
By assigning a small outer radius and excluding the effects of chopping, these authors should derive a somewhat steeper power law 
for the density distribution than that presented here.
The combination of these two effects (i.e., exclusion of the ISRF and truncated outer radius) seem to cancel in 
their analysis, rendering similar results to our own.

\subsection{Caveats} 

As presented in the previous sections, we have discussed several caveats that hinder direct interpretation of the dust continuum
emission:  beam convolution of our model, complex temperature distributions due to external heating, different assumed dust opacities,
contribution to the observed flux from a disk component, and the inner and outer radii assumed in the model.  We do not analyze the 
effects of outflows because, for most of our sources, the outflows are either very weak or undetected (e.g., Bontemps et al. 1997).
Our modeling is also limited by its one-dimensional nature, but we point out objects for which three-dimensional modeling will
prove fruitful (e.g., IRAS 04248+2612).  Our assertion that asymmetric sources exhibit shallower density distributions 
suggests that these objects are forming in truly different environments.  However, this issue will not be fully resolved until
models of radiative transfer for three dimensions are completed.

\section{Conclusions}

Our conclusions are as follows:

1) We find that $3/16$  (20\%) of the sources in this sample have distinct, neighboring condensations (IRAS 03256+3055, 
04016+2610 and 04361+2547).  These condensations could be the result of complex structure (03256+3055), neighboring preprotostellar
cores (04016+2610), or the heating due to outflows (04361+2547).  

2) We have fit power law models for the density distribution in the outer envelopes of nine Class I cores and find 
$\langle p\rangle=1.6\pm0.4$ and a median of $p=1.8$.  The existence of a disk or point-like component could cause $p$ to be 
much \it shallower\rm---by about 0.5.  Table~\ref{bestpower} has a summary of parameters for the best-fit models.

3) We test the Shu inside-out collapse model and find that, in most cases, this model provides a good fit for the intensity profile
but with an infall radius that is within the central resolution element of our observations.  
Most sources in this sample exhibit no evidence for a break in the 
power law density distribution on the scales we can probe with SCUBA ($\sim1000-8000$ AU).

4)  We compare our analysis with a simpler analysis put forth by Adams (1991) and as used in Paper I.  This method assumes a temperature
power law (with index $q=0.4$) and $h\nu/kT \ll 1$.  These assumptions are invalid for these low luminosity sources, 
and this simple analysis overestimates $p$ by $\sim$0.6.    

5) In modeling the density distribution, we have determined several sources of uncertainty: possible existence of a disk 
($\delta(p)=-0.5$), variations in the ISRF ($\delta(p)=\pm0.4$), and uncertainty in the outer radius ($\delta(p)=\pm0.3$).
Contributions by the disk to the observed submillimeter flux will be addressed by such instruments as SMA and ALMA while the
ISRF may be constrained by future observations in the mid- to far-infrared (e.g., SIRTF).   We find no evidence for a truncated 
outer radius in models completed over a wide range of radii.

6) For our modeled sources, we have calculated the mass of the envelope surrounding the protostar: 
$\langle M_{env} \rangle = 6.6\pm2.4$ M$_\odot$
for the two most luminous sources and $\langle M_{env} \rangle =0.2\pm0.1$ M$_\odot$ for those modeled sources in Taurus.     
We find that these masses are clearly less than those found for Class 0 cores (Paper III) and, further,  
contradict the virial masses predicted by N$_2$H$^+$ linewidths.  Molecular line tracers such as N$_2$H$^+$ simply do
not trace the mass in these Class I cores.

7) The standard classification criteria, $T_{bol}$ and $L_{obs}/L_{smm}$,  are sometimes inconsistent for Class 0 and I cores.
The mode of classification is important for studies that attempt to estimate core lifetimes based on the Class 0 and Class I source 
counts (e.g. Visser, Richer, \& Chandler 2002).   

8)  Sources with  higher aspect ratios are better fitted with shallower density power laws.  While this effect is partially 
due to our one-dimensional models, we propose that, indeed, these sources are forming in physically different initial conditions.  
Three-dimensional  modeling will help to unravel this question.

9) We note a trend between deconvolved FWHM source size and $p$---smaller cores have higher values for $p$.  This trend is expected
since steeper density power law distributions produce less well-resolved intensity profiles.  
It is also possible that we are observing the outer portions of the envelope in these less well-resolved
sources.

10) We have  also discussed several sources individually.
In particular, IRAS 03256+3055, with its unique morphology,  is an ideal object on which to test theories of fragmentation 
in the formation of low-mass protostars.   
Also, we note the possibility, through some simple calculations, that IRAS 04385+2550 is a young, forming substellar object 
with a mass of $\sim10$ M$_{Jup}$.

\section*{Acknowledgments}

We are grateful to L. Mundy for providing the computer code used 
for beam convolution and solution of Equation 1 (in Paper III) and to Steve Doty for
useful discussions and consistency checks of our 1D models.  Also, we thank
Iain Coulson for his invaluable input concerning observations and data reduction.
To our anonymous referee, we are greatly indebted for their thoughtful and thorough
reading of this work;  Paper IV is  better because of their comments.
We thank the State of Texas and NASA (Grants NAG5-7203 and 
NAG5-10488) for support.
NJE thanks the Fulbright Program and PPARC for support while at University
College London and NWO and NOVA for support in Leiden.
The JCMT is operated by the Joint Astronomy Centre on behalf of the Particle
Physics and Astronomy Research Council of the United Kingdom, The Netherlands 
Organization for Scientific Research and the National Research Council of 
Canada.


\clearpage

\clearpage

\clearpage

\clearpage

\clearpage

\begin{deluxetable}{lcccccl}
\tablecolumns{7}
\footnotesize
\tablecaption{Observed Sources \label{srclist}}
\tablewidth{0pt} 
\tabletypesize{\footnotesize}
\tablehead{
\colhead{Source}                &
\colhead{$\alpha$ (B1950.0)}    &
\colhead{$\delta$ (B1950.0)}    &
\colhead{$\alpha$ (J2000.0)}    &
\colhead{$\delta$ (J2000.0)}    &
\colhead{Date}              &
\colhead{Source}                     \\
\colhead{}				&
\colhead{($^h$~~$^m$~~$^s$~)~}        &  
\colhead{($^o$ ~\arcmin\ ~\arcsec)}          &  
\colhead{($^h$~~$^m$~~$^s$~)~}        &
\colhead{($^o$ ~\arcmin\ ~\arcsec)}          &
\colhead{}                      &
\colhead{environment}	
}
\startdata 
 CB230		 & 21 16 55.0 & 68 04 52 & 21 17 40.0 & 68 17 31.9 & 11/19/99   &IRAS 21169+6804	\\
 L1251B		 & 22 37 40.8 & 74 55 50 & 22 38 47.2 & 75 11 28.8 & 11/19/99   &IRAS 22376+7455	\\
 IRAS 03256+3055 & 03 25 39.2 & 30 55 20 & 03 28 44.5 & 31 05 39.7 & 02/02/00    &Per	\\
 IRAS 04016+2610& 04 01 40.0 & 26 10 49 & 04 04 42.5 & 26 18 57.9 & 02/03/00   	&L1489	\\
 IRAS 04108+2803 & 04 10 49.3 & 28 03 57 & 04 13 54.9 & 28 11 30.5 & 02/03/00   	&L1495	\\
 IRAS 04113+2758& 04 11 20.9 & 27 58 30 & 04 14 26.4 & 28 06 01.4 & 11/19/99   	&L1495 	\\ 
 IRAS 04166+2706\tablenotemark{b} & 04 16 37.8 & 27 06 29 & 04 19 42.5 & 27 13 39.7 & 08/30/98	&		\\
 IRAS 04169+2702 & 04 16 54.0 & 27 02 52 & 04 19 58.6 & 27 10 01.7 & 02/03/00   	&B213 	\\
 IRAS 04239+2436 & 04 23 54.5 & 24 36 54 & 04 26 56.4 & 24 43 35.9 & 02/02/00   	&L1524  	\\
 IRAS 04248+2612 & 04 24 53.2 & 26 12 39 & 04 27 57.2 & 26 19 16.9 & 02/01/00   	&HH31 IRS/B217	\\
 IRAS 04264+2433 & 04 26 28.1 & 24 33 24 & 04 29 30.0 & 24 39 55.6 & 02/03/00   	&L1524  	\\
 IRAS 04295+2251 & 04 29 32.2 & 22 51 11 & 04 32 32.1 & 22 57 30.3 & 02/01/00   	&L1536  	\\
 IRAS 04302+2247 & 04 30 16.6 & 22 47 05 & 04 33 16.4 & 22 53 21.3 & 02/02/00   	& L1536	\\
 IRAS 04361+2547 & 04 36 09.4 & 25 47 27 & 04 39 13.4 & 25 53 19.2 & 11/19/99   	&TMR1	\\	
 IRAS 04381+2540 & 04 38 07.6 & 25 40 48 & 04 41 11.6 & 25 46 32.1 & 02/01/00   	&TMC1   	\\   
 IRAS 04385+2550 & 04 38 34.2 & 25 50 43	& 04 41 38.4 & 25 56 25.3 & 02/03/00   	&TMC1 	\\
\enddata
\tablenotetext{a}{CB230 is at a distance of 450 pc (Launhardt \& Henning 1997), and L1251B is at 300 pc (Kun \& Prusti 1993).  
		 Perseus sources ($\alpha=03$ h) are located at a distance of 320 pc (de Zeeuw et al.  1999) while
		 Taurus sources ($\alpha=04$ h) are at 140 pc (Elias 1978).}
\tablenotetext{b}{The observation of IRAS 04166+2706 was reported in Paper I, but we reanalyze it here.}
\end{deluxetable}

\begin{deluxetable}{lcccccccc}
\footnotesize
\tablecolumns{9}
\tablecaption{Observing and Calibration Summary \label{opacity}}
\tablewidth{0pt} 
\tablehead{
\colhead{Date}                  &
\colhead{$\tau_{850}$}          &
\colhead{$\tau_{450}$}          &
\colhead{$\tau_{cso}$}          &
\colhead{$C_{40}^{850}$}        &
\colhead{$C_{120}^{850}$}       &
\colhead{$C_{40}^{450}$}        &
\colhead{$C_{120}^{450}$}       \\ 
\colhead{}                      &
\colhead{}                      &
\colhead{}                      &
\colhead{}                      &
\colhead{Jy/V\tablenotemark{a}} &
\colhead{Jy/V\tablenotemark{b}} &
\colhead{Jy/V\tablenotemark{a}} &
\colhead{Jy/V\tablenotemark{b}}          
}
\startdata
\multicolumn{8}{l}{November}\\                  
11/19/99 & 0.27 (0.08)  & 1.24(0.09)    & 0.061(0.006)  &  0.94 (0.08) & 0.78 (0.03)    &2.73 (0.78)    & 2.02 (0.46)   \\
\tableline
\multicolumn{8}{l}{ February} \\
02/01/00 & 0.23 (0.03)  & 1.23(0.18)    & 0.048(0.007)  &  0.83 (0.04) & 0.68 (0.03)    &1.87(0.63)     &1.44(0.95)     \\
02/02/00 & 0.17 (0.03)  & 0.79(0.27)    & 0.044(0.002)  &  0.84 (0.01) & 0.67 (0.01)    &2.54(0.26)     &1.71(0.31)     \\
02/03/00 & 0.21 (0.02)  & 1.02(0.16)    & 0.050(0.006)  &  0.83 (0.03) & 0.69 (0.03)    &2.88(0.09)     &2.36(0.06)     \\
        &               &               &               &\multicolumn{4}{l}{ February Average}\\ 
        &               &               &               &0.83(0.03)    & 0.68(0.03)     &2.35(0.61)     &1.78(0.71)             \\
\enddata

\tablenotetext{a}{Calibration Factor for a 40$\arcsec$ diameter aperture}
\tablenotetext{b}{Calibration Factor for a 120$\arcsec$ diameter aperture}

\end{deluxetable}

\begin{deluxetable}{llccccccc}
\footnotesize
\tablecolumns{9}
\tablecaption{Observed Flux Densities of Sources \label{flux}}
\tablewidth{0pt} 
\tabletypesize{\footnotesize}
\tablehead{
\colhead{Source}              &
\colhead{Centroid}            &
\colhead{$S_{\nu}$ (Jy)}      &
\colhead{$S_{\nu}$ (Jy)}      &
\colhead{$S_{\nu}$ (Jy)}      &
\colhead{$S_{\nu}$ (Jy)}      &
\colhead{$\alpha_{450/850}$\tablenotemark{a}}  &
\colhead{$\alpha_{450/850}$\tablenotemark{b}}  &
\colhead{S/N\tablenotemark{c}}			\\
\colhead{}                      &
\colhead{($\Delta\alpha\arcsec$,$\Delta\delta\arcsec$)}                  &
\colhead{850 $\mu$m\tablenotemark{a}}           &
\colhead{850 $\mu$m\tablenotemark{b}}           &
\colhead{450 $\mu$m\tablenotemark{a}}           &
\colhead{450 $\mu$m\tablenotemark{b}}           &
\colhead{}
}
\startdata 
 CB230		 & (-8,2)   &  2.74(0.33)  & 3.03(0.28) & 5.13(1.76) & 7.76(2.44)	&1.0	&1.5	&110	\\
 L1251B		 & (4,3)    &  3.66(1.50)  & 6.69(2.70) & 15.04(14.25)&24.34(7.07)	&2.2	&2.0	&70 		   \\
 IRAS 03256+3055 & \nodata  &  0.45(0.05)  & 2.45(0.29) & 1.90(1.93) & 8.08(3.89)	&2.3	&1.9	&30		   \\
 IRAS 04016+2610 & (10,0)   &  0.59(0.03)  & 1.92(0.10) & 4.23(1.38) & 11.13(6.65)	&3.1	&2.8	&40	\\
 		 & (74,16)  &  0.78(0.03)  & \nodata    & 3.53(1.38) &\nodata		&2.4	&\nodata&\nodata	\\
 IRAS 04108+2803 & (-3,-9)  &  0.17(0.02)  & 0.61(0.06) & 1.13(0.92) & 5.38(2.31)	&2.9	&3.4	&14	  \\
 IRAS 04113+2758 & (0,-1)   & 1.06(0.12)   & 1.65(0.15) & 1.86(0.63) & 4.36(1.27)	&0.9	&1.5	&180	 \\
 IRAS 04166+2706 & (1,-4)   & 1.08(0.06)   & 1.9(1.0)   & 4.2(1.0)   & 12.9(6.0)        &2.1	&3.0	&18	\\
 IRAS 04169+2702 & (4,2)    & 1.14(0.05)   & 2.53(0.12) & 6.09(1.88) & 11.75(5.05)	&2.6	&2.4	&60		 \\
 IRAS 04239+2436 & (2,-2)   & 0.42(0.05)   & 0.78(0.10) & 2.14(2.26) & 3.72(1.79)	&2.6	&2.5	&50		 \\
 IRAS 04248+2612 & (-2,1)   & 0.56(0.10)   & 1.95(0.33) & 2.96(3.03) & 8.97(3.93)	&2.6	&2.4	&20		 \\
 IRAS 04264+2433 & (3,-3)   & 0.13(0.01)   & 0.37(0.04) & 0.63(0.47) & 1.32(0.57)	&2.5	&2.0	&10		 \\
 IRAS 04295+2251 & (1,-6)   & 0.42(0.02)   & 0.78(0.05) & 2.66(1.01) & 5.34(2.34)	&2.9	&3.0	&40		 \\
 IRAS 04302+2247 & (5,1)    & 0.57(0.08)   & 0.62(0.08) & 2.09(2.50) & 1.88(0.91)	&2.0	&1.8	&90		 \\
 IRAS 04361+2547 & (5,4)    & 0.64(0.08)   & 1.21(0.11) & 2.35(0.80) & 4.55(1.32)	&2.1	&2.1	&60		 \\
 		 & (63,-57) & 0.48(0.08)   & \nodata    & 1.69(0.80) &\nodata		&2.0	&\nodata&\nodata		\\
 IRAS 04381+2540 & (15,6)   & 0.56(0.04)   & 1.35(0.11) & 2.82(1.43) & 6.41(2.80)	&2.5	&2.5	&40  \\
 IRAS 04385+2550 & (6,1.7)  & 0.08(0.01)   & 0.29(0.04) & 0.14(0.14) & 0.37(0.16) 	&0.8	&0.4	&10	 \\
\enddata

\tablenotetext{a}{In a 40$\arcsec$ aperture}
\tablenotetext{b}{In a 120$\arcsec$ aperture}
\tablenotetext{c}{S/N is given for the 850 $\mu$m maps.}

\end{deluxetable}


\begin{deluxetable}{lcccc}
\footnotesize
\tablecolumns{5}
\tablecaption{Spectral Energy Distributions of Sources \label{sed}}
\tablewidth{0pt} 
\tabletypesize{\footnotesize}
\tablehead{
\colhead{Source}                  &
\colhead{$\lambda$ ($\mu$m)}     &
\colhead{$S_{\nu}$ (Jy)}          &
\colhead{$\theta_{mb}$ (\arcsec)} &
\colhead{Ref.}                    
}
\startdata 
CB230	           & 2.2   & 0.005(0.001) & 6.3  & 8\tablenotemark{a} \\
                   & 12    & $<$0.25        & 300$\times$45  & 10 \\
                   & 25    & 0.68(0.07)    & 300$\times$45  & 10\tablenotemark{a} \\
                   & 60    & 11.75(0.71)   & 90$\times$300 & 10\tablenotemark{a} \\
                   & 100   & 33.53(2.01)   & 180$\times$300 & 10\tablenotemark{a} \\
                   & 350   & 12.0(0.2)     & 19.5 & 5 \\
                   & 450   & 4.4(0.5)      & 18.5 & 5 \\
                   & 450   & 7.76(2.44)    & 120  & 11\tablenotemark{a} \\
		   & 450   & 7.0(\nodata)  & \nodata& 6 \\
                   & 800   & 1.00(0.16)    & 16.5 & 5 \\
                   & 850   & 3.03(0.28)    & 120  & 11\tablenotemark{a} \\
		   & 850   & 1.2(\nodata)  &\nodata & 6 \\
                   & 1100  & 0.37(0.05)    & 18.5 & 5 \\
                   & 1300  & 0.23(0.01)    & 16.5 & 5 \\
                   & 1300  & 0.221(0.005)  & 12   & 12 \\
		   & 36000 & 0.2           & 11.1$\times$8.6  & 7 \\
                   & 60000 & 0.1           & 90$\times$300 & 7 \\
L1251B	           & 12    & 0.80(0.11)    & 300$\times$45   & 10\tablenotemark{a}  \\
		   & 25    & 5.55(0.33)    & 300$\times$45   & 10\tablenotemark{a}  \\
		   & 60    & 32.34(0.32)   & 90$\times$300   & 10\tablenotemark{a}  \\
		   & 100   & 66.84(9.36)   & 180$\times$300   & 10\tablenotemark{a}  \\
		   & 450   & 24.34(7.07)   & 120    & 11\tablenotemark{a}  \\
		   & 850   & 6.69(2.70)    & 120    & 11\tablenotemark{a}  \\
IRAS 03256+3055     & 60    & 1.43(0.14)    & 90$\times$300   & 10\tablenotemark{a}   \\
		   & 450   & 8.08(3.89)    & 120    & 11\tablenotemark{a}  \\
		   & 850   & 2.45(0.29)    & 120    & 11\tablenotemark{a}  \\
IRAS 04016+2610     & 2.2   & 0.13(0.02)    & 10.7   & 3\tablenotemark{a} \\
		   & 2.2   & 0.12	   & 3.8  & 16 \\
                   & 7.8   & 2.8(0.3)      &  12    & 1 \\
                   & 8.7   & 2.1(0.1)      & 12     & 1 \\
                   & 9.5   & 1.7(0.1)      & 12     & 1 \\
                   & 10.3  & 2.5(0.1)      & 12     & 1 \\
                   & 11.6  & 3.4(0.1)      & 12     & 1 \\ 
		   & 12	   & 3.64(0.07)	   & 300$\times$45   & 10\tablenotemark{a}\\
                   & 12.5  & 4.9(0.2)      & 12     & 1 \\
		   &25	   &15.81(0.95)    &300$\times$45    & 10\tablenotemark{a}	\\
		   &60	   &48.79(4.9)     &90$\times$300    & 10\tablenotemark{a}	\\
		   &100	   &55.69(7.8)	   &180$\times$300    & 10\tablenotemark{a}	\\
		   & 450   &11.13(6.65)     & 120    & 11\tablenotemark{a}  \\
                   &450	   &3.25(0.46)	   &17.5    & 13 \\
		   &800	   &0.29(0.042)	   &13.5    & 13 \\
                   &800	   &0.31(0.061)	   &16.8    & 13 \\
		   & 800   & 0.582(0.079)  & 16.8   & 2 \\
		   & 850   & 1.92(0.10)    & 120    & 11\tablenotemark{a}  \\
		   & 850   & 5.78()	   &\nodata & 15 \\	
                   &1100   & 0.055(0.016)  & 18.5   & 13\\
                   & 1100  & 0.180(0.021)  & 18.5     & 2 \\
		   &1300   & 0.130(0.005)  & 11     & 14 \\
		   & 1300  &0.150(\nodata) & 60     & 14\tablenotemark{a}\\ 
IRAS 04108+2803    &2.2   &0.029(0.004)   &10.7     & 3\tablenotemark{a} \\
		   &7.8	   &0.31(0.06)	   &12       & 1 \\
		   &8.7	   &0.50(0.08)	   &12       & 1 \\
		   &9.5	   &0.49(0.09)	   &12       & 1 \\
		   &10.3   &0.57(0.09)	   &12       & 1 \\
		   &11.6   &0.44(0.09)      &12       & 1 \\
		   &12.5   &1.3(0.2)       &12       & 1 \\
		   &12     & 0.87(0.05)    &300$\times$45     &10\tablenotemark{a}   \\ 
		   &25	   & 3.88(0.39)	   &300$\times$45    &10\tablenotemark{a}	\\
		   &60	   & 7.49(0.75)	   &90$\times$300    &10\tablenotemark{a}	\\
		   &100	   & 10.86(0.11)   &180$\times$300    &10\tablenotemark{a}	\\
		   & 450   & 5.38(2.31)    & 120    & 11\tablenotemark{a}  \\
		   &800	   & 0.085(0.033)  & 16.8   & 2 \\
		   & 850   & 0.61(0.06)    & 120    & 11\tablenotemark{a}  \\
		   &1100   & $<$0.1	   & 18.5   & 2 \\
		   &1300   & $<$0.02	   & 60     & 14 \\
IRAS 04113+2758     & 2.2   & 0.32 (0.05)   & 10.7 & 3\tablenotemark{a} \\
		   & 2.2   & 0.66(0.02)	   & 9	  & 13 \\
                   & 12    & 2.03 (0.28)   & 300$\times$45 & 10\tablenotemark{a} \\
                   & 60    & 12.75 (0.13)  & 90$\times$300 & 10\tablenotemark{a} \\
                   & 100   & 15.82 (2.21)  & 180$\times$300 & 10\tablenotemark{a} \\
		   & 450   & 4.36(1.27)    & 120    & 11\tablenotemark{a}  \\
                   & 800   & 0.98(0.12)    & 16.8 & 2 \\
		   & 850   & 1.65(0.15)    & 120    & 11\tablenotemark{a}  \\
                   & 1100  & 0.461(0.053)  & 18.5 & 2 \\
		   & 1300  & 0.410(0.040)  & 11   & 14 \\
		   & 1300  & 0.750(\nodata)& 60   & 14\tablenotemark{a} \\
		   & 3400  & 0.052(0.006)  & 8    & 17 \\
IRAS 04166+2706    & 1.6   & 0.00010(0.00002) & 10 & 4 \\
		   & 2.2   & 0.00019(0.00009) & 10 & 4\\
		   & 12    & 0.07(0.007)   & 300$\times$45   & 10\tablenotemark{a} \\
                   & 25    & 0.58(0.058)    & 300$\times$45   & 10\tablenotemark{a} \\
                   & 60    & 5.9(0.59)    & 90$\times$300   & 10\tablenotemark{a} \\
                   & 100   & 9.5(0.95)    & 180$\times$300   & 10\tablenotemark{a} \\
		   & 450   & 12.9(6.0)    & 120    & 11\tablenotemark{a}  \\
		   & 850   & 1.9(1.0)     & 120    & 11\tablenotemark{a}  \\
IRAS 04169+2702     & 2.2   & 0.02(0.003)  & 10.7   & 3\tablenotemark{a} \\
		   & 2.2   & 0.02(0.004)  & 9	    & 13 \\
                   & 12    & 0.75(0.05)   & 300$\times$45   & 10\tablenotemark{a} \\
                   & 25    & 5.21(0.31)    & 300$\times$45   & 10\tablenotemark{a} \\
                   & 60    & 17.00(1.7)    & 90$\times$300   & 10\tablenotemark{a} \\
                   & 100   & 17.46(2.4)    & 180$\times$300   & 10\tablenotemark{a} \\
		   & 450   & 11.75(5.05)   & 120    & 11\tablenotemark{a}  \\
                   & 800   & 0.75(0.18)    & 16.8   & 2 \\ 
                   & 800   &0.52(0.069)    & 13.5   & 13 \\
                   &800	   &0.73(0.051)	   &16.8    & 13 \\
		   & 850   & 2.53(0.12)    & 120    & 11\tablenotemark{a}  \\
                   &1100   &0.280(0.025)   &18.5    & 13 \\                   
		   & 1100  & 0.281(0.053)  & 18.5   & 2 \\
		   & 1300  & 0.190(0.009)  & 11     & 14\\
		   & 1300  & 0.730(\nodata)& 60     & 14\tablenotemark{a} \\
		   & 3400  & $<$0.026	   & 8      & 17 \\	
IRAS 04239+2436     & 2.2   & 0.07 (0.01)   & 10.7   & 3\tablenotemark{a} \\
		   &12	   & 1.71(0.1)	   &300$\times$45    &10\tablenotemark{a}	\\	
		   &25	   & 6.98(0.42)	   &300$\times$45    &10\tablenotemark{a}	\\
		   &60	   & 15.24(0.15)   &90$\times$300    &10\tablenotemark{a}	\\
		   &100	   & 15.88(2.22)   &180$\times$300    &10\tablenotemark{a}	\\
		   & 450   & 3.72(1.79)    & 120    & 11\tablenotemark{a}  \\
                   & 800   & 0.333(0.043)  & 16.8   & 2 \\
                   &800	   &0.23(0.025)	   & 13.5   & 13 \\
		   & 850   &0.78(0.10)     & 120    & 11\tablenotemark{a}  \\
                   & 1100  &0.077(0.023)   &18.5    & 13 \\
                   & 1100  & 0.114(0.021)  & 18.5   & 2   \\
		   &1300   & 0.080(0.01)   & 11     & 14\\
		   &1300   & 0.170         & 60     & 14\tablenotemark{a}\\
		   & 3400  & $<$0.024      & 8      & 17 \\
IRAS 04248+2612     & 2.2   & 0.04(0.006)   & 10.7 & 3\tablenotemark{a} \\
		   & 2.2   & 0.04()        & 3.8  & 16 \\
		   &25	   &1.33(0.13)	   &300$\times$45    &10\tablenotemark{a}	\\
		   &60	   &4.62(0.46)	   &90$\times$300    &10\tablenotemark{a}	\\
		   &100	   &9.26(0.93)	   &180$\times$300    &10\tablenotemark{a}	\\
		   & 450   &8.97(3.93)     & 120    & 11\tablenotemark{a}  \\
                   & 800   & 0.252(0.046)  & 16.8 & 2 \\  
                   &800	   &0.12(0.033)    &13.5  & 13 \\
                   &800	   &0.21(0.039)	   &16.8  & 13 \\
		   & 850   &1.95(0.33)     & 120    & 11\tablenotemark{a}  \\
                   &1100   &0.100(0.019)   &18.5    & 13 \\
                   & 1100  & 0.099(0.015)  & 18.5 & 2 \\
		   & 1100  &0.15(0.045)    &\nodata&18\\
		   &1300   &0.060(0.007)   &11     & 14\\
		   &1300   &0.450          &60     & 14\tablenotemark{a}\\
IRAS 04264+2433     & 2.2   & 0.026(0.001)  & 9	  & 13\tablenotemark{a} \\
		   & 12    & 0.39(0.02)    & 300$\times$45 & 10\tablenotemark{a} \\
                   & 25    & 2.94(0.18)    & 300$\times$45 & 10\tablenotemark{a} \\
                   & 60    & 5.21(0.52)    & 90$\times$300 & 10\tablenotemark{a} \\
		   & 450   &1.32(0.57)     & 120    & 11\tablenotemark{a}  \\
		   & 850   &0.37(0.04)     & 120    & 11\tablenotemark{a}  \\
		   &1300   & 0.031(0.008)  &11    & 14\\
		   &1300   & 0.031(\nodata)&60    & 14\tablenotemark{a}\\
IRAS 04295+2251     & 2.2   & 0.12 (0.02)   & 10.7 & 3\tablenotemark{a} \\
		   &12	   & 0.60(0.11)	   &300$\times$45  &10\tablenotemark{a}	\\
		   &25	   & 1.82(0.18)	   &300$\times$45  &10\tablenotemark{a}	\\
		   &60	   & 3.53(0.35)	   &90$\times$300  &10\tablenotemark{a}  \\
		   &100 & 7.41(1.04)	   &180$\times$300  &10\tablenotemark{a}	\\
		   & 450   & 5.34(2.34)    & 120    & 11\tablenotemark{a}  \\
                   & 800   & 0.241(0.050)  & 16.8 & 2 \\
		   & 850   & 0.78(0.05)    & 120    & 11\tablenotemark{a}  \\
                   & 1100  & 0.094(0.018)  & 18.5 & 2 \\
		   &1300   & 0.115(0.010)  & 11   & 14\\
		   &1300   & 0.115(\nodata)& 60   & 14\tablenotemark{a}  \\
IRAS 04302+2247     & 2.2   & 0.03(0.005)   & 10.7 & 3\tablenotemark{a} \\
		   & 2.2   & 0.025(0.001)  & 9	  & 13 \\
                   & 2.2   & 0.025(...)    & 3.8& 16 \\
		   & 25    & 0.44(0.06)    & 300$\times$45 & 10\tablenotemark{a} \\
                   & 60    & 6.40(0.13)    & 90$\times$300 & 10\tablenotemark{a} \\
                   & 100   & 9.43(0.19)    & 180$\times$300 & 10\tablenotemark{a} \\
		   & 450   & 1.88(0.91)    & 120    & 11\tablenotemark{a}  \\
                   & 800   & 0.342(0.057)  & 16.8 & 2 \\
		   & 850   & 0.62(0.08)    & 120    & 11\tablenotemark{a}  \\
                   & 1100  & 0.149(0.019)  & 18.5 & 2 \\
		   &1300   & 0.180(0.010)  & 11   & 14\\
		   &1300   & 0.180(\nodata)& 60   & 14\tablenotemark{a}\\
IRAS 04361+2547     & 2.2   & 0.04(0.006)   & 10.7 & 3\tablenotemark{a} \\ 
		   & 2.2   & 0.04(0.001)   & 9	  & 13 \\	
                   & 12    & 1.81(0.11)    & 300$\times$45 & 10\tablenotemark{a} \\
                   & 25    & 18.87(1.13)   & 300$\times$45 & 10\tablenotemark{a} \\
                   & 60    & 44.75(6.27)   & 90$\times$300 & 10\tablenotemark{a} \\
                   & 100   & 35.43(4.96)   & 180$\times$300 & 10\tablenotemark{a} \\
		   & 450   & 4.55(1.32)    & 120    & 11\tablenotemark{a}  \\
                   & 800   & 0.634(0.067)  & 16.8 & 2 \\
		   & 850   & 1.21(0.11)    & 120    & 11\tablenotemark{a}  \\
                   & 1100  & 0.188(0.027)  & 18.5 & 2  \\
		   &1300   & 0.110(0.008)  & 11   & 14\\
		   &1300   & 0.440(\nodata)& 60   & 14\tablenotemark{a}\\
IRAS 04381+2540     & 2.2   & 0.01 (0.002)  & 10.7 & 3\tablenotemark{a} \\
		   &12	   &0.44(0.06)	   &300$\times$45  &10\tablenotemark{a}	\\
		   &25	   &2.69(0.27)	   &300$\times$45  &10\tablenotemark{a}	\\
		   &60	   &10.29(1.0)	   &90$\times$300  &10\tablenotemark{a}	\\
		   &100	   &13.91(1.4)	   &180$\times$300  &10\tablenotemark{a}	\\
		   & 450   &6.41(2.80)     & 120    & 11\tablenotemark{a}  \\
		   &450    & 12.5()        &\nodata & 15 \\
                   & 800   & 0.289(0.053)  & 16.8 & 2  \\
		   & 850   & 1.35(0.11)    & 120    & 11\tablenotemark{a}  \\
		   & 850   & 2.60()        &\nodata &15\\
                   & 1100  & 0.116(0.013)  & 18.5 & 2    \\
		   &1300   & 0.070(0.009)  & 11   & 14\\
		   &1300   & 0.300(\nodata)& 60   & 14\tablenotemark{a}\\
IRAS 04385+2550	   &12	   & 0.54(0.03)	   &300$\times$45    &10\tablenotemark{a}	\\
		   &25	   & 1.55(0.16)	   &300$\times$45    &10\tablenotemark{a}	\\
		   &60	   & 2.88(0.29)	   &90$\times$300    &10\tablenotemark{a}	\\
		   & 450   &0.37(0.16)     & 120    & 11\tablenotemark{a}  \\
		   & 850   & 0.29(0.04)    & 120    & 11\tablenotemark{a}  \\
		   &1300   &0.030(0.005)   & 11   & 14\\
		   &1300   &0.030(\nodata) & 60   & 14\tablenotemark{a}\\
\enddata

\tablenotetext{a}{Flux value used in the calculation of $T_{bol}$ and $L_{obs}$.}
\tablerefs{1. Myers et al. 1987;2. Moriarty-Scheiven et al. 1994; 3. Tamura et al. 1991; 4. Kenyon et al. 1990; 
5. Launhardt et al. 1997; 6. Huard et al. 1999; 
7.Moreira et al. 1997; 8. Yun et al. 1995; 
9.Kun \& Prusti 1993; 10. IRAS PSC, 1988; 11. this paper; 12. Launhardt \& Henning, 1997; 13. Barsony \& Kenyon, 1992; 
14. Motte \& Andr\'e, 2000; 15. Hogerheijde \& Sandell, 2000; 16. Padgett et al., 1999; 17. Saito et al., 2001; 
18. Dent et al., 1998.}
\end{deluxetable}

\begin{deluxetable}{lc|ccc|ll}
\footnotesize
\tablecolumns{7}
\tablecaption{Observed Source Properties\label{src_prop}}
\tablewidth{0pt} 
\tablehead{
\colhead{Source}                &
\colhead{$T_{bol}$\tablenotemark{a}}                 &
\colhead{$T_{bol}$}                 &
\colhead{$L_{obs}$}                 &
\colhead{$L_{smm}/L_{obs}$}   &
\colhead{Aspect} 		&
\colhead{$\theta_{dec}/\theta_{mb}$}				\\
\colhead{}                      &
\colhead{K}                 	&
\colhead{K}                   	&
\colhead{$L_\odot$}			&
\colhead{}			&
\colhead{$\frac{a}{b}$}			&
\colhead{}			
}
\startdata 
\tableline
        CB230					&47	&74		& 11.5	&	100	&1.2$\pm$0.2  	&1.3$\pm$0.3 	\\
        L1251B\tablenotemark{b}			&64	&\nodata	&10.8	&	100 	&1.7$\pm$0.2 	&1.7$\pm$0.4	\\	
        IRAS 03256+3055\tablenotemark{b}  	&16	&\nodata	&0.7	&	\nodata &\nodata 	&6.9$\pm$0.9	\\	
        IRAS 04016+2610  			&92	&167		& 4.6	&	1000	&\nodata 	&0.7$\pm$0.3	\\
        IRAS 04108+2803  			&98	&179		& 1.0	&	500	&\nodata 	&0.0$\pm$0.3 	\\
        IRAS 04113+2758  			&82	&274		& 2.0	& 	333	&\nodata	&0.7$\pm$0.3	\\
	IRAS 04166+2706  			&50     &75             & 0.5   &       50      &1.2$\pm$0.2 	&1.4$\pm$0.3	\\
        IRAS 04169+2702  			&83	&133		& 1.4	&	111	&1.2$\pm$0.2 	&1.9$\pm$0.2	\\
        IRAS 04239+2436  			&103	&194		& 1.8	&	500	&1.4$\pm$0.3 	&0.8$\pm$0.3	\\
        IRAS 04248+2612  			&49	&136		& 0.9	&	142	&2.3$\pm$0.4 	&3.0$\pm$0.3	\\
        IRAS 04264+2433  			&109	&186		& 0.5	&	500	&\nodata 	&0.0$\pm$0.3	\\
        IRAS 04295+2251  			&91	&270		& 0.6	&	125	&1.1$\pm$0.2 	&0.8$\pm$0.2    \\
        IRAS 04302+2247  			&53	&122		& 0.5	&	111	&\nodata 	&0.0$\pm$0.3	\\
        IRAS 04361+2547  			&98	&141		& 3.6	& 	1000	&1.0$\pm$0.2 	&1.1$\pm$0.2	\\
        IRAS 04381+2540  			&78	&126		& 0.9	&	142	&1.2$\pm$0.2 	&1.7$\pm$0.2    \\
        IRAS 04385+2550\tablenotemark{b}  	&122	&\nodata	&0.2	&	500	&\nodata 	&0.0$\pm$0.3	\\
\enddata
\tablenotetext{a}{Near-infrared data is not included in the calculation of this value.}
\tablenotetext{b}{No near-infrared emission has been detected towards these objects.}
\end{deluxetable}

\begin{deluxetable}{lcccccccccl}
\footnotesize
\tablecolumns{11}
\tablecaption{Test Case, CB230\label{cb230}}
\footnotesize
\tablewidth{0pt} 
\tablehead{
\colhead{Source}      	&
\colhead{$p$}     	&
\colhead{$n_f$}          &
\colhead{$r_i$} 			&
\colhead{$r_o$}                   &
\colhead{$\kappa_\nu$}			&
\colhead{$L_{int}$}	&
\colhead{$s_{ISRF}$}	&
\colhead{$\chi^2_{450}$}	&
\colhead{$\chi^2_{850}$}	&
\colhead{$\chi^2_{SED}$}	\\
\colhead{}            		&
\colhead{}     		&
\colhead{($\times10^5cm^{-3}$)}       	   	&
\colhead{(AU)} 		&
\colhead{($\times10^5$AU)}                 &
\colhead{}		&
\colhead{$L_\odot$}			&
\colhead{}			&
\colhead{}			&
\colhead{}			&
\colhead{}			
}
\startdata 
\multicolumn{11}{l}{\bf ISRF}		\\
	&  1.9 	&22     &100	&1		&OH5	&6.7	&\bf 0.3 &0.4	&28	&91		\\
	&  2.0  &22.5	&100	&1		&OH5	&5.8	&\bf 0.6 &0.9    &31	&105	      \\
	&  2.2	&25	&100	&1		&OH5	&4.5	&\bf 1   & 0.6   &53	&113	      \\
        &  2.2  &$<25$	&100	&1		&OH5	&\nodata&\bf 3   &\nodata&\nodata&\nodata	        \\
\multicolumn{11}{l}{\bf Inner Radius}										\\
	&  2.2  &30	&\bf 50     &1		&OH5	&4.5	& 1     & 0.6	&53	&113	        \\
	&  2.2	&25	&\bf 100    &1		&OH5	&4.5	& 1     & 0.6   &53	&113	      \\
	&  2.2  &25	&\bf 200    &1		&OH5	&4.75	& 1     & 0.5	&49	&120	       \\
	&  2.2  &25	&\bf 400    &1		&OH5	&6.75	& 1     & 0.3	&43	&117	        \\
	&  2.2  &25	&\bf 800    &1		&OH5	&6.8	& 1     & 1 	&38	&200	        \\
\multicolumn{11}{l}{\bf Outer Radius}										\\
	&  1.9  &35	&100    &\bf 0.25	&OH5	&4.8	& 1     &1.6 	&33	&32	       \\
	&  2.1  &33	&100    &\bf 0.5  	&OH5	&4.7	& 1     & 1.9	&50	&28	       \\
	&  2.1  &27	&100    &\bf 0.75	&OH5	&4.4	& 1     &2.9 	&48	&88	       \\
	&  2.2	&25	&100	&\bf 1		&OH5	&4.5	&1	& 0.6   &53	&113	      \\
	&  2.2  &0.5	&100    &\bf 2		&OH5	&6.0	& 1     & 33	& 560	&591	\\
\multicolumn{11}{l}{\bf Dust Type\tablenotemark{a}}										\\
	&  2.2	&25	&100	&1		&\bf OH5 &4.5	&1      & 0.6   &53	&113	      \\
	&  2.2	&\nodata&100	&\nodata	&\bf OH2 &\nodata&1	&\nodata&\nodata&\nodata  	      \\
\multicolumn{11}{l}{\bf Beam\tablenotemark{b}}										\\
	&  2.2	&25	&100	&1		&OH5	&4.5	&1	&2    	&20	&113	     \\

\enddata
\tablenotetext{a}{We were unable to fit the luminosity or fiducial density of CB230 with OH2 dust, but the intensity profiles 
			were fit well with $p=2.2$.}
\tablenotetext{b}{We use the beam profile observed from the February observations (CB230 was observed in November).}

\end{deluxetable}

\begin{deluxetable}{lcccccccccl}
\footnotesize
\footnotesize
\tablecolumns{11}
\tablecaption{Test Case, IRAS 04295+2251\label{iras04295}}
\tablewidth{0pt} 
\tablehead{
\colhead{Source}      	&
\colhead{$p$}     	&
\colhead{$n_f$}          &
\colhead{$r_i$} 			&
\colhead{$r_o$}                   &
\colhead{$\kappa_\nu$}			&
\colhead{$L_{int}$}	&
\colhead{$s_{ISRF}$}	&
\colhead{$\chi^2_{450}$}	&
\colhead{$\chi^2_{850}$}	&
\colhead{$\chi^2_{SED}$}	\\
\colhead{}            		&
\colhead{}     		&
\colhead{($\times10^5cm^{-3}$)}       	   	&
\colhead{(AU)} 		&
\colhead{($\times10^4AU$)}                 &
\colhead{}		&
\colhead{$L_\odot$}			&
\colhead{}			&
\colhead{}			&
\colhead{}			&
\colhead{}			
}
\startdata 
\multicolumn{11}{l}{\bf ISRF}		\\
	&  2.3	&2	&100	&3.3	&OH5	&0.25	&\bf 0.3	&0.3	&2	&12	   \\
	&  2.5	&1.5	&100	&3.3	&OH5	&0.19	&\bf 0.6	&0.3	&4	&12	    \\
	&  2.7	&0.8	&100	&3.3	&OH5	&0.1	&\bf 1		&0.3    &5	&18	    \\
	&  2.7	&$<0.8$	&100	&3.3	&OH5	&0.1	&\bf 3		&\nodata&\nodata&\nodata	    \\
\multicolumn{11}{l}{\bf Inner Radius}	\\
	&  2.7	&1	&\bf 50	&3.3	&OH5	&0.06	&1	&0.3	&5	&20	  \\
	&  2.7	&0.8	&\bf 100&3.3	&OH5	&0.1	&1	&0.4    &4	&23	    \\
	&  2.6	&1	&\bf 200&3.3	&OH5	&0.1	&1	&0.6	&3	&21	     \\
	&  2.6	&1	&\bf 400&3.3	&OH5	&0.4	&1	&0.6	&4	&25	      \\
\multicolumn{11}{l}{\bf Outer Radius}										\\
	&  2.5	&3.6	&100	&\bf 1.1&OH5	&0.19	&1	&0.4	&2	&16  	   \\
	&  2.5	&2.2	&100	&\bf 2.2&OH5	&0.12	&1	&0.9	&2	&16	    \\
	&  2.3	&0.8	&100	&\bf 3.3&OH5	&0.1	&1	&3.3    &3.5	&21	   \\
	&  2.3	&\nodata&100	&\bf 6.6&OH5	&\nodata&1	&\nodata&\nodata&\nodata	     \\
\multicolumn{11}{l}{\bf Dust Type}										\\
	&  2.3	&0.9	&100	&3.3	&\bf OH2&0.05	&1	&6    	&5	&24	     \\
	&  2.3	&0.8	&100	&3.3	&\bf OH5&0.1	&1	&3.3    &3.5	&21	    \\
\multicolumn{11}{l}{\bf Beam\tablenotemark{a}}										\\
	&  2.7	&0.8	&100	&3.3	&OH5	&0.1	&1	&2    	&5	&19	     \\

\enddata
\tablenotetext{a}{We use the beam profile observed from the November observations (IRAS 04295+2251 was observed in February).}
\end{deluxetable}

\begin{deluxetable}{lccc}
\footnotesize
\footnotesize
\tablecolumns{4}
\tablecaption{Modeling Uncertainty Analysis\label{error}}
\tablewidth{0pt} 
\tablehead{
\colhead{Variable}			&
\colhead{Range}				&
\colhead{$\Delta p$}			&
\colhead{$\Delta L_{int}$}		
}
\startdata 
ISRF		& 0.3-3\tablenotemark{a}		& $\pm$0.3	& 2.2		\\
		& 0.3-3\tablenotemark{b}		& $\pm$0.4	& 0.15		\\
Inner Radius	& 50-400AU\tablenotemark{a}		& $<0.1$	& 2.3		\\
		& 50-400AU\tablenotemark{b}		& $<0.1$	& 0.3		\\
Outer Radius	& 25,000-200,000AU\tablenotemark{a}	& $\pm$0.3	& 1.2	 	\\
		& 11,000-66,000AU\tablenotemark{b}	& $\pm$0.2	& 0.1		\\
Dust Type	& OH2-OH5				& $<0.1$	&\nodata	\\
Disk		& 20-80$\%$ of total flux		& $-0.5$	&\nodata	\\
Beam		& Nov99-Feb00				& $<0.1$	&\nodata	\\
\tableline
\multicolumn{2}{r}{Uncertainty:\tablenotemark{a}} &$^{+0.4}_{-0.7}$  &3.4$L_\odot$	\\
\multicolumn{2}{r}{Uncertainty:\tablenotemark{b}} &$^{+0.4}_{-0.7}$  &0.4$L_\odot$	\\
\enddata
\tablenotetext{a}{For CB230. Outer radii range corresponds to $55-440\arcsec$.  The uncertainty
is simply the quadratic sum of the contributing factors and, since each is not entirely independent, this
uncertainty is probably an overestimate.}
\tablenotetext{b}{For IRAS 04295+2251.  Outer radii range corresponds to $80-430\arcsec$.}
\end{deluxetable}

\begin{deluxetable}{lccccccl}
\tabletypesize{\footnotesize}
\tablecolumns{8}
\tablecaption{Power Law Models\tablenotemark{a}\label{power}}
\tablewidth{0pt} 
\tablehead{
\colhead{Source}      	&
\colhead{$p$}     	&
\colhead{$n_f$}          &
\colhead{$L_{int}$}	&
\colhead{$\chi^2_{450}$}\tablenotemark{b}	&
\colhead{$\chi^2_{850}$}	&
\colhead{$\chi^2_{SED}$}	&
\colhead{Comments}\\
\colhead{}            		&
\colhead{}     		&
\colhead{($\times10^5cm^{-3}$)}       	   	&
\colhead{$L_\odot$}			&
\colhead{}			&   
\colhead{}			&
\colhead{}		&
\colhead{}	
}
\startdata 
CB230		&2.0    &30	&6.6	        &1 	&45	&37		&      \\
		&1.9 	&22     &6.7	 	&0.4	&28	&91		&Best-fit	\\
		&1.8 	&18     &7.2	 	&1	&37	&119		&	\\
		&2.4 	&55     &6.8	 	&10	&242	&169	 	&	\\
		&1.4 	&6.0    &12	 	&34	&455	&251	 	&	\\
IRAS04166+2706	&1.7	&3.2	&0.4		&0.2	&3	&64		&	\\
		&1.6	&3.0	&0.4		&0.2	&2	&72		&Best-fit	\\
		&1.5	&2.4	&0.4		&0.3	&3	&92		&	\\
		&2.1	&6.6	&0.2		&2	&16	&57		&	\\
		&1.1	&1.0	&0.4		&6	&26	&28		&	\\
IRAS 04169+2702	&1.6 	&3.5    &0.73	 	&11	&50	&35		&	\\
		&1.5	&2.8    &0.77	 	&10	&57	&39	 	&Best-fit	\\
		&1.4 	&2.5    &0.80	 	&10	&71	&40		&	\\
		&2.0	&7.0    &0.55	 	&50	&94	&18	 	&	\\
		&1.0	&0.8    &0.9	 	&155	&308	&57	 	&	\\
IRAS 04239+2436	&2.0	&1.0	&1.0		&1.7	&1.6	&94		&	\\
		&1.9	&1.0    &1.0	 	&0.8	&0.7	&18		&	\\
		&1.8	&0.9    &0.9	 	&0.1	&2	&6	 	&Best-fit	\\
		&1.7	&0.75   &0.9	 	&0.2	&5	&12	 	&	\\
		&2.3	&1.5    &0.75	 	&5	&9	&3	 	&	\\
		&1.3	&0.3    &1.7		&18	&66	&150		&	\\
IRAS 04248+2612	&0.9	&0.5    &0.5		&0.7	&1	&11	 &	\\
		&0.8	&0.35   &0.6		&0.5	&2	&12	 	&Best-fit	\\
		&0.7	&0.25   &0.95		&1.4	&6	&13	 	&	\\
		&1.3	&1.5    &0.4		&7	&20	&5	 	&	\\
		&0.3	&0.09   &1.0		&15	&67	&22	 &	\\
		&1.0\tablenotemark{a}&1.0 &0.5  &1 	&3 	&8		&	\\
IRAS 04295+2251	&2.4	&2.0 	&0.25		&0.3	&4	&11		&	\\
		&2.3	&2.0 	&0.25		&0.3	&2	&12	     &Best-fit	\\	
		&2.2	&2.0 	&0.25	 	&0.9	&2	&12		&	\\		
		&2.8	&2.0    &0.2	 	&2	&13	&7	 	&	\\
		&1.8	&0.9    &0.4	 	&9	&14	&29	 	&	\\
IRAS 04361+2547	&1.9	&1.9    &2.2	 	&9	&17	&16	 	&	\\
		&1.8	&1.8    &2.2	 	&6	&14	&17		&Best-fit	\\
		&1.7	&1.5    &2.2	 	&4	&16	&20	 	&	\\
		&2.3	&3.0    &1.5	 	&17	&49	&6	 	&	\\
		&1.3	&0.5    &3.0	 	&8	&125	&33	 	&	\\
IRAS 04381+2540	&1.7	&2.0    &0.53	 	&2	&17	&4		&	\\
		&1.6	&1.8    &0.53	 	&4	&9	&4		&Best-fit	\\
		&1.5	&1.4    &0.53	 	&7	&4	&6	 	&	\\
		&1.4	&1.0    &0.6	 	&12	&1	&8		&	\\
		&1.3	&0.8    &0.7	 	&22	&1	&9		&	\\
		&2.0	&3.2    &0.6	 	&1	&54	&1	 	&	\\
		&1.0	&0.35   &1.1		&78	&42	&15	 	&	\\
L1251B		&1.4	&8.3    &13		&22	&183	&646	 	&	\\
		&1.5	&12     &10		&41	&111	&73	 	&Best-fit	\\
		&1.6	&16     &10		&61	&172	&82	 	&	\\
		&2.0	&44     &7.5		&138	&1050	&1261	 	&	\\
		&1.0	&2.3    &14.0		&36	&2033	&570	 	&	\\
\enddata
\tablenotetext{a}{We set an inner radius of 100AU for all sources.  The outer radius is assigned as follows: CB230, 100,000AU;
		L1251B, 72,000AU; Taurus sources, 33,000AU.  The ISRF is 0.3 times the standard as discussed in the text.}
\tablenotetext{b}{We excluded the extended emission for this model.  See text for a more detailed discussion.}
\end{deluxetable}

\clearpage

\begin{deluxetable}{lccccccc}
\footnotesize
\tablecolumns{8}
\tablecaption{Derived Masses \& Velocities\label{mass}}
\tablewidth{0pt} 
\tablehead{
\colhead{Source}                &
\colhead{$\Delta v_{FWHM}$}                 &
\colhead{$a_{eff}$}		&
\colhead{$M_v^{\theta_{ap}}$}                 &
\colhead{$M_{env}$\tablenotemark{b}}   &
\colhead{$M_{env}^{iso}$}	&
\colhead{$T_{iso}$} 	&
\colhead{$\Delta v$}\\
\colhead{}                      &
\colhead{km s$^{-1}$}                &
\colhead{km s$^{-1}$}                &
\colhead{$M_\odot$}                 &
\colhead{$M_\odot$}                 &
\colhead{$M_\odot$}                 &
\colhead{K}        		&
\colhead{Ref.}           	
}
\startdata 
        CB230		&0.8			&0.39&	15	&4.60	&\nodata	&11	&2	\\
        L1251B		&1.00			&0.46&	18	&4.95	&\nodata	&14	&1	\\
        IRAS 03256+3055  &0.34			&0.23&	6.1	&\nodata&1.74		&16	&1	\\
        IRAS 04016+2610  &0.23			&0.21&	2.4	&\nodata&0.26		&16	&3	\\
        IRAS 04108+2803  &0.22			&0.20&	2.3	&\nodata&0.08		&16	&4	\\
        IRAS 04113+2758  &0.22\tablenotemark{c}	&0.20&	2.3     &\nodata&0.22		&16	&4	\\
	IRAS 04166+2706  &0.34			&0.23& 	2.3	&0.34   &\nodata	&13	&1	\\
        IRAS 04169+2702  &0.29			&0.22&	2.4	&0.37	&\nodata	&15	&1	\\
        IRAS 04239+2436  &0.26\tablenotemark{c}	&0.21&	2.4	&0.08	&\nodata	&19	&4	\\
        IRAS 04248+2612  &0.34\tablenotemark{c}	&0.23&	4.3	&0.14	&\nodata	&23	&4	\\
        IRAS 04264+2433  &0.26\tablenotemark{c}	&0.21&	2.4	&\nodata&0.05		&16	&4	\\
        IRAS 04295+2251  &0.3\tablenotemark{c}	&0.22&	1.1	&0.10	&\nodata	&16	&4	\\
        IRAS 04302+2247  &0.3\tablenotemark{c}	&0.22&	2.5	&\nodata&0.08		&16	&4	\\
        IRAS 04361+2547  &0.36\tablenotemark{c}	&0.24&	2.5	&0.16	&\nodata	&16	&4	\\
        IRAS 04381+2540  &0.36\tablenotemark{c}	&0.24&	2.8	&0.18	&\nodata	&15	&4	\\
        IRAS 04385+2550  &0.27\tablenotemark{c}  &0.22&	2.4	&\nodata&0.04		&16	&4	\\
\enddata

\tablenotetext{a}{$M_v^{\theta_{ap}}$ is calculated by Equation 8 in Paper III.}
\tablenotetext{b}{The envelope mass, $M_{env}$,  is based on the
		best-fit density distribution (per Equation 6 in Paper III) from which an isothermal temperature is 
		derived.  The average isothermal temperature (16K)
		is then used for the other sources to calculate the envelope mass, $M_{env}^{iso}$, via 
		the method derived by Hildebrand (1983)(Equation 4, Paper I).}
\tablenotetext{c}{These velocity linewidths are for the region in which these cores reside.  However,
		their positions are often several beamwidths from the IRAS source ($\sim$3-10$\arcmin$).}
\tablerefs{1. Mardones et al. 1997 (N$_2$H$^+$); 	2. Wang et al. 1995 (C$^{18}$O); 
		3. Fuller \& Myers 1993 (HC$_3$N); 4. Caselli et al. 2002 (N$_2$H$^+$). }
\end{deluxetable}

\begin{deluxetable}{lccccccl}
\tabletypesize{\footnotesize}
\tablecolumns{8}
\tablecaption{Shu Collapse Models\label{shu}}
\tablewidth{0pt} 
\tablehead{
\colhead{Source}      	&
\colhead{$r_{infall}$}     	&
\colhead{$a_{eff}$}          &
\colhead{$L_{int}$}	&
\colhead{$\chi^2_{450}$}	&
\colhead{$\chi^2_{850}$}	&
\colhead{$\chi^2_{SED}$}	&
\colhead{Comments}		\\
\colhead{}            		&
\colhead{AU}     		&
\colhead{km s$^{-1}$}       	   	&
\colhead{$L_\odot$}			&
\colhead{}			&
\colhead{}			&
\colhead{}			&
\colhead{}			
}
\startdata 
CB230		&1000	&0.39	&6.7	&2	&21	&4600	  &	\\
		& 2000& 0.39	& 6.7	& 0.8	& 18	& 2089	& Best-fit 	\\
		&3000	&0.39	&6.7	&0.7	&78	&1251	&	\\
		&4000	&0.39	&6.7	&1	&170	&880	&	\\
IRAS 04166+2706 &1000	&0.23	&0.4	&0.3	&3	&43	&	\\
		& 2000	& 0.23	& 0.4	& 0.2	& 0.8	& 36	& Best-fit	\\
		&3000	&0.23	&0.4	&0.8	&4	&33	&	\\
IRAS 04169+2702	&1000	&0.22	&0.8	&23	&58	&92	&	\\
		&2000	&0.22	&0.8	&6	&29	&72	 	&	\\
		& 3000	& 0.22	& 0.8	& 6	& 29	& 64 	 & Best-fit	\\
		&4000	&0.22	&0.8	&12	&49	&60	 &	\\
IRAS 04239+2436	& 500	& 0.21	& 0.9	& 0.5	& 1	& 682	& Best-fit	\\
		&1000	&0.21	&0.9	&0.3	&5	&488	 &	\\
		&2000	&0.21	&0.9	&3	&20	&328		&	\\
IRAS 04248+2612	& 10,000	& 0.23	& 0.6	& 6	& 13	& 146	  & Best-fit	\\
		&15,000	&0.23	&0.6	&7	&17	&140	  &	\\
IRAS 04295+2251	& 500	& 0.22	& 0.25	& 3	& 12	& 928	 & Best-fit	\\
		&1000	&0.22	&0.25	&10	&23	&759	&	\\
IRAS 04361+2547	&500	&0.24	&2.2	&8	&14	&359	&	\\
		& 1000	& 0.24	& 2.2	& 4	& 12	& 655	 & Best-fit	\\
		&1500	&0.24	&2.2	&2	&18	&546	&	\\
IRAS 04381+2540	& 3000	& 0.24	& 0.5	& 10	& 3	& 187	& Best-fit	\\
		&4000	&0.24	&0.5	&14	&2	&152	&	\\
		&5000	&0.24	&0.5	&17	&4	&131	&	\\
L1251B		&2000	&0.46	&10	&52	&161	&2316	&	\\
		& 3000	& 0.46	& 10	& 36	& 23	& 1989	& Best-fit	\\
		&4000	&0.46	&10	&28	&40	&1678	&	\\
\enddata
\end{deluxetable}

\begin{deluxetable}{lcccccccl}
\footnotesize
\tablecolumns{9}
\tablecaption{Power Law Models: Best-Fit Summary\label{bestpower}}
\tablewidth{0pt} 
\tablehead{
\colhead{Source}        &
\colhead{p}             &
\colhead{$n_f$}          &
\colhead{$r_i$}                         &
\colhead{$r_o$}                   &
\colhead{$\kappa_\nu$}                  &
\colhead{$L_{int}$}     &
\colhead{$s_{ISRF}$}    &
\colhead{$p_m$}         \\
\colhead{}                      &
\colhead{}              &
\colhead{($\times10^5cm^{-3}$)}                 &
\colhead{(AU)}          &
\colhead{($\times10^4$AU)}                 &
\colhead{}              &
\colhead{$L_\odot$}                     &
\colhead{}              
}
\startdata                                                              
CB230           &1.9    &22  &100       &10     &OH5    &6.7    &0.3    &2.5    \\
IRAS 04166+2706 &1.6    &3.0 &100       &3.3    &OH5    &0.4    &0.3    &2.4    \\
IRAS 04169+2702 &1.5    &2.8 &100       &3.3    &OH5    &0.77   &0.3    &2.4            \\
IRAS 04239+2436 &1.8    &0.9 &100       &3.3    &OH5    &0.9    &0.3    &2.1            \\
IRAS 04248+2612 &0.8    &0.35&100       &3.3    &OH5    &0.6    &0.3    &1.5                    \\
IRAS 04295+2251 &2.3    &2.0 &100       &3.3    &OH5    &0.25   &0.3    &\nodata                \\      
IRAS 04361+2547 &1.8    &1.8 &100       &3.3    &OH5    &2.2    &0.3    &2.3            \\
IRAS 04381+2540 &1.6    &1.8 &100       &3.3    &OH5    &0.53   &0.3    &1.7            \\
L1251B          &1.5    &12  &100       &7.2    &OH5    &10     &0.3    & 2.3           \\

\enddata

\end{deluxetable}

\end{document}